\documentclass{pasj02} 

\usepackage[switch,mathlines]{lineno} 

\jyear{2026}
\Received{2025/6/28}
\Accepted{2025/12/25}
\usepackage{threeparttable}
\usepackage{graphicx}
\usepackage{amsmath}
\usepackage{comment}
\usepackage{txfonts}
\newcommand{\ko}[1]{\textcolor{blue}{\bf#1}}
\newcommand{\rev}[1]{{\textcolor{black}{#1}}}
\newcommand{\revrev}[1]{{\textcolor{black}{#1}}}
\usepackage[rightcaption]{sidecap}
\usepackage{tablefootnote}
\usepackage{natbib}
\usepackage{booktabs}
\usepackage{multirow}
\usepackage{comment}
\usepackage{graphicx}
\usepackage{txfonts}
\usepackage{multicol}

\begin{document} 
\title{A Dichotomy of the Mass--Metallicity Relation of Exoplanetary Atmospheres Demarcated by their Birthplace}
\author{
 Kazumasa \textsc{Ohno},\altaffilmark{1,4}\altemailmark\orcid{0000-0003-3290-6758} \email{ohno.k.ab.715@gmail.com} 
 Masahiro \textsc{Ikoma},\altaffilmark{1,2,3,4}
 Satoshi \textsc{Okuzumi},\altaffilmark{5}
 and 
 Tadahiro \textsc{Kimura}\altaffilmark{6,7}
}
\altaffiltext{1}{Division of Science, National Astronomical Observatory of Japan, 2-21-1 Osawa, Mitaka, Tokyo 181-8588, Japan}
\altaffiltext{2}{Astrobiology Center, 2-21-1 Osawa, Mitaka, Tokyo 181-8588, Japan}
\altaffiltext{3}{Department of Earth and Planetary Science, The University of Tokyo, Hongo, Bunkyo-ku, Tokyo 113-0033, Japan}
\altaffiltext{4}{Graduate Institute for Advanced Studies, SOKENDAI, 2-21-1 Osawa, Mitaka, Tokyo 181-8588, Japan}
\altaffiltext{5}{Department of Earth and Planetary Sciences, Institute of Science Tokyo, Meguro, Tokyo 152-8551, Japan}
\altaffiltext{6}{UTokyo Organization for Planetary Space Science (UTOPS), University of Tokyo, Hongo, Bunkyo-ku, Tokyo 113-0033, Japan}
\altaffiltext{7}{Kapteyn Astronomical Institute, University of Groningen Landleven 12, 9747 AD, Groningen, Netherlands}



\maketitle

\begin{abstract} 
   Atmospheric observations by JWST raise a growing evidence that atmospheric metallicity exhibits an anti-correlation with masses of giant exoplanets.
   While such a trend was anticipated by planetesimal-based planet formation models, it remains unclear what kind of atmospheric metallicity trends emerge from pebble-based planet formation. 
   {Moreover, while recent studies of solar system Jupiter suggest that uppermost observable atmosphere may not represent the bulk envelope composition, it remains uncertain how the envelope inhomogeneity influences the atmospheric metallicity trend.}
   In this study, we develop disk evolution and planet formation models to investigate the possible atmospheric metallicity trends of giant exoplanets formed via pebble accretion and how they depend on the metallicity inhomogeneity within the envelope.
   We find that pebble-based planet formation produces two distinct mass--metallicity relations depending on planetary birthplace.
   Planets formed beyond the H$_2$O snowline exhibit a mass--metallicity anti-correlation similar to that predicted by planetesimal-based models if their atmospheres are fully convective.
   This anti-correlation disappears if the convective mixing is inefficient.
   In contrast, planets formed inside the H$_2$O snowline show a shallower mass--metallicity anti-correlation, regardless of the efficiency of atmospheric mixing.
   We test different initial disk properties and fragmentation threshold velocity of dust particles, demonstrating that the dichotomy of mass--metallicity relation is robust against these uncertainties.
   Many gas giants observed by JWST observations lie around the mass--metallicity relation predicted for formation at close-in orbits, although some planets with sub-stellar atmospheric metallicity appear to require unmixed envelopes and formation beyond the H$_2$O snowline.
   We also examine the relationship between bulk and atmospheric metallicity and find a clear correlation that closely follows atmospheric metallicity that is comparable to bulk metallicity.
   Our findings will help future survey of exoplanetary atmospheres by JWST and Ariel to shed light on where close-in giants come from on the basis of the mass--metallicity relation.   
\end{abstract}


\section{Introduction}
Atmospheric compositions of exoplanets are believed to provide insights into planetary formation and evolution. 
The James Webb Space Telescope (JWST) has demonstrated an unprecedented ability to constrain exoplanetary atmospheric compositions \citep[for review, see][]{Kempton&Knutson24,Espinoza&Perrin25} and already allowed us to search for population-level trends in atmospheric properties \citep[][]{Fu+25}.
The ESA's Ariel mission, scheduled for launch in 2029, also aims to observe approximately $1000$ exoplanetary atmospheres to identify population-level trends in atmospheric compositions \citep[e.g.,][]{Tinetti+21,Edwards&Tinetti22}. 
Many studies proposed the potential metric of atmospheric compositions to explore planet formation, including the C/O ratio \citep[e.g.,][]{Oberg+11,Madhusudhan+14,Cridland+19_CtoO}, the N/O ratio \citep[e.g.,][]{Piso+16,Cridland+20,Ohno&Fortney23a}, sulfur abundances \citep{Turrini+21,Pacetti+22,Crossfield23}, and the abundances of refractory elements \citep{Lothringer+21,Knierim+22,Chachan+23}.
Exoplanetary atmospheres would serve as one of the primary tools for investigating the origins and diversity of planetary systems in the coming decades.

Atmospheric metallicity---the abundance of elements heavier than helium---is a fundamental atmospheric property that encapsulates information about a planet's formation history.
Recent spectroscopic studies with JWST have revealed that many close-in gas giants possess atmospheres with super-stellar metallicities \citep[e.g.,][]{Alderson+23,Feinstein+23,Powell+24,Bean+23,Bell+23,Schlawin+24,Xue+23,Kirk+24}, \rev{though sub-stellar metallicities have also been reported for several exoplanets \citep[e.g.,][]{August+23,Fournier+24,Fournier+24_HAT-P-18b}.}
Several studies suggested that such enhanced metallicities may result from planetesimal bombardment followed by ablation within the protoatmosphere \citep[e.g.,][]{Pollack+86,Fortney+13,Mordasini+16,Venturini+16,Turrini+21,Shibata&Helled22,Pacetti+22}. 
Alternatively, recent studies proposed that super-stellar metallicity may also arise from the accretion of vapor-enriched disk gases, produced by the sublimation of inward-drifting pebbles \citep[e.g.,][]{Booth+17,Booth&Ilee19,Schneider&Bitsch21,Schneider&Bitsch21b,Danti+23,Penzlin+24}.

Previous studies suggested a possible anti-correlation between atmospheric metallicity and planetary mass \citep[e.g.,][]{Kreidberg+14_mass-metal,Wakeford+17,Welbanks+19,Sun+24}. 
This anti-correlation, called the mass--metallicity relation, is known to exist for giant planets in the Solar System \citep{Kreidberg+14_mass-metal}\footnote{Unless otherwise stated, we use the term ``mass--metallicity relation'' in this paper to refer to the relationship between planetary mass and {\it atmospheric} metallicity. This should not be confused with the relation between the planetary mass and {\it bulk} metallicity \citep[e.g.,][]{Thorngren+16}.}.
Before the launch of JWST, atmospheric observations struggled to compellingly identify this relation in exoplanets  \citep[e.g.,][]{Guillot+22} \rev{; for instance, \citet{Edwards+22} performed uniform atmospheric retrievals on the transmission spectra of 70 giant planets observed by the Hubble Space Telescope (HST) WFC3 G141 and found no clear relation between mass and atmospheric metallicity in their dataset. It remained unclear whether the lack of trend is real or due to the limited wavelength coverage of HST/WFC3}. 
Now, JWST enables us to constrain atmospheric metallicity with significantly improved precision \rev{and wavelength coverage}, hinting at the putative correlation \citep{Kempton&Knutson24}.
By comparing atmospheric models with JWST/NIRSpec transmission spectra of eight giant planets, \citet{Fu+25} demonstrated that the presence of a mass--metallicity correlation is statistically favored over the hypothesis of uniform atmospheric metallicity across all planets.
\rev{
Based on the compilation of JWST measurements, \citet{Lothringer+26} also reported significant Bayesian evidence for the mass-depending metallicity fit compared to constant-with-mass metallicity fit, though the measured metallicity exhibits a large scatter.
While the explorations with JWST is still ongoing, it is now vital to examine what we can learn about planet formation from the possible mass-metallicity relation.}

Previous studies suggested that planetesimal accretion plays a key role in shaping the mass--metallicity relation.
Using a population synthesis model, \citet{Fortney+13} predicted that atmospheric metallicity peaks at Earth to super-Earth masses and declines with increasing planetary mass (see also \citealp{Mordasini+16}).
Such anti-correlation emerges because the rate of disk gas accretion tends to overwhelm the planetesimal accretion rate as a planetary mass increases.
\citet{Cridland+19_CtoO} further showed that the vertical offset of the mass--metallicity relation depends on the planet's formation location due to radial variations in the surface density of planetesimals.
The mass–metallicity relation predicted by \citet{Fortney+13} appears to be consistent with current JWST observations \citep{Kempton&Knutson24}.

Although the aforementioned studies appear to explain the mass--metallicity relation, those studies adopted multiple simplifications that require further investigations to assess the possible metallicity trends.
For example, previous studies ignored the radial transport of dust in protoplanetary disks when calculating disk gas compositions, whereas it has been shown that inward-drifting pebbles can greatly enrich disk gases in vapors through sublimation, producing super-stellar metallicity gas inside snowlines \citep[e.g.,][]{Cuzzi&Zahnle04,Oberg&Bergin+16,Booth+17,Booth&Ilee19,Schneider&Bitsch21}.
Such vapor enrichment due to pebble drift has recently been suggested by observations of bright H$_2$O and CO$_2$ emission in several protoplanetary disks 
\citep[e.g.,][]{Banzatti+23,Gasman+23,Vlasblom+25}.
Although several recent studies started to investigate possible trends in planetary composition emerging from the pebble-driven gas enrichment \citep[][]{Schneider&Bitsch21,Schneider&Bitsch21b,Danti+23,Penzlin+24}, it remains unclear whether the pebble accretion scenario also produces a mass--metallicity anti-correlation.

Another aspect of the mass--metallicity relation less explored to date is the role of metallicity inhomogeneity within planetary envelope.
Previous studies have postulated that the envelope is fully convective, resulting in uniform metallicity throughout the envelope.
However, measurements of Jupiter's gravitational moments by the {\it JUNO} mission have revealed the presence of a diluted core accompanied by a significant metallicity gradient \citep[e.g.,][]{Wahl+17,Debras&Chabrier19}.
Moreover, \citet{Debras&Chabrier19} suggested that the bulk of the Jovian envelope possesses a lower metallicity than that constrained by {\it Voyager} observations for the upper atmosphere.
The existence of a detached radiative zone \citep{Guillot+94} may inhibit convective mixing between the upper atmosphere and the deeper envelope, potentially stabilizing such an inverted metallicity gradient  \citep{Howard+23,Muller&Helled23}.
If this universally occurs on giant planet interiors, the observable upper atmospheres may only record the final stages of planetary accretion, capturing just a fraction of the total heavy-element inventory.

In this study, we develop models of protoplanetary disk evolution and giant planet formation to investigate atmospheric metallicity trends expected under the pebble accretion paradigm.
We show that planets originating from close-in orbits and outer disk regions exhibit distinct mass--metallicity relations.
We also consider both fully mixed and unmixed planetary interiors as two end-member scenarios, thereby investigating how metallicity inhomogeneity within the envelope affects the mass--metallicity relation.
The organization of this paper is as follows.
In Section \ref{sec:method_disk}, we describe our disk model to simulate the evolution of disk compositions.
Section \ref{sec:method_planet} introduces our pebble-based planet formation model, which is coupled with the disk model to predict atmospheric elemental abundances.
In Section \ref{sec:result_disk}, we present basic results of the disk and planet formation models.
In Section \ref{sec:result_population}, we perform atmospheric population synthesis to investigate possible atmospheric metallicity trends in the pebble accretion scenario.
In Section \ref{sec:discussion}, we discuss the possible trends of other elemental abundances and model caveats.
Section \ref{sec:summary} summarizes our main findings.

\begin{figure*}[t]

\centering
\includegraphics[clip,width=0.85\hsize]{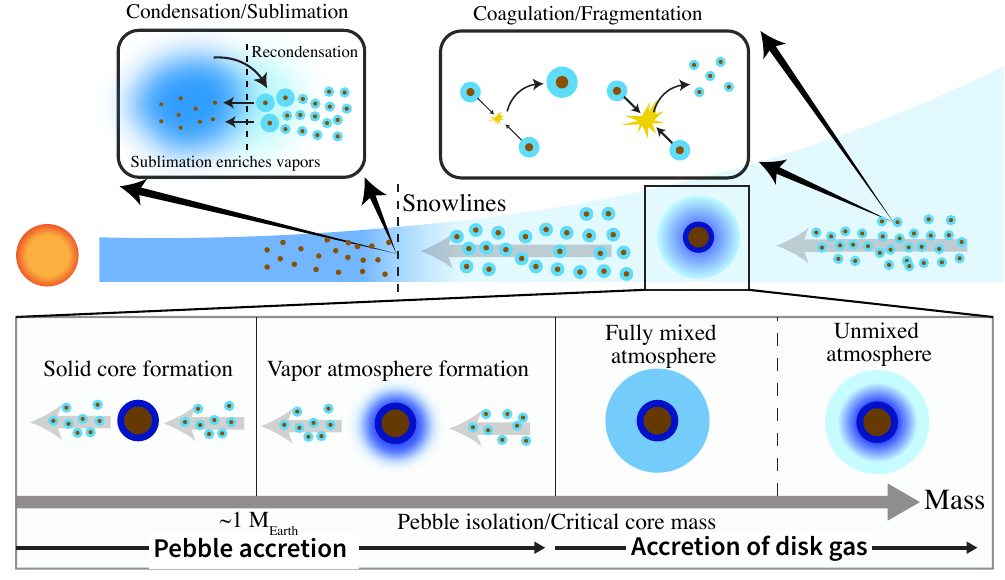}

\caption{Cartoon illustrating physical processes taken into account in our disk evolution and planet formation models.
}
\label{fig:Cartoon}
\end{figure*}
\section{Disk Model}\label{sec:method_disk}
We develop a 1D viscous accretion disk model that simulates the evolution of gas and dust surface densities.
The model takes into account collision growth and fragmentation of dust particles as well as condensation and sublimation of vapors and solid materials (ices and rocks), respectively, in a self-consistent manner (see Figure~\ref{fig:Cartoon} for a schematic illustration of our disk evolution and planet formation models).
We describe the model details in the following subsections. 
We discuss similarities and differences from previous pebble-based planet formation models \citep{Booth+17,Schneider&Bitsch21,Penzlin+24} in Section \ref{sec:discussion}.

\subsection{Temperature structure} 

The temperature of a protoplanetary disk is controlled by viscous heating and irradiation from the central star.
The disk mid-plane temperature is expressed by
\begin{equation}
T = \left(T_{\rm visc}^4 + T_{\rm irr}^4\right)^{1/4},
\end{equation}
where $T_{\rm visc}$ and $T_{\rm irr}$ are the mid-plane temperature determined by the viscous heating \citep[e.g.,][]{Nakamoto&Nakagawa94,Hueso&Guillot05,Garaud&Lin07,Oka+11} and stellar irradiation, respectively. 
The former is given by \citep{Mori+21}
\begin{equation}
T_{\rm visc} = \left[\left(\frac{9\dot{M}\Omega^2}{32\pi\sigma}\right)\left(\frac{\tau_{\rm mid}}{2} + \frac{1}{\sqrt{3}}\right)\right]^{1/4},
\end{equation}
where $\dot{M}$ is the disk gas accretion rate, $\Omega=\sqrt{GM_{\rm *}/r^3}$ is the Keplerian frequency, $M_{\rm *}$ is the mass of the central star, $G$ is the gravitational constant, $r$ is the radial distance from the central star, and $\sigma$ is the Stefan–Boltzmann constant. 
$\tau_{\rm mid}$ is the vertical optical depth to the disk mid-plane given by $\tau_{\rm mid} = \kappa_{\rm R}\Sigma_{\rm g,tot}/2$, where $\Sigma_{\rm g,tot}$ is the total gas surface density, and $\kappa_{\rm R}$ is the Rosseland mean opacity.  
As done in \citet{Kunitomo+20}, we adopt an analytic disk opacity given by
\begin{eqnarray}\label{eq:kappa}
    \nonumber
    \kappa_{\rm R}&=&2.25~{\rm cm^2~g^{-1}}\times \min{\left[ 1,\left(\frac{T}{150~{\rm K}}\right)^2\right]}\\
    &&\times \left[ 1-{\rm tanh}{\left(\frac{\max{(T-2000~{\rm K},0)}}{150~{\rm K}} \right)}\right]
\end{eqnarray}
We have slightly modified the bracket part from the original expression to ensure that the opacity drops after dust particles are completely sublimated.
Equation \eqref{eq:kappa} was derived to reproduce the Rosseland mean opacity which \citet{Nakamoto&Nakagawa94} obtained from the calculation for interstellar dust that follows the MRN size distribution with dust-to-gas mass ratio of $\sim0.01$ \citep{Pollack+85}.
We ignore the effects of grain growth on disk opacity, which affect the disk thermal structures \citep{Savvidou+20,Kondo+23}, for the sake of simplicity.

For the irradiation-dominated temperature, we assume an optically thick disk where the disk mid-plane is heated only by the thermal emission reprocessed from the heated dust particles at the disk surfaces \citep[e.g.,][]{Kusaka+70,Chiang&Goldreich97,Okuzumi+22}.
We follow \citet{Ida+16} to describe the irradiation-dominated temperature as
\begin{equation}\label{eq:Tirr_thick}
    T_{\rm irr} = 150 \left(\frac{r}{1~\rm{au}} \right)^{-3/7}\left(\frac{L_{\rm *}}{L_\odot}\right)^{2/7}\left(\frac{M_{\rm *}}{M_\odot}\right)^{-1/7}~\rm{K},
\end{equation}
where $L_{\rm *}$ is the stellar luminosity. 
\citet{Ida+16} derived Equation \eqref{eq:Tirr_thick} by fitting the radiative-equilibrium temperature structure obtained by a non-gray radiative transfer calculation of \citet{Oka+11}.
We ignore the stellar luminosity evolution and set to $L_{\rm *}=L_{\rm \odot}$ for simplicity.
In addition, we introduce a floor value of $T_{\rm 0}=10~{\rm K}$ for the midplane temperature following \citet{Hueso&Guillot05}.

\subsection{Gas disk evolution} 
In this study, we separately simulate the evolution of surface densities for hydrogen-helium gas and other vapors sublimated from dust particles. 
The time evolution of the hydrogen-helium surface density is described by
\begin{eqnarray}\label{eq:gas_viscous}
    \frac{\partial \Sigma_{\rm H}}{\partial t}=-\frac{1}{r}\frac{\partial}{\partial r}\left(ru_{\rm gas}\Sigma_{\rm H} \right)
\end{eqnarray}
where $u_{\rm gas}$ is the radial advection velocity of the disk gas driven by the angular momentum transport by shearing stress, given by \citep{Lynden-Bell&Pringle74}
\begin{equation}
    u_{\rm gas}=-\frac{3\nu_{\rm ss}}{r}\frac{\partial \ln{(r^{1/2}\nu_{\rm ss}\Sigma_{\rm g,tot}})}{\partial \ln{r}}
\end{equation}
where $\Sigma_{\rm g,tot}$ is the total gas (i.e., hydrogen-helium and other vapors) surface density, and $\nu_{\rm ss}$ is the effective viscosity due to disk turbulence.
Following \citet{Shakura+73}, we parameterize the turbulence viscosity as $\nu_{\rm ss} = \alpha_{\rm ss} c_{\rm s} h_{\rm g}$, where $c_{\rm s}=\sqrt{k_{\rm B}T/m_{\rm g}}$ is the isothermal sound speed, $k_{\rm B}$ is the Boltzmann constant, $m_{\rm g}$ is the mean molecular mass, and $h_{\rm g}=c_{\rm s}/\Omega$ is the scale height of the gas disk. 

\subsection{Vapor disk evolution}\label{sec:method_vapor}
We simulate the surface density evolution of vapor species for the predominant reservoirs of O, C, N, S, and other refractory elements such as P, K, and Na (see Appendix \ref{sec:appendix_A} for included chemical species and their material properties).
The evolution of each surface density is described by \citep[e.g.,][]{Schoonenberg&Ormel17}
\begin{equation}\label{eq:vap_evolve}
    \frac{\partial \Sigma_{{\rm vap},i}}{\partial t}=\frac{1}{r}\frac{\partial}{\partial r}\left[rD_{\rm gas}\Sigma_{\rm g,tot} \frac{\partial}{\partial r}\left(\frac{\Sigma_{{\rm vap},i}}{\Sigma_{\rm g,tot}} \right)-ru_{\rm gas}\Sigma_{{\rm vap},i} \right]-\left(\frac{d\Sigma_{{\rm ice},i}}{dt}\right)_{\rm cond/subl}
\end{equation}
where $\Sigma_{{\rm vap},i}$ is the vapor surface density for species $i$, $D_{\rm gas}$ is the gas diffusion coefficient which we set to $D_{\rm gas}=\nu_{\rm ss}$.
The second term on the right-hand side expresses the loss or production of vapors through condensation and sublimation, which is introduced in Section \ref{sec:condensation}.
At every time step, we update the total surface density $\Sigma_{\rm g,tot}=\Sigma_{\rm H}+\sum_{\rm i} \Sigma_{{\rm vap},i}$ and mean molecular mass $m_{\rm g}=\Sigma_{\rm g,tot}/[\Sigma_{\rm H}/m_{\rm H}+\sum_{\rm i} (\Sigma_{{\rm vap},i}/m_{{\rm vap},i})]$, where $m_{\rm H}=2.34~{\rm amu}$, assuming the solar H/He ratio, and $m_{{\rm vap},i}$ are the mean masses of hydrogen-helium mixture and each trace vapor molecule.

\subsection{Dust disk evolution}
We simulate the evolution of the surface density and the mean particle mass of dust particles with the characteristic size approximation \citep{Sato+16}. 
In this method, the dust mass distribution is assumed to follow a narrowly peaked mass distribution with a mean mass $m_{\rm p}$ at each orbital distance.
The mean particle mass is then given by $m_{\rm p}=\Sigma_{\rm d,tot}/\mathcal{N}_{\rm d}$, where $\Sigma_{\rm d,tot}$ and $\mathcal{N}_{\rm d}$ are the total surface density and the column number density of dust particles.
This methodology has also been utilized in cloud models for atmospheric science \citep[e.g.,][]{Ziegler85,Ferrier94,Ohno&Okuzumi18,Ormel&Min19} and well reproduces the evolution of the dust surface density in a simulation that fully solves a size distribution \citep{Sato+16}.

The evolution of the column number density is described by 
\begin{equation}\label{eq:Ndust_evolve}
\frac{\partial \mathcal{N}_{\rm d}}{\partial t}=\frac{1}{r}\frac{\partial}{\partial r}\left[rD_{\rm dust}\Sigma_{\rm g,tot}\frac{\partial}{\partial r}\left(\frac{\mathcal{N}_{\rm d}}{\Sigma_{\rm g,tot}}\right)-ru_{\rm dust}\mathcal{N}_{\rm d}\right] -\frac{\mathcal{N}_{\rm d}}{\tau_{\rm coll}}\left( \frac{\Delta m}{m_{\rm p}}\right),
\end{equation}
where $u_{\rm dust}$ is the radial velocity of dust particles given by \citep[e.g.,][]{Takeuchi&Lin02}
\begin{equation}\label{eq:v_dust}
u_{\rm dust} = -\frac{2\textrm{St}}{1+\textrm{St}^2}\eta r\Omega+\frac{1}{1+{\rm St}^2}u_{\rm gas},
\end{equation}
where ${\rm St}$ is the Stokes number of dust particle, which is the ratio of the stopping time to Kepler time given by
\begin{equation}
    {\rm St}=\frac{\pi}{2}\frac{\rho_{\rm p}a_{\rm p}}{\Sigma_{\rm g,tot}}\max{\left(1,\frac{4a_{\rm p}}{9\lambda_{\rm g}} \right)},    
\end{equation}
where $a_{\rm p}$ and $\rho_{\rm p}$ are the radius and internal density of dust particles, $\lambda_{\rm g}=m_{\rm g}/\sigma_{\rm g}\rho_{\rm g}$ is the mean free path of ambient gases, $\rho_{\rm g}=\Sigma_{\rm g,tot}/\sqrt{2\pi}h_{\rm g}$ is the gas mass density at a disk mid-plane, and $\sigma_{\rm g}$ is the molecular collision cross-section which we set $\sigma_{\rm g}=2\times{10}^{-19}~{\rm m^{2}}$ following \citet{Sato+16}.
$\eta$ is a dimensionless parameter that quantifies the sub-Kepler motion of the gas disk, defined as \citep{Adachi+76}
\begin{equation}\label{eq:eta}
    \eta \equiv -\frac{1}{2}\left(\frac{h_{\rm g}}{r}\right)^2 \left( \frac{d\ln{P}}{d\ln{r}}\right).
\end{equation}
$D_{\rm dust}$ is the diffusion coefficient for dust particles given by \citep{Youdin&Lithwick07}
\begin{equation}
    D_{\rm dust}=\frac{D_{\rm gas}}{1+{\rm St}^2}.
\end{equation}
The last term in Equation \eqref{eq:Ndust_evolve} stands for the decrease or increase in the number density through collisional growth or fragmentation, which we introduce in Section \ref{sec:collision}.

For the mass surface density of the dust particles, we simulate the evolution of each condensed ice and rock separately, as done for the vapor surface density.
The evolution of each ice (and rock) surface density is described by
\begin{equation}\label{eq:ice_evolve}
\frac{\partial \Sigma_{{\rm ice},i}}{\partial t} = \frac{1}{r}\frac{\partial}{\partial r}\left[rD_{\rm dust}\Sigma_{\rm g,tot}\frac{\partial}{\partial r}\left(\frac{\Sigma_{{\rm ice},i}}{\Sigma_{\rm g,tot}}\right)-ru_{\rm dust}\Sigma_{{\rm ice},i}\right] +\left(\frac{d\Sigma_{{\rm ice},i}}{dt}\right)_{\rm cond/subl}
\end{equation}
where $\Sigma_{{\rm ice},i}$ is the ice (or rock) surface density for species $i$.
The total dust surface density is given as $\Sigma_{\rm d,tot}=\sum_{\rm i} \Sigma_{{\rm ice},i}$, and particle internal density is given as $\rho_{\rm p}=\Sigma_{\rm d,tot}/\sum_{\rm i} (\Sigma_{{\rm ice},i}/\rho_{{\rm ice},i})$, where $\rho_{{\rm ice},i}$ is the material densities of the condensed ices/rocks for species $i$.
We ignore particle porosity for simplicity, since millimeter-wave polarimetry of protoplanetary disks suggests relatively compact dust aggregates \citep{Tazaki+19}.

\subsection{Condensation and sublimation}\label{sec:condensation}
Condensation and sublimation exchange molecules between the vapor and solid phases.
The rate of change in local ice mass density for species $i$ is given by
\begin{equation}\label{eq:drho_ice}
    \frac{d\rho_{{\rm ice},i}}{dt}= n_{\rm d}\frac{dm_{{\rm ice},i}}{dt},
\end{equation}
where the rate of change in individual particle mass is given by \citep{Ciesla&Cuzzi06}
\begin{equation}\label{eq:dm_ice}
    \frac{dm_{{\rm ice},i}}{dt}= \pi a_{\rm p}^2 v_{\rm th,i}\left( \rho_{\rm v,i}-\frac{m_{\rm v,i}P_{\rm sat}}{k_{\rm B}T}\right),
\end{equation}
where $a_{\rm p}$ and $n_{\rm d}$ are the radius and number density of dust particles, $v_{\rm th,i}=\sqrt{8k_{\rm B}T/\pi m_{\rm v,i}}$ and $m_{\rm v,i}$ are the mean thermal velocity and masses of vapor molecules, $\rho_{\rm v,i}$ is the vapor mass density, $P_{\rm sat}$ is the saturation vapor pressure. 
Assuming that the vapor is uniformly mixed in vertical, the local vapor mass density in a hydrostatic gas disk is given by
\begin{equation}\label{eq:rhov}
    \rho_{\rm v,i}=\frac{\Sigma_{{\rm vap},i}}{\sqrt{2\pi}h_{\rm g}}\exp{\left( -\frac{z^2}{2h_{\rm g}^2}\right)}.
\end{equation}
Similarly, the dust number density is given by
\begin{equation}\label{eq:nd}
    n_{\rm d}=\frac{\Sigma_{\rm d,tot}}{\sqrt{2\pi}h_{\rm d}m_{\rm p}}\exp{\left( -\frac{z^2}{2h_{\rm d}^2}\right)},
\end{equation}
where $h_{\rm d}$ is the dust-disk scale height given by \citep{Youdin&Lithwick07,Okuzumi+12}
\begin{equation}
    h_{\rm d}=h_{\rm g}\left( 1+\frac{{\rm St}}{\alpha_{\rm ss}}\frac{1+2{\rm St}}{1+{\rm St}}\right)^{-1/2}.
\end{equation}
Integrating Equation \eqref{eq:drho_ice} vertically with Equations \eqref{eq:rhov} and \eqref{eq:nd}, we obtain the rate of change in the ice surface density as
\begin{equation}\label{eq:stammler_ice}
\left( \frac{d\Sigma_{{\rm ice},i}}{dt}\right)_{\rm cond/subl}=\frac{\Sigma_{\rm d,tot}a_{\rm p}^2}{m_{\rm p}}(R_{\rm c}-R_{\rm e}),
\end{equation}
where $R_{\rm c}$ and $R_{\rm e}$ are condensation and sublimation rates given by
\begin{equation}
R_{\rm c}=2\frac{\Sigma_{{\rm vap},i}}{h_{\rm g}}\sqrt{\frac{k_{\rm B}T}{m_{\rm v,i}}}\left( 1+\frac{h_{\rm d}^2}{h_{\rm g}^2}\right)^{-1/2}
\end{equation}
\begin{equation}
R_{\rm e}=2\sqrt{2\pi}\sqrt{\frac{m_{\rm v,i}}{k_{\rm B}T}}P_{\rm sat}.
\end{equation}
This formula is equivalent to that adopted by \citet{Hyodo+21} in the limit of $(h_{\rm d}/h_{\rm g})^2\rightarrow 0$. 
\rev{We note that our model calculates the condensation/sublimation rate for the mass-dominating large grains, whereas actual condensation/sublimation should preferentially occur on smaller grains that dominate total surface areas.
Since condensation/sublimation timescale is faster at smaller grains, condensation/sublimation presumably takes place in more confined orbits around snowlines than what our model predicts in reality. 
However, it would not change the picture of the global disk gas enrichment through the sublimation of drifting dust.
}

\subsection{Collision growth and fragmentation}\label{sec:collision}
Collision growth and fragmentation play a major role in controlling the dust particle sizes.
The collision rate is expressed by the collision timescale given by \citep{Sato+16}
\begin{equation}
    \tau_{\rm coll}=\frac{h_{\rm d}}{2\sqrt{\pi} a_{\rm p}^2 \Delta v \mathcal{N}_{\rm d}},
\end{equation}
where $\Delta v$ is the relative collision velocity, which is given by the root-sum-square of multiple-velocity components, as
\begin{equation}
    \Delta v=\sqrt{\Delta v_{\rm B}^2+\Delta v_{\rm r}^2+\Delta v_{\rm \phi}^2+\Delta v_{\rm z}^2+\Delta v_{\rm t}^2},
\end{equation}
where $\Delta v_{\rm B}$, $\Delta v_{\rm r}$, $\Delta v_{\rm \phi}$, $\Delta v_{\rm z}$, and $\Delta v_{\rm t}$ are the relative velocity driven by Brownian motion, radial drift, azimuthal drift, vertical settling, and turbulence, respectively \citep[see e.g.,][]{Okuzumi+12}. 
For the pairwise collision between similar-sized particles, the Brownian motion gives a relative velocity of $\Delta v_{\rm B}=\sqrt{16k_{\rm B}T/\pi m_{\rm p}}$.
The radial and azimuthal drift induced velocities are given by \citep[e.g.,][]{Adachi+76}
\begin{equation}
    \Delta v_{\rm r}=\left|\frac{2{\rm St}}{1+{\rm St}^2}-\frac{2\epsilon{\rm St}}{1+(\epsilon{\rm St})^2}\right|\eta v_{\rm K}
\end{equation}
\begin{equation}
    \Delta v_{\rm \phi}=\left|\frac{1}{1+{\rm St}^2}-\frac{1}{1+(\epsilon{\rm St})^2}\right|\eta v_{\rm K}
\end{equation}
where $\epsilon=0.5$ is the dimensionless parameter that accounts for the dispersion of the particle size distribution \citep{Sato+16}.
For the velocity driven by vertical settling, we first consider the vertical-dependent relative velocity given by \citep{Brauer+08}
\begin{equation}\label{eq:vz1}
    \Delta v_{\rm z}(z)=\left(\frac{{\rm St}}{1+{\rm St}}-\frac{\epsilon{\rm St}}{1+\epsilon{\rm St}}\right)\Omega_{\rm k}z .
\end{equation}
Taking a weighting average of Equation \eqref{eq:vz1} with dust mass density in vertical, we obtain
\begin{eqnarray}
    \Delta v_{\rm z}&=&\left(\frac{{\rm St}}{1+{\rm St}}-\frac{\epsilon{\rm St}}{1+\epsilon{\rm St}}\right)\frac{\Omega_{\rm k}h_{\rm d}}{\sqrt{\pi}}.
\end{eqnarray}
We calculate the turbulence-driven collision velocity $\Delta v_{\rm t}$ based on Equations (16)--(18) of \citet{Ormel&Cuzzi07}.

We take fragmentation into account by calculating the change in particle mass through collision $\Delta m$ that depends on the collision velocity.
Following \citet{Okuzumi+16}, we model $\Delta m$ as \citep{Okuzumi&Hirose12}
\begin{equation}
    \frac{\Delta m}{m_{\rm p}}=\min{\left[ 1, -\frac{\ln{(\Delta v/v_{\rm frag})}}{\ln{(5)}}\right]},
\end{equation}
where $v_{\rm frag}$ is the fragmentation threshold velocity.
The exact value of the fragmentation threshold velocity is controversial.
Conventionally, icy particles are supposed to be sticky and avoid fragmentation at collision velocities up to $\sim 10$--$50~{\rm m~s^{-1}}$ \citep[e.g.,][]{Wada+13,Gundlach&Blum15}.
However, recent experimental studies suggest that H$_2$O ice is much more fragile than previously thought \citep[][]{Musiolik&Wurm19}, and other ices such as CO$_2$ are also fragile \citep{Musiolik&Wurm16,Fritscher&Teiser22}.
Such fragile icy particles are also suggested by comparisions of several disk observations with grain growth models \citep{Okuzumi&Tazaki19,Jiang+24,Ueda+24,Yoshida+25}.
Unless otherwise indicated, we assume $v_{\rm frag}=1~{\rm m~s^{-1}}$ as the fiducial parameter in this study.

\subsection{Numerical Procedures and Initial conditions}
We solve Equations \eqref{eq:gas_viscous}, \eqref{eq:vap_evolve}, \eqref{eq:Ndust_evolve}, and \eqref{eq:ice_evolve} with the first-order implicit method.
Following \citet{Okuzumi+16}, we set the initial profile of the hydrogen-helium gas surface density to be a power-law function with an exponential taper given by 
\begin{equation}\label{eq:sigma_ini}
    \Sigma_{\rm H}=\frac{M_{\rm disk}}{2\pi r_{\rm c}^2}\left( \frac{r}{r_{\rm c}}\right)^{-1}\exp{\left[ -\left( \frac{r}{r_{\rm c}}\right)\right]},
\end{equation}
where $r_{\rm c}$ is the characteristic radius and $M_{\rm disk}$ is the initial disk mass.
For the trace chemical components, we partition all volatiles into their respective solid phases at the beginning of the simulation.
The initial dust surface density for each chemical component is given by
\begin{eqnarray}
    \nonumber
    \Sigma_{{\rm ice},i}&=&{10}^{\rm [Fe/H]}\times m_{\rm v,i}f_{\rm i}\frac{N_{\rm H}}{N_{\rm H}/2+N_{\rm He}}\frac{\Sigma_{\rm H}}{m_{\rm H}},
\end{eqnarray}
where ${\rm [Fe/H]}$ is the base-10 log of initial disk metallicity normalized by the solar metallicity, and $f_{\rm i}$ is the number of each molecule per hydrogen atom, which is summarized in Appendix \ref{sec:appendix_A}.
We remind the reader that $\Sigma_{\rm H}$ and $m_{\rm m}$ stand for the surface density and mean mass of H$_2$-He mixture.
For the initial dust surface density, we introduce a truncation radius $r_{\rm d}=3r_{\rm c}$ to help computational stability.
We adopt the zero-flux outer boundary condition, while gas and dust are allowed to flow out from the computation domain with inward velocity of $3\nu/2r$ and $u_{\rm dust}$ at the inner boundary.
We assume a solar-mass star with solar luminosity throughout this paper.

\section{Planet Formation Model}\label{sec:method_planet}
Giant planet formation takes place with multiple steps including core formation, envelope formation, and orbital migration \citep[e.g.,][]{Youdin&Zhu25,Ikoma&Kobayashi25}.
In this study, we simulate core formation through pebble accretion followed by subsequent gas accretion and orbital migration (Figure \ref{fig:Cartoon}).
Based on the disk model in Section \ref{sec:method_disk}, we simulate the evolution of the atmospheric elemental abundances, namely, O/H, C/H, N/H, S/H, P/H, K/H, N/H, Si/H, Mg/H, Fe/H, Ti/H, and V/H.
For computing the observable atmospheric compositions, we consider two end-member scenarios: the envelope is fully mixed or unmixed.
Note that this study ignores the feedback of planet formation on the disk structure, including the pebble trap at the pressure bump produced by the planet and the consumption of disk gas through the envelope formation, as discussed in in Section \ref{sec:model_caveat}.

\subsection{Pebble Accretion}
The planetary core grows by capturing aerodynamically small (${\rm St}{\lesssim}1$) solid particles---often called pebbles. 
The core growth rate can be expressed by
\begin{equation}\label{eq:dMcore/dt}
    \frac{dM_{\rm peb}}{dt}=P_{\rm coll}\Sigma_{\rm d,tot},
\end{equation}
where $P_{\rm coll}$ is the accretion area per time, which can be in general expressed as \citep[][]{Guillot+14,Ida+16}
\begin{equation}\label{eq:P_coll_general}
    P_{\rm coll}=\min{\left( \sqrt{\frac{8}{\pi}}\frac{h_{\rm d}}{b},1 \right)}\times \frac{\pi}{2}\frac{b^2}{h_{\rm d}}\Delta v_{\rm peb}
\end{equation}
where $\Delta v_{\rm peb}$ is the approach velocity of the pebbles to the planet, and $b$ is the linear cross section of the collision.
In Equation \eqref{eq:P_coll_general}, the prefactor accounts for the transition between the 2D and 3D pebble accretion, which is controlled by the ratio of the pebble scale height to the linear cross section.

Pebble accretion can be demarcated into the Bondi regime and the Hill regime, depending on the main driver of the pebble approach velocity $\Delta v_{\rm peb}$ \citep[][]{Ormel&Klahr10,Lambrechts&Johansen12,Ormel17}.
We calculate the cross section as $b=\min{\left[b_{\rm Bondi}, b_{\rm Hill}\right]}$.
The Bondi regime applies when the deviation from the Keplerian motion ($\Delta v_{\rm peb}\sim \eta r\Omega$) dominates $\Delta v_{\rm peb}$, which yields the linear cross section given by \citep{Lambrechts&Johansen12,Guillot+14,Ida+16}
\begin{equation}
    b_{\rm Bondi}\approx \sqrt{ \frac{12{\rm St}R_{\rm H}^3}{\eta r}},
\end{equation}
where $R_{\rm H}=r(M_{\rm p}/3M_{\rm *})^{1/3}$ is the Hill radius of the planet.
The Hill regime applies when the relative velocity is dominated by Keplerian shear ($\Delta v_{\rm peb}\sim b \Omega$).
For the Hill regime, we adopt the formula of \citet{Okamura&Kobayashi21} who provides the accretion area that smoothly connects the 3D and 2D accretion regimes as
\begin{equation}\label{eq:P_coll_3D}
    P_{\rm Hill}=\left[ P_{\rm Hill,2D}^{-2}+\left( P_{\rm Hill,2D}\frac{b_{\rm Hill}}{0.65h_{\rm d}}\right)^{-2}\right]^{-1/2}.
\end{equation}
\citet{Okamura&Kobayashi21} demarcated  the accretion area into 3 regimes as
\begin{equation}
    P_{\rm coll,2D}=\min{(P_{\rm coll,set},P_{\rm coll,ss},P_{\rm col,ho})}.
\end{equation}
Based on an analytical argument, \citet{Okamura&Kobayashi21} derived a 2D collision rate in each regime that is calibrated by the orbital simulations of pebbles coupled with hydrodynamic simulations for gas flows around the planet.
When the relative velocity of pebbles is slower than the sound speed, the accretion area is
\begin{equation}\label{eq:Pcoll_set}
    \frac{P_{\rm coll,set}}{R_{\rm H}^2\Omega}=3(2C_{\rm 1}{\rm St})^{2/3},
\end{equation}
where $C_{\rm 1}=1.5$ is an order unity calibration factor.
If the relative velocity is supersonic, the change of the gas drag law yields an alternative formula given by
\begin{equation}
    \frac{P_{\rm coll,ss}}{R_{\rm H}^2\Omega}=2\sqrt{6}C_{\rm 1}\left(\frac{\rho_{\rm p}}{\rho_{\rm g}}\frac{c_{\rm s}}{R_{\rm H}\Omega}\frac{\lambda_{\rm mfp}}{R_{\rm H}}{\rm St}\right)^{1/4}.
\end{equation}
Recent hydrodynamic simulations coupled with the orbital simulations of pebbles suggested that the core induces outflow to hinder the pebble accretion, called outflow barrier \citep{Kuwahara+19,Kuwahara&Kurokawa20,Kuwahara&Kurokawa20b}.
\citet{Okamura&Kobayashi21} obtained an analytic formula for the collision area including this effect as
\begin{equation}\label{eq:Pcoll_flow}
    \frac{P_{\rm coll,ho}}{R_{\rm H}^2\Omega}=2\frac{R_{\rm B}}{R_{\rm H}}\left[ 3{\rm St}\left( \frac{R_{\rm H}}{R_{\rm B}}\right)^{2} -\xi \sqrt{\frac{3R_{\rm H}}{R_{\rm B}}}\right]
\end{equation}
where $R_{\rm B}=GM_{\rm p}/c_{\rm s}^2$ is the Bondi radius 
, and $\xi c_{\rm s} =0.1(R_{\rm B}/h_{\rm g})$ is the core-driven outflow velocity, for which the parameter dependence comes from the analytic argument of \citet{Kuwahara+19}.
The impact parameter is then given by
\begin{equation}\label{eq:xss}
    b_{\rm Hill} =
\left\{
\begin{array}{ll}
{\displaystyle \sqrt{P_{\rm coll,ss}/3\Omega} } & (\min{(P_{\rm col,ss},P_{\rm col,set})}< P_{\rm col,ho}) \\[1.5ex]
{\displaystyle 2R_{\rm B} } & (\min{(P_{\rm col,ss},P_{\rm col,set})}> P_{\rm col,ho}),
         \end{array}
\right.
\end{equation}
Note that \citet{Okamura&Kobayashi21} also summarized the formula applicable to ${\rm St}>1$, but we have only used the formula for ${\rm St}<1$ in this study, as our current model does not produce particles with ${\rm St}>1$ due to radial drift and fragmentation.

Pebble accretion ceases once the planetary mass exceeds the pebble isolation mass, above which the core creates a pressure maxima around the planet that filters the pebbles to prevent further accretion \citep{Lambrechts+14}.
We set $P_{\rm coll}=0$ when the planetary mass exceeds the pebble isolation mass given by \citep{Bitsch+18}
\begin{equation}
    M_{\rm iso}=25M_{\rm \oplus}~\left( \frac{h_{\rm g}/r}{0.05}\right)^{3}\left[ 0.34\left(\frac{-3}{\log{\alpha}} \right)^4 + 0.66 \right]\left[ 1-\frac{1}{6}\left(\frac{d\ln{P}}{d\ln{r}}+2.5\right)\right].
\end{equation}
We can also estimate the mass above which the planet-driven outflow prevents further pebble accretion, which is proposed by \citet{Kuwahara+19}.
Based on Equation \eqref{eq:Pcoll_flow}, we solve $3{\rm St}(R_{\rm H}/R_{\rm B})^2=\xi \sqrt{3R_{\rm H}/R_{\rm B}}$ with respect to $M_{\rm p}$, thereby obtaining the isolation mass due to the outflow barrier as
\begin{equation}\label{eq:M_flow}
    M_{\rm flow}=M_{\rm *}\left( \frac{{\rm St}}{0.1}\right)^{1/2}\left( \frac{h_{\rm g}}{r}\right)^{3}
    \approx4M_{\rm \oplus} \left( \frac{{\rm St}}{{10}^{-3}}\right)^{1/2}\left( \frac{h_{\rm g}/r}{0.05}\right)^{3}\left( \frac{M_{\rm *}}{M_{\rm \odot }}\right).
\end{equation}
Equation \eqref{eq:M_flow} is equivalent to the threshold of $R_{\rm B}/h_{\rm g}\sim \sqrt{{\rm St}}$ proposed by \citet{Kuwahara+19}.
Importantly, the outflow barrier halts the pebble accretion especially when pebbles have a small Stokes number.

\subsection{Envelope accretion}



Once the core mass reaches the critical mass or the isolation mass, we allow the core to capture the surrounding disk gases to form an envelope.
The critical core mass is given by \citep{Ikoma+00}
\begin{equation}
    M_{\rm cri}=7M_{\rm \oplus}~\left( \frac{d{M}_{\rm peb}/dt}{10^{-7}M_{\rm\oplus}~{\rm yr}^{-1}}\right)^{0.25}\left( \frac{\kappa_{\rm env}}{1~{\rm cm}^2~{\rm g}^{-1}}\right)^{0.25},
\end{equation}
where $\kappa_{\rm env}$ is the averaged envelope opacity, which we set to $\kappa_{\rm env}=0.03~{\rm cm^2~g^{-1}}$ by accounting for the opacity reduction due to grain growth \citep{Ormel14,Mordasini14}. 
Once the core is isolated from the pebble accretion so that $dM_{\rm peb}/dt\rightarrow0$, the pebble isolation mass corresponds to the critical core mass.
For planetary mass of $M_{\rm p}>M_{\rm cri}$, we simulate the evolution of the envelope mass $M_{\rm env}$ as
\begin{equation}\label{eq:dMdt}
    \frac{dM_{\rm env}}{dt}=\min{[ \dot{M}_{\rm KH},\dot{M}_{\rm hydro} ]}
\end{equation}
where $\dot{M}_{\rm KH}$ and $\dot{M}_{\rm hydro}$ are the envelope accretion rate regulated by the Kelvin-Helmholtz contraction and hydrodynamic gas flow, as explained below.

The rate of envelope accretion is controlled by the bottleneck process of gas supply.
At the beginning of the runaway gas accretion, the rate is regulated by how fast the envelope can cool to contract \citep[e.g.,][]{Bodenheimer&Pollack86,Pollack+96,Ikoma+00}.
The accretion rate can be then estimated as
\begin{equation}
    \dot{M}_{\rm KH}=\frac{M_{\rm p}}{\tau_{\rm KH}},
\end{equation}
where $\tau_{\rm KH}$ is the Kelvin-Helmholtz timescale given by \citep{Ikoma+00}
\begin{equation}\label{eq:Mdot_KH}
    \tau_{\rm KH}\approx 1~{\rm Myr}~\left( \frac{M_{\rm p}}{M_{\rm \oplus}}\right) ^{-2.5}\left( \frac{\kappa_{\rm env}}{0.01~{\rm {cm}^2~g^{-1}}}\right).
\end{equation}
Since the accretion rate rapidly increases with increasing planetary mass, the gas supply from the surrounding disk medium eventually becomes the next bottleneck of envelope accretion.
The accretion rate regulated by the gas supply from the disk can be described by \citep{Tanigawa&Ikoma07,Tanigawa&Tanaka16}
\begin{equation}\label{eq:Mdot_hydro}
    \dot{M}_{\rm hydro}=D\Sigma_{\rm gap},
\end{equation}
where $D$ is the accretion area per unit time, and $\Sigma_{\rm gap}$ is the gas surface density at the accretion channel in the protoplanetary disk.
The gas accretion rate in this regime can be demarcated into three sub-regimes in terms of the ratio of the planet's Hill radius to disk scale height \citep{Ginzburg&Chiang19,Rosenthal+20,Choksi+23}.
We model the accretion area per time as
\begin{equation}
    D=\frac{1}{1/D_{\rm Bondi}+1/D_{\rm Hill,3D}+1/D_{\rm Hill,2D}},
\end{equation}
where $D_{\rm Bondi}$ is the accretion area for $R_{\rm H}\ll h_{\rm g}$ (Bondi regime, \citealt{Ginzburg&Chiang19}), $D_{\rm Hill,3D}$ is the area for $R_{\rm H}\sim h_{\rm g}$ (3D Hill regime, \citealt{Rosenthal+20}), and $D_{\rm Hill,2D}$ is the area for $R_{\rm H}\gg h_{\rm g}$ (2D Hill regime, \citealt{Rosenthal+20}).
We obtain the accretion area for each regime from the semi-analytical gas accretion rate of \citet{Choksi+23} calibrated by a 3D isothermal hydrodynamical simulation, given by
\begin{equation}
    D_{\rm Bondi}=\frac{3.5}{\sqrt{2\pi}}\left(\frac{M_{\rm p}}{M_{\rm *}}\right)^{2}\left( \frac{h}{r_{\rm p}}\right)^{-4}r_{\rm p}^2\Omega_{\rm K}
\end{equation}
\begin{equation}
    D_{\rm Hill,3D}=\frac{4}{3\sqrt{2\pi}}\left(\frac{M_{\rm p}}{M_{\rm *}}\right)\left( \frac{h}{r_{\rm p}}\right)^{-1}r_{\rm p}^2\Omega_{\rm K}
\end{equation}
\begin{equation}
    D_{\rm Hill,2D}=\frac{9}{3^{2/3}\sqrt{2\pi}}\left(\frac{M_{\rm p}}{M_{\rm *}}\right)^{2/3}r_{\rm p}^2\Omega_{\rm K}.
\end{equation}
As the planet grows, it opens a gap within a gas disk and reduces the envelope accretion rate \citep{Tanigawa&Ikoma07,Tanigawa&Tanaka16}.
The gap surface density can be expressed by \citep{Kanagawa+18}
\begin{equation}\label{eq:sigma_gap}
    \Sigma_{\rm gap}=\frac{\Sigma_{\rm g,tot}}{1+0.04K}\left[ 1+\frac{D/(1+0.04K)}{3\pi \nu_{\rm SS}}\right]^{-1},
\end{equation}
with
\begin{equation}\label{eq:Kanagawa_parameter}
    K=\left( \frac{h_{\rm g}}{r_{\rm p}}\right)^{-5}\left( \frac{M_{\rm p}}{M_{\rm *}}\right)^2\alpha^{-1}.
\end{equation}
The second term in the bracket in Equation \eqref{eq:sigma_gap} takes into account the gas consumption by envelope accretion onto the planet following \citet{Tanaka+20}.
This term ensures that envelope accretion cannot exceed the rate of global disk accretion, i.e., $\dot{M}_{\rm hydro}\leq 3\pi \Sigma_{\rm g,tot}\nu_{\rm ss}$.
To clarify this, one can substitute Equation \eqref{eq:sigma_gap} into \eqref{eq:Mdot_hydro} with $D/(1+0.04K)\gg 3\pi \nu_{\rm ss}$ and find
\begin{equation}
    \dot{M}_{\rm hydro}\approx D\frac{\Sigma_{\rm g,tot}}{1+0.04K}\frac{3\pi \nu_{\rm ss}(1+0.04K)}{D}=3\pi \Sigma_{\rm g,tot}\nu_{\rm ss}.
\end{equation}

\subsection{Evaluation of Atmospheric Compositions}\label{sec:method_calc_composition}
We simulate the evolution of planetary compositions through both pebble and disk gas accretion.
To this end, we simulate the time evolution of the total masses of species $i$ within the core ($M_{\rm core, i}$) and the envelope ($M_{\rm env, i}$) as
\begin{equation}\label{eq:dMcore_i}
    \frac{dM_{\rm core,i}}{dt}=\frac{dM_{\rm peb}}{dt}\left(\frac{\Sigma_{{\rm ice},i}}{\Sigma_{\rm d,tot}}\right)\mathcal{H}(M_{\rm \oplus}-M_{\rm core}),
\end{equation}
\begin{equation}\label{eq:dMenv_H}
    \frac{dM_{\rm env,H}}{dt}=\frac{dM_{\rm env}}{dt}\left(\frac{\Sigma_{\rm H}}{\Sigma_{\rm g,tot}}\right),
\end{equation}
\begin{equation}\label{eq:dMenv_i}
    \frac{dM_{\rm env,i}}{dt}=\frac{dM_{\rm peb}}{dt}\left(\frac{\Sigma_{{\rm ice},i}}{\Sigma_{\rm d,tot}}\right)\mathcal{H}(M_{\rm core}-M_{\rm \oplus})+\frac{dM_{\rm env}}{dt}\left(\frac{\Sigma_{{\rm vap},i}}{\Sigma_{\rm g,tot}}\right),
\end{equation}
where $M_{\rm core}=\sum_i {M_{{\rm core},i}}$, and $\mathcal{H}(x)$ is the step function defined as $\mathcal{H}(x)=1$ for $x>0$ and  $\mathcal{H}(x)=0$ for $x<0$.
In Equation \eqref{eq:dMenv_i}, the first term  expresses the deposition of vapors sublimated from accreted pebbles, and the second term expresses the deposition of vapors contained in the disk gas.
We assume that pebbles fully sublimate and dissolve into the envelope once the core mass exceeds the Earth mass, which is motivated by \citet{Brouwers+18} who showed that pebbles cannot reach the core due to ablation at $M_{\rm core}\gtrsim0.5M_{\rm \oplus}$.
\rev{We also tested a threshold mass of $0.1M_{\rm \oplus}$ and confirmed that it only has a minor impact on the simulated mass-metallicity relation.}

For comparisons with atmospheric observations, we calculate the atmospheric elemental abundances as follows.
For fully mixed envelope, we first calculate the total number of hydrogen $N_{\rm H}$ and other heavy elements within the envelope $N_{\rm Z}$ as
\begin{equation}
    N_{\rm H} = \frac{2M_{\rm env,H}}{(m_{\rm H2}+2m_{\rm He}N_{\rm He}/N_{\rm H})}=\frac{M_{\rm env,H}}{1.3~{\rm amu}},
\end{equation}
\begin{equation}
    N_{\rm Z} = \sum_{i\in Z} x_{i}\frac{M_{{\rm env},i}}{m_{i}},
\end{equation}
where $x_{\rm i}$ is the stoichiometric coefficient of molecule $i$ for heavy element $Z$. 
For example, the total number of oxygen is calculated as
\begin{eqnarray}
    \nonumber
    N_{\rm O} &=& \frac{M_{\rm env,H_2O}}{m_{\rm H_2O}}+\frac{M_{\rm env,CO}}{m_{\rm CO}}+2\frac{M_{\rm env,CO_2}}{m_{\rm CO_2}}\\
    \nonumber
    &&+4\frac{M_{\rm env,Mg_2SiO_4}}{m_{\rm Mg_2SiO_4}}+\frac{M_{\rm env,SiO}}{m_{\rm SiO}}+\frac{M_{\rm env,TiO}}{m_{\rm TiO}}+\frac{M_{\rm env,VO}}{m_{\rm VO}}
\end{eqnarray}
Then, one can readily calculate the heavy elemental abundances as ${\rm Z/H}=N_{\rm Z}/N_{\rm H}$.

For the unmixed envelope, the present-day atmosphere only records the compositions of planetary building blocks that accreted at the end of the planetary assembly.
In this limit, one can estimate the elemental abundance from the ratio of accretion rates of heavy elements to hydrogen as
\begin{equation}    
    \frac{N_{\rm Z}}{N_{\rm H}}
    =
    \left(\sum_{i \in Z}{ x_{i}\frac{\dot{M}_{{\rm env},i}}{m_{i}} }\right)/\left(\frac{\dot{M}_{\rm env,H}}{1.3~{\rm amu}}+2\frac{\dot{M}_{\rm env,H_2O}}{18~{\rm amu}}\right),
\end{equation}
where we have taken into account the contribution of hydrogen supplied by H$_2$O so that Z/H value takes a finite value even if the gas accretion does not take place.

\subsection{Orbital Migration}
In addition to planetary growth, we simultaneously simulate the orbital migration of the planet driven by the disk-planet interaction.
The evolution of the planet's orbital distance can be expressed by the evolution of the planet's angular momentum as \citep[e.g.,][]{Tanaka+20}
\begin{equation}\label{eq:drdt}
    \frac{1}{r_{\rm p}}\frac{dr_{\rm p}}{dt}=\frac{2\Gamma}{M_{\rm p}r_{\rm p}^2\Omega_{\rm K}},
\end{equation}
where $\Gamma$ is the torque exerted on the planet.
Following the prescription of \citet{Kanagawa+18}, we take into account the effect of gap opening by the reduction of gas surface density and express the exerted torque as
\begin{equation}\label{eq:torque}
    \Gamma = \frac{\Gamma_{\rm L}+\Gamma_{\rm C}\exp{(-K/20)}}{\Sigma_{\rm g,tot}/\Sigma_{\rm gap}},
\end{equation}
where $\Gamma_{\rm L}$ and $\Gamma_{\rm C}$ stand for the Lindblad and corotation torque, respectively.
We adopt the Lindblad and corotation torque formula from \citet{Jimez&Masset17} that was calibrated by their 3D numerical simulations.
Equation \eqref{eq:torque} smoothly connects the type I migration for small planets to type II migration for massive planets that open a deep gas gap.

\subsection{Numerical Procedures}
Along with the simulations of disk evolution (Section \ref{sec:method_disk}), we solve Equations \eqref{eq:dMcore_i}, \eqref{eq:dMenv_H}, \eqref{eq:dMenv_i}, and \eqref{eq:drdt} with a first-order forward Euler method.
For the initial conditions, we implant the core with a mass of $0.1M_{\rm \oplus}$ at certain time and orbital distance.
Here, we suppose that photoevaporation and/or disk wind drive the disk dissipation, leading the gas disk to suddenly disappear at a certain time \citep[e.g.,][]{Kunitomo+20,Kunitomo+21}.
Since our current model has not explicitly included disk dissipation processes, we run the simulation until the time reaches a prescribed disk lifetime $t_{\rm life}$ that is treated as a free parameter.


\section{Results from Disk and Planet Formation Models}\label{sec:result_disk}

In this section, we present the general trends of disk and planet evolution obtained from our simulations.
As a fiducial model, we adopt a turbulence strength of $\alpha = 10^{-3}$, a fragmentation threshold velocity of $v_{\rm frag} = 1~{\rm m~s^{-1}}$, an initial characteristic disk radius of $r_{\rm c} = 30~{\rm au}$, and an initial disk mass of $M_{\rm disk}/M_* = 0.1$.
We then systematically vary each of these parameters to assess the sensitivity of the results.

\subsection{General trend of metal enrichment in disk gas}
\begin{figure*}[t]

\centering
\includegraphics[clip,width=0.9\hsize]{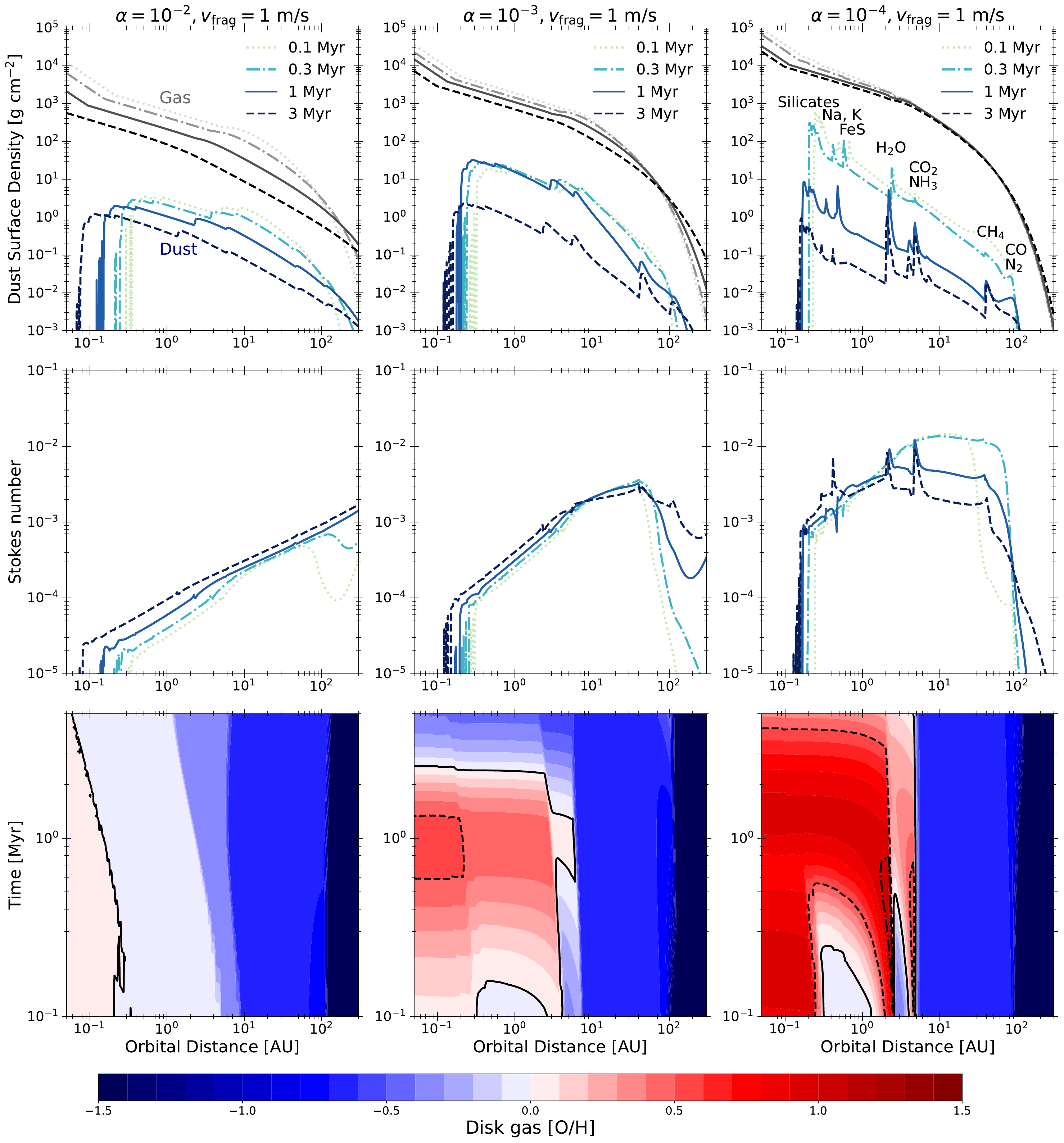}
\caption{Time evolution of disk radial profiles under various turbulence strengths.
From left to right, each column shows the results for the turbulence strength of $\alpha={10}^{-2}$, ${10}^{-3}$, and ${10}^{-4}$, respectively.
From top to bottom, each row shows the profiles of dust and gas surface densities, stokes number of dust, and O/H ratio of disk gases (color scale), respectively.
In the top and the middle rows, different colored lines shows the radial profiles at different time.
In the bottom row, the black solid, dashed, and dash-dot lines denote the contour of ${\rm[O/H]}=0.0$, $0.5$ and $1.0$, respectively.
}
\label{fig:basic}
\end{figure*}

\begin{figure*}[ht]

\centering
\includegraphics[clip,width=0.9\hsize]{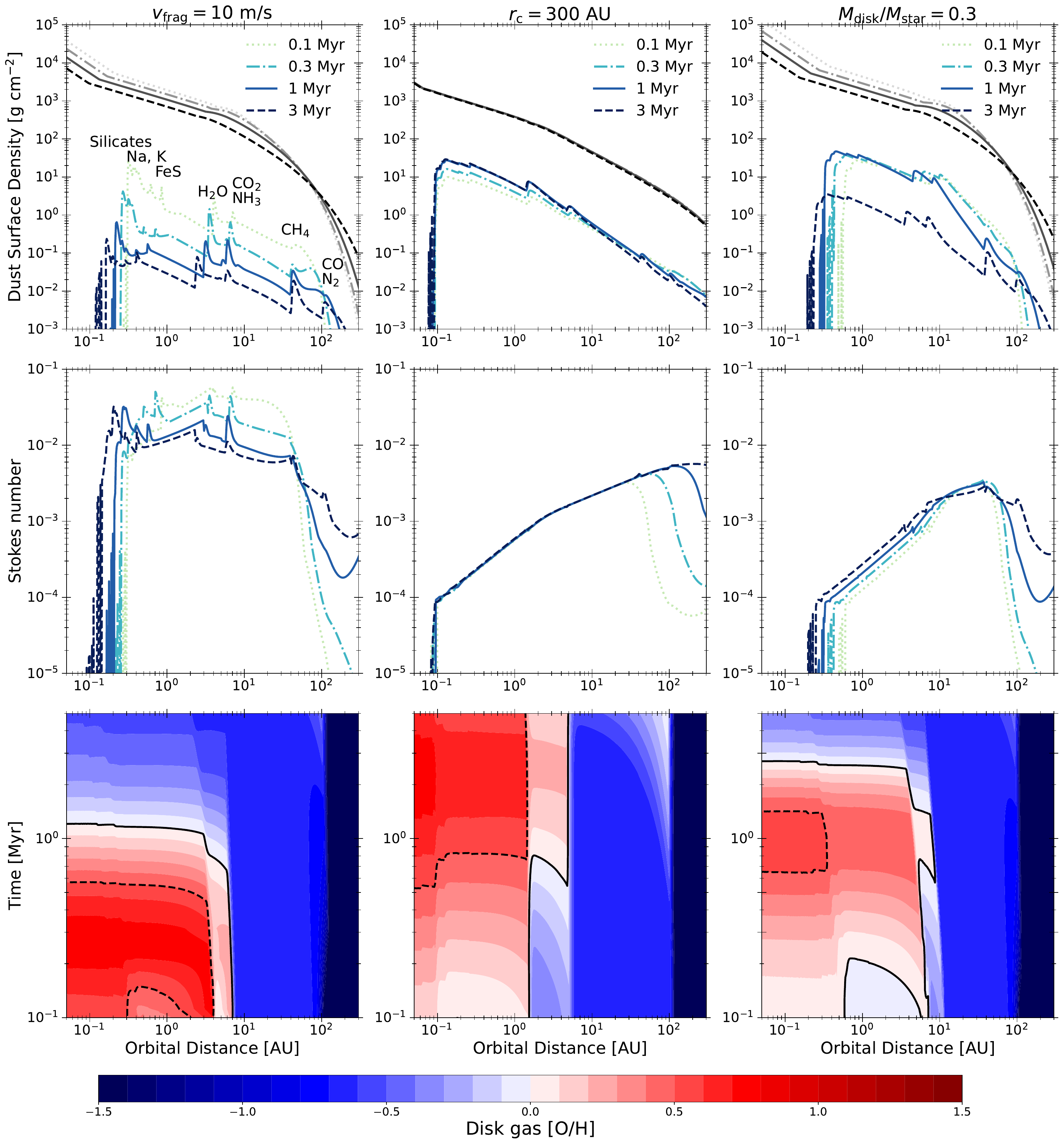}
\caption{Same as Figure \ref{fig:basic} but showing the impacts of various model parameters.
We adopt $\alpha=10^{-3}$, $v_{\rm frag}=1~{\rm m~s^{-1}}$, $r_{\rm c}=30~{\rm au}$, and $M_{\rm disk}/M_{\rm *}=0.1$ as fiducial parameters and perturb the fragmentation threshold velocity to $v_{\rm frag}=10~{\rm m~s^{-1}}$ in the left column, initial characteristic radius to $r_{\rm c}=300~{\rm au}$ in the middle column, and initial disk mass to $M_{\rm disk}/M_{\rm *}=0.3$ in the right column. 
}\label{fig:basic2}
\end{figure*}

\begin{figure*}[t]
{
\centering
\includegraphics[clip,width=\hsize]{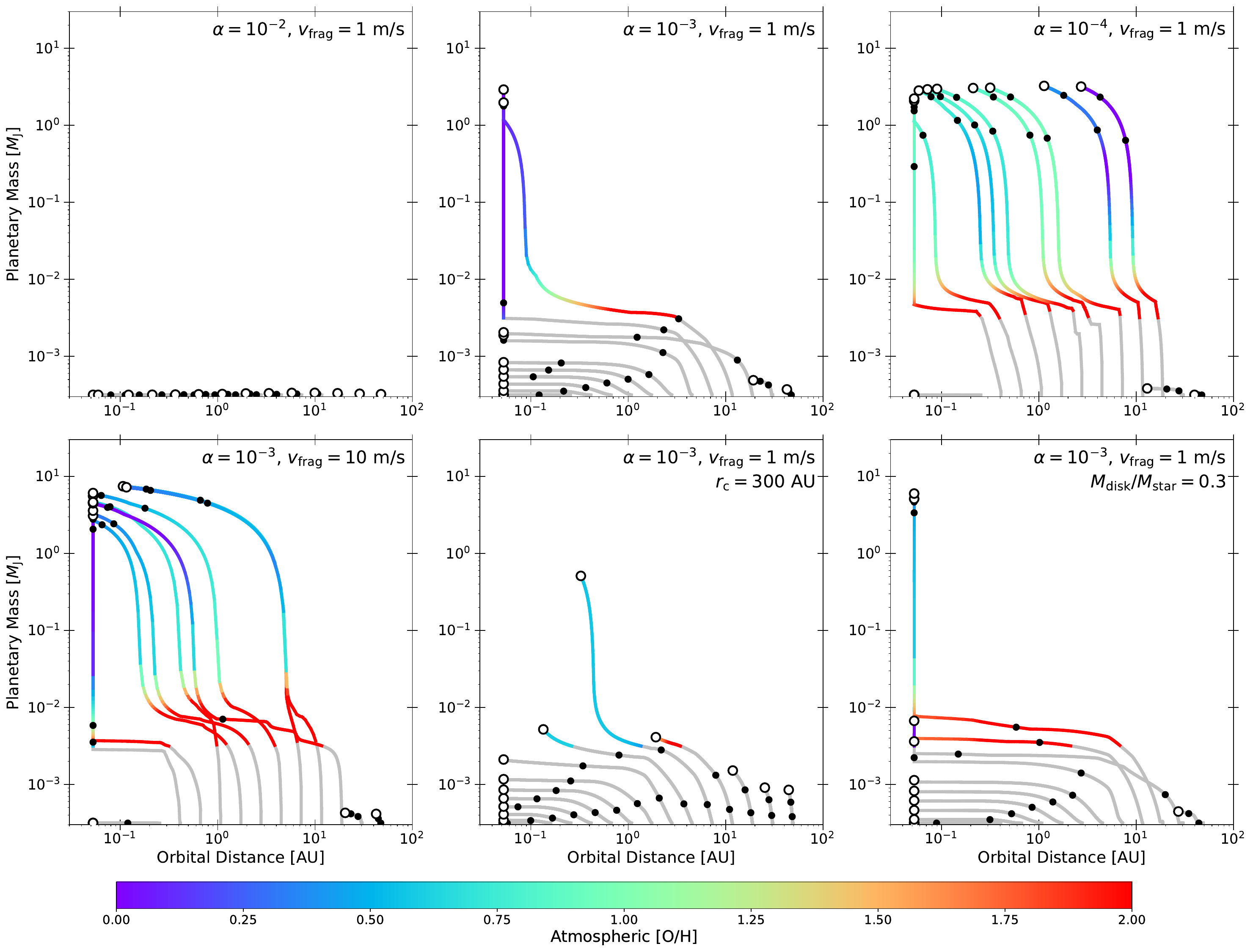}
}
\caption{Evolution tracks of planetary mass and orbital distance in planet formation simulations over $5$ Myr. In the upper row, the left, center, and right panels show the evolution tracks for $\alpha=10^{-2}$, ${10}^{-3}$, and ${10}^{-4}$, respectively, where we have assumed $M_{\rm disk}/M_{\rm *}=0.1$, $r_{\rm c}=30~{\rm au}$, $v_{\rm frag}=1~{\rm m~s^{-1}}$ as fiducial parameters. In the lower row, we assume $\alpha={10}^{-3}$ and change each parameter to $v_{\rm frag}=10~{\rm m~s^{-1}}$ (left), $r_{\rm c}=300~{\rm au}$ (middle), and $M_{\rm disk}/M_{\rm *}=0.3$ (right). The colors along each track indicate the atmospheric [O/H] at each planetary age for planets with $M_{\rm p}>M_{\rm \oplus}$. 
The black dots on each track denote the mass-orbit relations at $t=1~{\rm Myr}$ and $3~{\rm Myr}$, and the white-filled dots denote the final mass-orbit relations.
}
\label{fig:planet_evolve}
\end{figure*}

Turbulence strength plays a central role in controlling the evolution of protoplanetary disks. 
Figure \ref{fig:basic} shows the time evolution of gas and dust surface densities for $\alpha={10}^{-2}$, ${10}^{-3}$, and ${10}^{-4}$.
Stronger turbulence results in more rapid viscous diffusion of the gas surface density.
In contrast, weaker turbulence leads to a faster depletion of the dust surface density.
This is because the Stokes number of dust particles increases with decreasing $\alpha$ when fragmentation limits particle sizes, as seen in the middle row \citep[see also e.g.,][]{Birnstiel+10}.

The dust surface density exhibits multiple spikes at specific orbital distances, which correspond to sublimation lines of various ices and rocks.
These spikes result from the recondensation of vapors that diffuse outward from the inner orbit, leading to local increases in the dust surface density, as seen in previous studies \citep[e.g.,][]{Schoonenberg&Ormel17,Schneider&Bitsch21}. 
At $r\lesssim0.2~{\rm au}$, the dust surface density drops sharply due to the complete sublimation of dust particles.

Disk gas in the inner orbits ($\lesssim3~{\rm au}$) is enriched in metals due to the sublimation of inward-drifting dust, and the extent of this enrichment strongly depends on the turbulence strength.
The bottom row of Figure \ref{fig:basic} shows the gas-phase O/H ratio---a representative heavy element abundance---as a function of orbital distance.
In the case of strong turbulence ($\alpha={10}^{-2}$), O/H ratio remains close to the stellar value throughout $5$ Myr, as the motion of dust is tightly coupled to the gas motion.
For moderate turbulence ($\alpha={10}^{-3}$), the sublimation of inward-drifting dust noticeably enhances the gas-phase O/H ratio, particularly inside H$_2$O snowline.
This occurs because weaker turbulence results in faster radial drift of dust and slower removal of sublimated vapors.
The gas-phase O/H ratio reaches a peak value of [O/H]$\sim0.5$ around $t\sim1~{\rm Myr}$ and subsequently declines as the outer disk becomes depleted of pebbles\footnote{Here, the square brackets conventionally expresses the elemental abundance as a base-10 log of element-to-hydrogen number ratio normalized by the solar value, i.e., ${\rm[O/H]}=\log_{\rm 10}[{\rm(O/H)/(O/H)_{\rm sun}]}$}.
The O/H ratio reaches an even higher peak value value of ${\rm[O/H]}\sim1.0$ for weaker turbulence ($\alpha={10}^{-4}$).
Although the dust disk is depleted more rapidly at $\alpha={10}^{-4}$ than at $\alpha={10}^{-3}$, the metallicity enrichment is persistent for a longer duration due to the slow turbulent diffusion.
Overall trends are broadly consistent with previous studies that examined pebble drift and sublimation \citep[e.g.,][]{Booth+17,Schneider&Bitsch21}.

\subsection{Roles of fragmentation threshold velocity, disk size, and disk mass}\label{sec:result_disk_sensitivity}
The fragmentation threshold velocity, $v_{\rm frag}$, also greatly influences the evolution of the dust disk and the degree of metallicity enrichment in the inner disk.
The left column of Figure \ref{fig:basic2} shows the results for $v_{\rm frag}=10~{\rm m~s^{-1}}$.  
A higher $v_{\rm frag}$ allows dust particles to grow larger sizes, which in turn leads to more rapid inward drift.
A comparison with the middle column of Figure \ref{fig:basic} demonstrates that metal enrichment in the inner disk is more pronounced for larger $v_{\rm frag}$. 
In this sticky dust case, metal enrichment takes place at earlier time ($\sim0.1$--$0.3~{\rm Myr}$) owing to rapid dust drift.
As a consequence, dust is depleted earlier, and the disk gas becomes metal-poor at $t\gtrsim1~{\rm Myr}$, in contrast to the case with $v_{\rm frag}=1~{\rm m~s^{-1}}$.

The initial characteristic radius of the gas disk, $r_{\rm c}$, also affects both the magnitude and duration of pebble-driven metal enrichment (middle column of Figure \ref{fig:basic2}).
Larger initial disk radii result in more substantial and longer-lasting metal enrichment in the inner disk. 
This is because dust requires a longer time to travel from outer disk to inner orbits and thus stays in the disk for a longer time.  
As a result, drifting dust can supply fresh sublimated vapors to the inner disk to sustain metal enrichment over an extended period.

The initial disk mass has only a minor influence on the degree of metal enrichment in the disk gas.
The right column of Figure \ref{fig:basic2} presents the results for $M_{\rm disk}/M_{\rm star}=0.3$, which closely resemble the evolution observed for $M_{\rm disk}/M_{\rm star}=0.1$ (middle column of Figure \ref{fig:basic}). 
The main difference is that metal enrichment occurs slightly further out, around $\sim5~{\rm au}$.
This shift originates from enhanced viscous heating in the more massive disk, which pushes the H$_2$O and CO$_2$ snowlines outward.

\subsection{Planetary Growth Track}\label{sec:evolve_track}
Here, we present general trends of planetary growth and migration.
Figure \ref{fig:planet_evolve} shows the evolution of planetary masses and orbital distances for various disk conditions, where we implant planetary cores at multiple orbits at $t=0.1$ Myr.
In general, planetary cores initially grow through pebble accretion without significant migration.
In our model, core growth via pebble accretion is most efficient around $\sim10~{\rm au}$ owing to high Stokes numbers (${\rm St}$).
Pebble accretion turns out to be inefficient at even larger distances due to the limited amount of pebbles available there and the onset of the drift-limited regime that makes ${\rm St}$ lower at larger distances.
As the cores become more massive, they begin to migrate inward.
Sufficiently massive cores also start runaway gas accretion and grow into gas giants. 
This general evolution behavior is in agreement with previous studies based on the pebble accretion framework \citep[e.g.,][]{Bitsch+15,Booth+17,Brugger+18,Johansen+19,Schneider&Bitsch21}.
It is worth noting that, in our model, \rev{significant orbital migration mainly occurs prior to the onset of runaway gas accretion.
Once the runaway gas accretion sets in, gas giants grow without experiencing significant migration until they reach super-Jupiter masses} because of rapid envelope accretion compared to type II migration, as discussed in \citet{Tanaka+20}.

The fate of planetary cores is highly sensitive to the turbulence strength $\alpha$, as illustrated in the top row of Figure~\ref{fig:planet_evolve}.
In general, weak turbulence (small $\alpha$) promotes the formation of gas giants.
Under strong turbulence ($\alpha = 10^{-2}$), planetary cores hardly grow to sufficiently high core mass through pebble accretion, inhibiting rapid envelope accretion to form gas giants. 
For weak turbulence ($\alpha \leq 10^{-3}$), in contrast, several cores grow massive enough to become gas giants.
This outcome stems from the dependence of the Stokes number ${\rm St}$ on $\alpha$: weaker turbulence allows pebbles to grow larger sizes with larger ${\rm St}$ (Figure~\ref{fig:basic}), which enhances the efficiency of pebble accretion.

The fragmentation threshold velocity $v_{\rm frag}$ also influences the efficiency of planetary growth.
The lower left panel of Figure \ref{fig:planet_evolve} shows the results for $\alpha={10}^{-3}$ and $v_{\rm frag}=10~{\rm m~s^{-1}}$.
Compared to the fiducial case with $v_{\rm frag}=1~{\rm m~s^{-1}}$ (upper middle panel), a larger number of cores grow massive enough to end up with gas giants. 
This result is attributed to the higher Stokes number associated with higher $v_{\rm frag}$, which enhances the efficiency of pebble accretion.


Although gas giants can form in both models—low $v_{\rm frag}$ with low $\alpha$, and high $v_{\rm frag}$ with high $\alpha$—the final planetary masses considerably differ between the two cases.
Namely, gas giants tend to reach higher final masses under stronger disk turbulence. 
This trend arises because strong turbulence facilitates the supply of disk gas to a planet's feeding zone.
To clarify this trend, we first consider the global disk accretion rate of the self-similar solution for $r{\ll}r_{\rm c}$, given by \citep[e.g.,][]{Hartmann+98,Manara+23}
\begin{equation}
    \dot{M}_{\rm disk}=\frac{M_{\rm disk}}{2\tau_{\rm ss}}\left( 1+\frac{t}{\tau_{\rm ss}}\right)^{-3/2},
\end{equation}
where $\tau_{\rm ss}=r_{\rm c}^2/3\nu_{\rm SS}(r_{\rm c})$ is the characteristic timescale for the viscous evolution.
Assuming that envelope accretion is regulated by global disk accretion, we can estimate the final mass of gas giants as \citep[for relevant discussion, see also][]{Tanigawa&Tanaka16}
\begin{equation}\label{eq:Mp_fin1}
    M_{\rm p}\sim \int_{\rm 0}^{t_{\rm life}} \dot{M}_{\rm disk}dt= M_{\rm disk}\left[ 1-\left(1+ \frac{t_{\rm life}}{\tau_{\rm ss}}\right)^{-1/2}\right]\approx \frac{M_{\rm disk}}{2}\frac{t_{\rm life}}{\tau_{\rm ss}},
\end{equation}
where we have assumed $t_{\rm life}\ll\tau_{\rm ss}$ for the last expression.
This analysis suggests $M_{\rm p}\propto\tau_{\rm ss}^{-1} \propto \alpha$, indicating that weaker turbulence results in lower final masses.

The initial disk characteristic radius also influences the fate of planetary cores, as shown in the lower middle panels of Figure~\ref{fig:planet_evolve}.
Planetary cores grow more slowly in disks with a large characteristic radius ($r_{\rm c} = 300~{\rm au}$) compared to those with a smaller radius ($r_{\rm c} = 30~{\rm au}$; upper middle panel).
This difference arises because the delivery of pebbles from the outer to the inner disk takes more time in larger disks.

It is intuitive to expect that massive disks facilitate planetary growth, which is indeed seen in our model.
The lower right panel of Figure \ref{fig:planet_evolve} shows the evolution tracks for a disk with an initial mass of $M_{\rm disk}/M_{\rm *}=0.3$.
Compared to the case with $M_{\rm disk}/M_{\rm *} = 0.1$ in the middle top panel, the more massive disk produces gas giants with larger final masses.
This result can be readily understood from Equation~\eqref{eq:Mp_fin1}, which shows that the final planetary mass increases with the initial disk mass.

Atmospheric metallicity also evolves throughout planet formation.
To illustrate this behavior, the colors along each evolution track in Figure \ref{fig:planet_evolve} indicate the atmospheric [O/H] ratio under the assumption of fully mixed envelopes (see Section \ref{sec:method_calc_composition} for the definition of fully mixed and unmixed envelopes).
In our model, planets initially acquire atmospheres with high metallicity of [O/H]$\gtrsim2.0$ at low planetary masses, which is due to the sublimation of accreted pebbles that deposits metal-rich vapors.
As the planets grow and enter the runaway gas accretion phase, envelope accretion rate ($dM_{\rm env}/dt$) overwhelms the pebble accretion rate ($dM_{\rm peb}/dt$) to reduce atmospheric metallicities.
In other words, the influx of surrounding disk gas dilutes the pre-existing vapor atmospheres.
We note that dilution through gas inflow is essentially the same mechanism with which previous planetesimal-based models produced a mass-metallicity anti-correlation \citep{Fortney+13,Mordasini+16,
Cridland+19_CtoO}.
During this stage, the atmospheric [O/H] ratio gradually approaches the metallicity of the accreting disk gas.
For example, gas giants formed in weakly turbulent disks tend to retain high atmospheric metallicities due to persistent vapor enrichment in the inner disk.
We also find that giant planets acquiring their envelopes inside the H$_2$O snowline generally exhibit elevated atmospheric [O/H] ratios, as the disk gas in this region is highly enriched by H$_2$O vapors delivered by inward-drifting pebbles.

\section{Atmospheric Population Synthesis}\label{sec:result_population}
\subsection{Method} 
To investigate possible trends in atmospheric metallicity, we perform simplified population synthesis calculations.
We fix the initial disk mass to be $M_{\rm disk}/M_{\rm *}=0.1$ and disk metallicity of [Fe/H] = 0.1, the typical metallicity of stars hosting hot Jupiters \citep{Osborn&Baylis20}.
For an initial disk mass of $M_{\rm disk}/M_{\rm *}=0.1$, we assume the initial characteristic radius of $r_{\rm c}=148~{\rm au}$, which is estimated from the observational disk mass-radius correlation derived by \citet{Tripathi+17}. 
While the observed disk radius actually refers to the flux-enclosing radius rather than the characteristic radius \citep[e.g.,][]{Rosotti+19}, the former is sensitive to assumptions imposed on dust opacity \citep[e.g.,][]{Zormpas+22}, which remain highly uncertain.
To avoid this complication, we treat $r_{\rm c}$ as an approximate disk radius.
We also set the fragmentation threshold velocity to $v_{\rm frag}=1~{\rm m~s^{-1}}$, in line with the fragile dust suggested by recent disk observations \citep{Okuzumi&Tazaki19,Jiang+24,Ueda+24} and experimental studies \citep{Musiolik&Wurm16,Musiolik&Wurm19,Fritscher&Teiser21}.
The sensitivity of our results to the choice of $M_{\rm disk}$, $r_{\rm c}$, and $v_{\rm frag}$ is discussed in Section \ref{sec:mass_meta2}.

Following previous population synthesis studies \citep[e.g.,][]{Ida&Lin04,Emsenhuber+21,Kimura&Ikoma22}, we randomly vary several model parameters: turbulence strength $\alpha$, disk lifetime $t_{\rm life}$, initial core orbit $r_{\rm p0}$, and the inner edge of the gas disk $r_{\rm in}$.
Since our goal is to explore potential trends in atmospheric metallicity rather than to reproduce planetary occurrence rates, we adopt log-uniform distributions for these parameters.
We vary the turbulence strength within the range $\alpha = 10^{-4}$--$10^{-3}$, which covers the turbulence strength suggested by various disk observations \citep[][]{Rosotti23}.
The disk lifetime is sampled from a range of $t_{\rm life}=0.5$--$5~{\rm Myr}$.
We set the location of the disk inner edge such that the corresponding orbital period lies between $2.34$ and $9.59$ Earth days, consistent with the typical rotation periods of protostars \citep{Venuti+17}.
To mimic early core formation potentially occurring during the Class 0 phase \citep{Manara+18}, we implant $0.1M_{\oplus}$ cores at randomly selected orbits at $t = 0.01{\rm Myr}$.
For each set of disk parameters, five planetary cores are implanted and evolved to enhance the statistical sample of synthetic planets.

\subsection{Mass--Metallicity Relations}\label{sec:atm_bulk_metal}
\begin{figure*}[t]
{
\centering
\includegraphics[clip,width=\hsize]{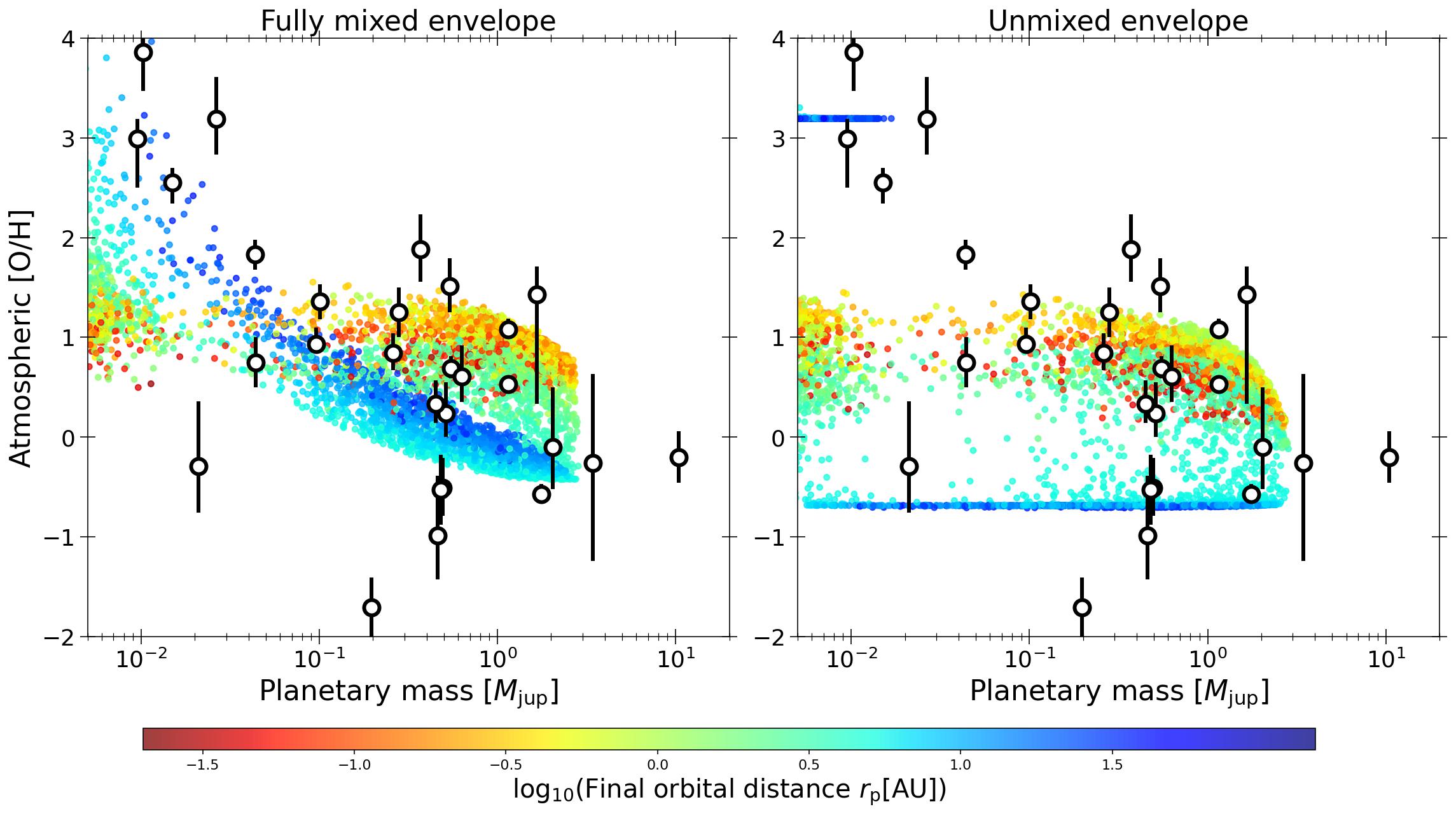}
}
\caption{Relation between planetary mass and atmospheric metallicity in our synthetic planets. 
The left panel shows results for fully mixed atmospheres, whereas the right panel shows results for unmixed atmospheres. The color of each dot denotes the orbital distance of planets at the time of disk dissipation. The data point indicate the atmospheric metallicity constrained by JWST observations relative to stellar [O/H], as summarized in Appendix \ref{appendix_JWST}. We note a caveat on the results at low-mass end for which we simplify atmospheric accretion for such low-mass planets (see Section \ref{sec:model_caveat}).
}
\label{fig:mass_metal}
\end{figure*}

\begin{figure*}[t]

\centering
\includegraphics[clip,width=0.8\hsize]{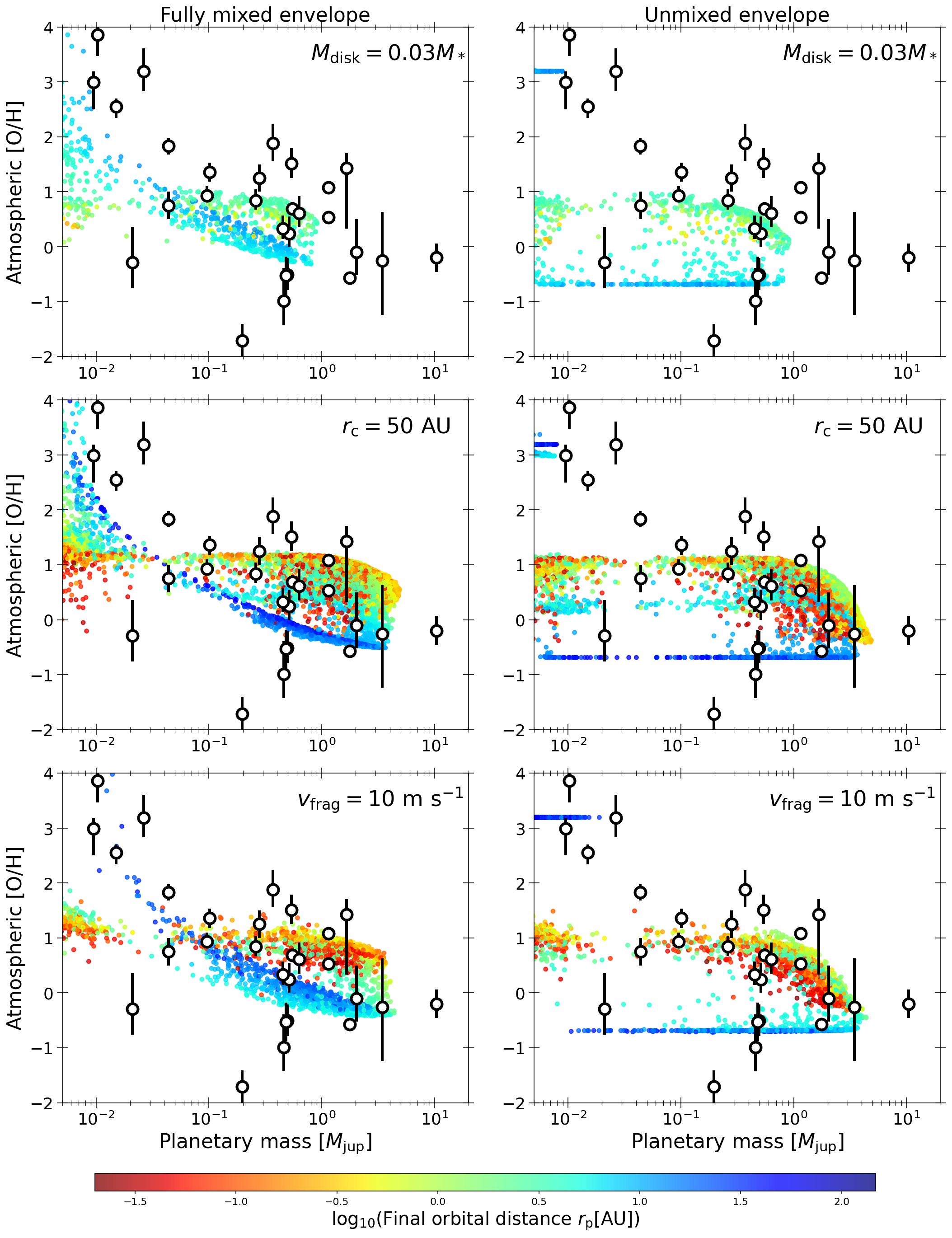}
\caption{Same as Figure \ref{fig:mass_metal} but for different initial disk mass $M_{\rm disk}$, characteristic radius $r_{\rm c}$, and fragmentation threshold velocity $v_{\rm frag}$. 
From top to bottom, each row presents the population synthesis results for parameter set of ($M_{\rm disk}/M_{\rm *}=0.03$, $r_{\rm c}=148~{\rm au}$, $v_{\rm frag}=1~{\rm m~s^{-1}}$), ($M_{\rm disk}/M_{\rm *}=0.1$, $r_{\rm c}=50~{\rm au}$, $v_{\rm frag}=1~{\rm m~s^{-1}}$), and ($M_{\rm disk}/M_{\rm *}=0.1$, $r_{\rm c}=148~{\rm au}$, $v_{\rm frag}=10~{\rm m~s^{-1}}$), respectively.
}
\label{fig:mass_metal2}
\end{figure*}

One of the most intriguing findings of this study is that planets formed at close-in versus distant orbits exhibit markedly different mass--metallicity correlations.
Figure \ref{fig:mass_metal} shows the atmospheric metallicity, measured by [O/H], as a function of the planetary mass for our synthetic planet population.
Our simulations reveal two distinct populations of planets with different metallicity trends.
Under the assumption of fully mixed envelopes (we refer readers to Section \ref{sec:method_calc_composition} to recall the definition of fully mixed and unmixed envelopes), planets that end up in $\lesssim1~{\rm au}$---inside the H$_2$O snowline---during disk lifetimes (reddish points in Figure \ref{fig:mass_metal}) exhibit atmospheric metallicity of [O/H]$\sim0.0$--$1.5$.
This ``\rev{inside-snowline (I-SL)} population'' exhibits atmospheric metallicity that moderately decreases with increasing mass, especially for $M_{\rm p}\gtrsim0.1M_{\rm J}$.
In contrast, planets ending up beyond $\gtrsim1~{\rm au}$---outside the H$_2$O snowline---exhibit a steeper anti-correlation between metallicity and mass.
This ``\rev{outside-snowline (O-SL) population}'' exhibits metallicity that is approximately inversely proportional to planetary mass.
While metallicities of low-mass planets with close-in final orbits plateaus at [O/H]$\sim1.5$, planets with distant final orbits exhibit metallicity that reach higher metallicities up to [O/H]$\sim3.0$.

The two distinct mass--metallicity correlations originate from fundamentally different processes for the \rev{I-SL} and \rev{O-SL populations}.
For the \rev{O-SL population}, for fully mixed envelopes, atmospheric enrichment is primarily driven by pebble accretion followed by sublimation within the proto-atmosphere, which generates metal-rich vapors on the low-mass planetary core for which envelope accretion is still slow.
As these planets grow, pebble accretion eventually halts---either due to pebble isolation or the exhaustion of pebbles---terminating the supply of sublimated vapors to their atmospheres.
Because pebble accretion tends to be effective at $\sim10~{\rm au}$ (see Figure \ref{fig:planet_evolve} and Section \ref{sec:evolve_track}), planets formed in these regions tend to obtain the highest atmospheric metallicities.
At higher planetary masses, runaway gas accretion dilutes the metal-rich vapor atmosphere with surrounding disk gas.
Since the disk gas at distant orbits is depleted in vapors, gas accretion reduces the atmospheric metallicity until it approaches a sub-stellar value.
We note that the process introduced here is essentially the same as the process by which planetesimal bombardment generates a mass--metallicity anti-correlation, which was studied in previous studies \citep{Fortney+13,Mordasini+16,Cridland+19_CtoO,Danti+23}.

The origin of the mass--metallicity correlation for the \rev{I-SL} population is related to the metal enrichment of the disk gas.
Because the disk gas inside H$_2$O snowline can be highly enriched in vapors, the metallicity of the disk gas mainly controls atmospheric metallicity of the \rev{I-SL population}.
The \rev{I-SL population} exhibits a moderate anti-correlation between planetary mass and metallicity at $\gtrsim0.3M_{\rm jup}$.
This is because disk turbulence acts on disk gas metallicity and the final planetary mass in an opposite way.
Strong turbulence reduces the metallicity of disk gas and thus atmospheric metallicity (see Figure \ref{fig:basic}), whereas it increases the final mass of giant planets (see Figure \ref{fig:planet_evolve}).
Consequently, strong turbulence leads to form gas giants with high planetary mass and low atmospheric metallicity, which produces the moderate mass--metallicity anti-correlation.

\subsection{Impacts of Inhomogeneous Envelope}
The mass--metallicity relation depends on the inhomogeneity within the envelope, especially for the \rev{O-SL population}.
The right panel of Figure \ref{fig:mass_metal} illustrates the mass--metallicity relation under the unmixed envelope scenario.
The \rev{O-SL population} exhibits a markedly different mass--metallicity relation: low-mass planets have atmospheric metallicity of [O/H]$\sim3.0$ that sharply drops to sub-stellar metallicity [O/H]$\sim-0.7$ as planetary mass increases.
This trend stems from the assumption that vapors produced by sublimation of accreted pebbles do not mix with the subsequently accreted disk gas.
In other words, the unmixed scenario forces atmospheric metallicity to follow the metallicity of surrounding disk gas once pebble accretion ceases.
The lowest metallicity of [O/H]$\sim-0.7$ reflects the initial abundance of CO (20\% of total oxygen budget, see Appendix \ref{sec:appendix_A}) that always exists as vapors at orbits where giant planets form in our model (but see Section \ref{sec:model_caveat} for caveats).

In contrast to the \rev{O-SL population}, the \rev{I-SL population} exhibits a similar mass--metallicity relation under both the fully mixed and unmixed envelope scenarios.
This is because atmospheric metallicity is dictated by the metallicity of the surrounding disk gas in both scenarios.
\rev{There is slight difference in shape of the mass-metallicity relation.
This difference arises because, in our model, the fully-mixed envelope exhibits time-integrated disk gas compositions while the unmixed envelope records the disk gas composition at the moment when envelope accretion completes.}

\subsection{Sensitivity to disk properties}\label{sec:mass_meta2}
We investigate the impacts of several disk properties, namely, the initial disk mass, the disk characteristic radius, and the fragmentation threshold velocity.
\rev{Our analysis resembles \citet{Savvidou&Bitsch23} that investigated how disk properties affect giant planet formation, while this study focuses on how the disk properties affect the mass-metallicity relation of giant planet atmospheres.}
In a nutshell, we find that the emergence of the two distinct mass--metallicity relations for \rev{I-SL} and \rev{O-SL population}s is robust against the choice of these parameters.
However, the vertical offset of the mass--metallicity relations does depend on these disk properties, as described below.

\subsubsection*{Initial disk mass}
The offset of the mass--metallicity relation depends on the initial disk mass, which was fixed to $M_{\rm disk}=0.1M_{\rm *}$ in the previous section.
Here, we repeated the population synthesis with a lower initial disk mass of $M_{\rm disk}=0.03M_{\rm *}$.
Although disk observations suggest a positive correlation between the total disk mass, which is diagnosed by millimeter continuum luminosity, and the disk radius \citep[][]{Andrews+10,Andrews+18,Tripathi+17,Hendler+20}, we fix the characteristic radius to $r_{\rm c}=148~{\rm au}$ as in Figure \ref{fig:mass_metal} to isolate the impacts of the initial disk mass.

A lower initial disk mass acts to shift the mass--metallicity relation downward, as shown in the upper row of Figure \ref{fig:mass_metal2}.
As expected from Equation \eqref{eq:Mp_fin1}, final planetary mass tends to be lower at lower initial disk mass: the maximum planetary mass reaches $\sim2$--$3M_{\rm jup}$ for $M_{\rm disk}=0.1M_{\rm *}$ (Figure \ref{fig:mass_metal}), while the maximum mass only reaches $\sim0.8M_{\rm jup}$ for $M_{\rm disk}=0.03M_{\rm *}$.
In both the \rev{I-SL} and \rev{O-SL population}s, planets tend to show atmospheric metallicity lower than that found for the model with $M_{\rm disk}=0.1M_{\rm *}$ (Figure \ref{fig:mass_metal}) by a factor of $2$--$3$ for a given mass.
The shift of the \rev{O-SL population} is attributed to the reduced total dust mass.
Since core assembly and associated vapor production via pebble sublimation terminate at lower planetary mass, dilution of vapor atmosphere by disk gas starts at a lower mass.

The decline in atmospheric metallicity for the \rev{I-SL population} is attributed to the timing effect.
Although an initial disk mass has a minor impact on the disk gas metallicity (Section \ref{sec:result_disk_sensitivity} and Figures \ref{fig:basic} and \ref{fig:basic2}), it affects the core growth timescale and the timing at which runaway gas accretion sets in.
Since the disk gas metallicity decreases with time once the pebbles at outer regions are exhausted (Figure \ref{fig:basic}), gas giants can acquire more metal-rich disk gas if runaway gas accretion starts earlier.
This results in a lower atmospheric metallicity of \rev{I-SL population} at a lower initial disk mass that delays core growth, and thus the onset of runaway gas accretion.
The timing effects discussed here can also be seen in the upper middle and lower right panels of Figure \ref{fig:planet_evolve}.

\subsubsection*{Disk characteristic radius}
Disk characteristic radius has a notable impact on the mass--metallicity relation.
The middle row of Figure \ref{fig:mass_metal2} presents the population synthesis results for $r_{\rm c}=50~{\rm au}$.
Compared to the model with $r_{\rm c}=148~{\rm au}$ shown in Section \ref{sec:atm_bulk_metal}, gas giants in the \rev{I-SL population} tend to grow to higher masses.
This trend arises because a smaller characteristic radius reduces the disk diffusion timescale, $\tau_{\rm ss}=r_{\rm c}^2/3\nu$, enabling gas giants to capture more gas during the disk lifetime (see also Equation \ref{eq:Mp_fin1}).
On the other hand, atmospheric metallicity of the \rev{I-SL population} is slightly lower than that found for $r_{\rm c}=148~{\rm au}$ (Figure \ref{fig:mass_metal}). 
This reduction is due to the less efficient enrichment of inner disk gas in smaller disks (see Figure \ref{fig:basic2}).

Atmospheric metallicity of the \rev{O-SL population} also tends to be lower at a smaller disk characteristic radius.
In the \rev{O-SL population}, the offset in the mass--metallicity relation is primarily controlled by the core mass at which pebble accretion halts.
\rev{The total dust mass available beyond a planetary orbit $r_{\rm pl}$ can be expressed as the product of initial dust-to-gas mass ratio and $M_{\rm disk}e^{(-r_{\rm pl}/r_{\rm c})}$ derived from the integration of Equation \eqref{eq:sigma_ini}.}
Thus, in general, a smaller $r_{\rm c}$ provides less pebble mass for a given orbital distance under the same initial disk mass.
\rev{While the disk size dependence of available pebble mass is subtle for most planets at $r_{\rm pl}\ll r_{\rm c}$, it can affect the final core mass for planets in the \rev{O-SL population}, as they orbit around a few tens au that is comparable to the characteristic disk radius.}
In our sensitivity test, where we vary $r_{\rm c}$ while keeping the initial disk mass fixed, pebble accretion terminates at a lower mass at smaller $r_{\rm c}$ \rev{for cores orbiting at distant orbits}.
As a result, vapor generation by pebble accretion is limited, leading to a downward shift in the mass--metallicity correlation for the \rev{O-SL population}.

\begin{figure*}[t]
{
\centering
\includegraphics[clip,width=\hsize]{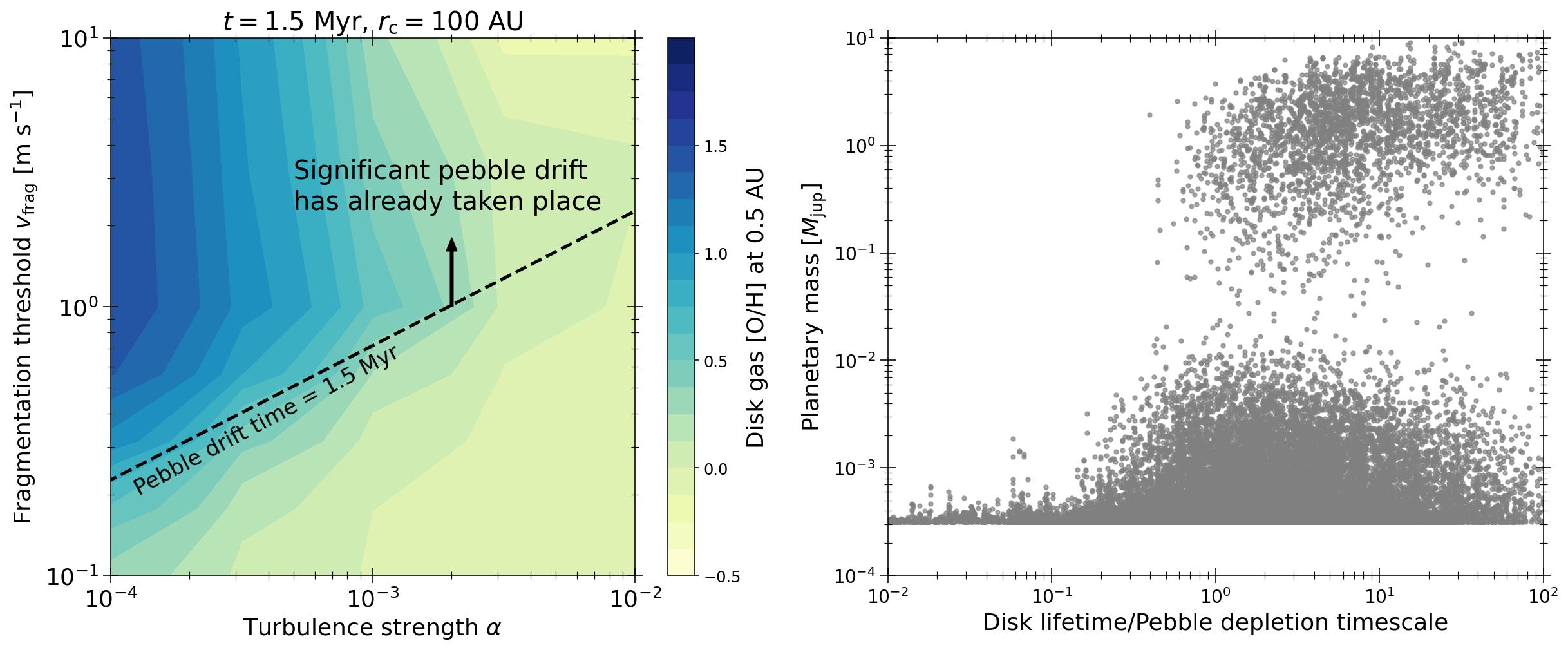}
}
\caption{
(Left) Disk gas [O/H] ratio at $0.5$ au and $t=1.5~{\rm Myr}$. The vertical and horizontal axes are fragmentation threshold velocity and turbulence strength. Here we assume $r_{\rm c}=100~{\rm au}$. The black dashed line denotes the $\alpha$--$v_{\rm frag}$ relation at which pebble depletion timescale (Equation \ref{eq:tau_dep}) satisfies $\tau_{\rm peb}=1.5~{\rm Myr}$. (Right) Final planet mass as a function of disk lifetime normalized by the pebble depletion timescale. In this panel, we numerically measure the e-folding time of the decay of total dust mass as the pebble depletion time.
This figure demonstrates that giant planets with $M_{\rm p}>0.1M_{\rm J}$ could form only when significant pebble drift takes place during the disk lifetime. 
}
\label{fig:vfrag_test}
\end{figure*}

\subsubsection*{Fragmentation threshold velocity}
Although we have fixed the fragmentation threshold velocity to $v_{\rm frag}=1~{\rm m~s^{-1}}$, we find that the mass--metallicity relation is insensitive to this parameter.
The lower row of Figure \ref{fig:mass_metal2} shows population synthesis results for $v_{\rm frag}=10~{\rm m~s^{-1}}$.
The model produces a mass--metallicity relation that closely resembles the relation found for $v_{\rm frag}=1~{\rm m~s^{-1}}$ (Figure \ref{fig:mass_metal}), both in shape and in vertical offset.
The minor role of fragmentation threshold velocity may appear counter intuitive, since the metallicity of disk gas seems to strongly depend on $v_{\rm frag}$ as shown in Figure \ref{fig:basic2}.

To understand the weak dependence on fragmentation threshold velocity, we first show the disk gas metallicity at $0.5$ au and $t=1.5$ Myr as a function of $v_{\rm frag}$ and turbulence strength in Figure \ref{fig:vfrag_test}.
\rev{At low $v_{\rm frag}$, the metallicity increases with $v_{\rm frag}$ for a given turbulence strength, until it eventually plateaus.}
The boundary of this transition can be understood from timescale argument.
The timescale with which pebbles travel from outer regions to inner disk may be estimated as $\tau_{\rm drift}\sim r_{\rm c}/u_{\rm dust}\approx 1/2\eta \Omega(r_{\rm c}){\rm St}$, where we have assumed ${\rm St}\ll 1$.
For fragmentation limited size, the pebble Stokes number is given by \citep[e.g.,][]{Birnstiel+10}
\begin{equation}
    {\rm St}_{\rm frag}=\frac{1}{3\alpha}\left( \frac{v_{\rm frag}}{c_{\rm s}}\right)^2.
\end{equation}
Combining the two equations, the pebble drift timescale can be crudely estimated as
\begin{equation}\label{eq:tau_dep}
    \tau_{\rm peb}=\frac{3\alpha}{\Omega(r_{\rm c})}\left( \frac{r_{\rm c}\Omega(r_{\rm c})}{v_{\rm frag}}\right)^2\left| \frac{d\ln{P}}{d\ln{r}}\right|^{-1}.
\end{equation}
We plot a $\alpha$--$v_{\rm frag}$ relation satisfying $\tau_{\rm peb}=t$ in Figure \ref{fig:vfrag_test}, demonstrating that $v_{\rm frag}$ dependence turns out to be weak once the pebble drift timescale falls short of the disk age.
This trend emerges because the disk gas metallicity is mainly controlled by how fast the disk turbulence drains vapors rather than how fast pebbles drift once the pebble supply from outer orbits has ceased.

The weak dependence of the atmospheric metallicity on fragmentation threshold velocity is attributed to the fact that giant planet formation requires rapid pebble drift.
The right panel of Figure \ref{fig:vfrag_test} shows the final planetary mass as a function of $\tau_{\rm life}/\tau_{\rm peb}$, where we have repeated the population synthesis with randomly varying $r_{\rm c}$, $M_{\rm disk}$, and $v_{\rm frag}$ in addition to $\alpha$, $r_{\rm p0}$, $t_{\rm life}$, and $r_{\rm in}$.
The figure demonstrates that gas giants with $M_{\rm p}>0.1M_{\rm J}$ could form only when the pebble drift timescale is shorter than the disk lifetime.
This is a natural consequence of pebble accretion: to form a massive enough core to trigger runaway gas accretion, pebbles must travel from outer regions to core's orbit during the disk lifetime.
This condition actually coincides with the condition under which disk gas metallicity is nearly independent of $v_{\rm frag}$ (left panel of Figure \ref{fig:vfrag_test}), explaining why the atmospheric metallicity trend of gas giants is insensitive to $v_{\rm frag}$.

\subsection{A Comparison with Atmospheric Metallicity Constrained by JWST}
We now compare our results with atmospheric metallicities of exoplanets reported from published JWST observations (see Appendix \ref{appendix_JWST} for the data compilation).
Focusing on giant planets with $M_{\rm p}\gtrsim0.1M_{\rm jup}$, we find that the majority of JWST results align well with the trend found for the \rev{I-SL population} (see Figure \ref{fig:mass_metal}).
This is intriguing, as the widely favored formation scenario for close-in giants involves formation at distant orbits followed by inward migration that is driven either by planet-disk interaction or dynamical events such as planet--planet scattering \citep[e.g.,][]{Dawson+18,Fortney+21}.
Simulated planets for perturbed $r_{\rm c}$ and $v_{\rm frag}$ also explain the observed metallicity, whereas the model with $M_{\rm disk}=0.03M_{\rm *}$ does a poor job to explain the observations, implying the need of the initial disk mass of $\sim0.1M_{\rm *}$ to explain the metal-rich atmospheres of close-in gas giants (see Figure \ref{fig:mass_metal2}).

Although the atmospheric metallicity of the \rev{I-SL population} aligns with those observed for many close-in giants, it struggles to explain gas giants exhibiting sub-stellar metallicities.
Within our modeling framework, such low metallicities can be explained only by the \rev{O-SL population}, since the sublimation of inward-drifting pebbles enrich the inner disk gas, leading the \rev{I-SL population} to inevitably acquire metal-enriched disk gases.
This suggest that gas-giants with sub-stellar atmospheric metallicity possibly formed at distant orbits.
This result is also consistent with \citet{Bitsch+22} who also suggested that WASP-77Ab---a gas giant with a sub-stellar atmospheric metallicity \citep{Line+21,August+23,Edwards&Changeat24,Smith+24}---originally formed beyond the CO$_2$ snowline.

Within the \rev{O-SL population}, the unmixed atmosphere scenario appears to better explain the gas giants with sub-stellar atmospheric metallicities.
The fully mixed envelope inherently incorporates vapors sublimated from accreted pebbles during core formation, making sub-stellar metallicity difficult to achieve. This is especially true for low planetary mass at which the pre-existing vapors are not well diluted by accreted disk gas.
Thus, low-mass planets with sub-stellar atmospheric metallicity, such as TOI-421b \citep[$M_{\rm p}=0.021M_{\rm J}$,][]{Davenport+25} and HAT-P-18b \citep[$M_{\rm p}=0.197M_{\rm J}$,][]{Fournier+24_HAT-P-18b}, are strong candidates that originate from distant orbits and retains envelopes with inefficient convective mixing.

Our model struggles to explain the highly metal-rich atmospheres observed in several low-mass ($<0.1M_{\rm jup}$) planets.
The \rev{I-SL population} reaches atmospheric metallicities only up to [O/H]$\sim1.5$, which is insufficient to explain [O/H]$\gtrsim2.0$ suggested for many sub-Neptunes, including TOI-270d \citep{Benneke+24,Holmberg&Madhusudhan24}, GJ3470b \citep{Beatty+24}, GJ436b \citep{Mukherjee+25}, GJ1214b \citep{Kempton+23,Gao+23,Schlawin+24,Ohno+25}, GJ9827d \citep{Piaulet+24}, and GJ3090b \citep{Ahrer+25a}.
While the \rev{O-SL population} can yield [O/H]$\gtrsim2.0$ at low planetary masses, the simulated metallicities are still systematically lower than those inferred from JWST observations of sub-Neptunes.
This discrepancy arises because pebble accretion ceases around $M_{\rm p}\sim0.01M_{\rm jup}$ in our current model, triggering disk gas accretion to prevent retaining the metal-rich atmospheres on higher mass planets.
We however discuss the caveats for the present results of sub-Neptunes in Section \ref{sec:discussion}.

\subsection{Mass-Bulk Metallicity Relation}
\begin{figure}[t]
{
\centering
\includegraphics[clip,width=\hsize]{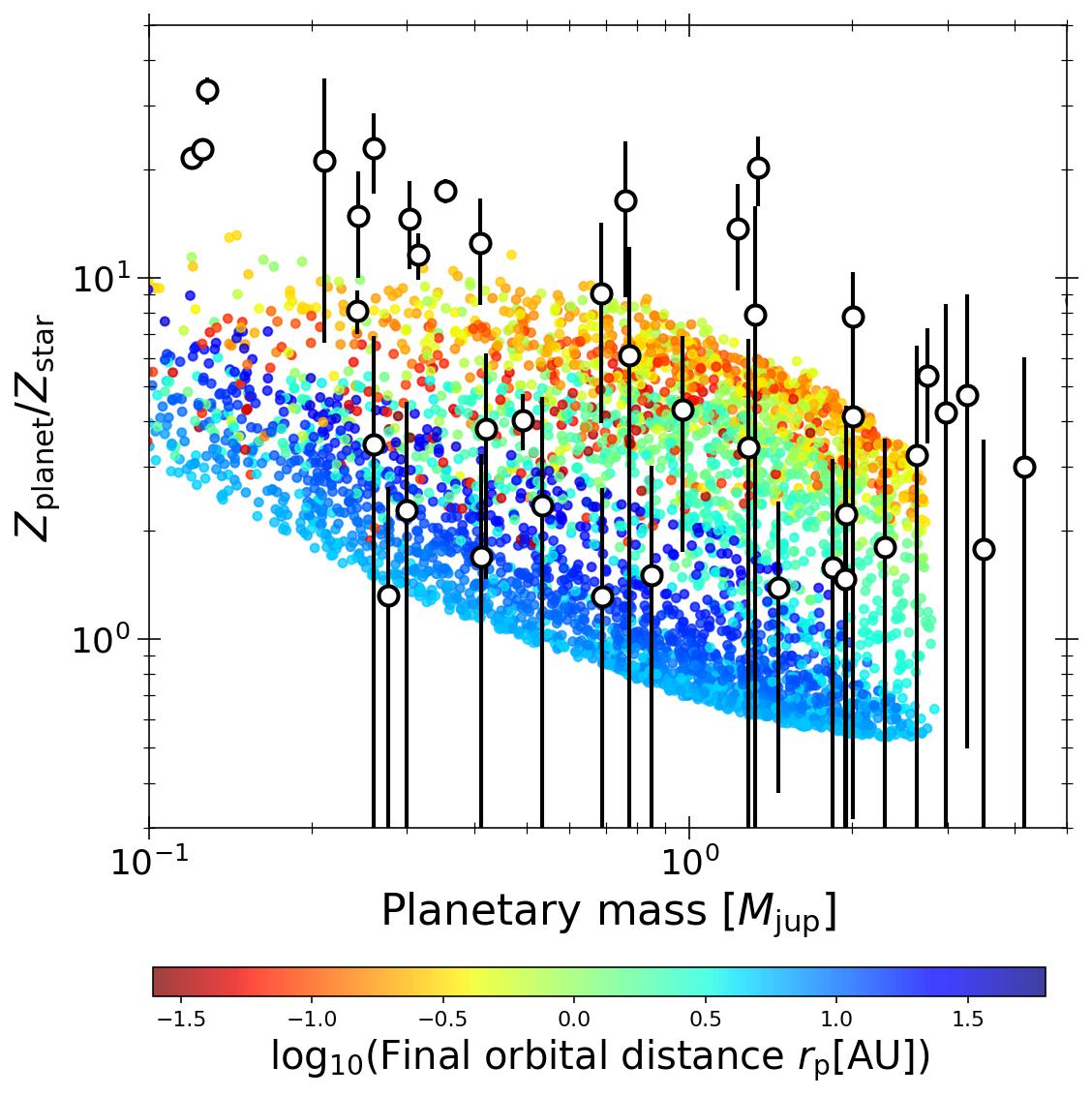}
}
\caption{
Planetary bulk metallicity normalized by stellar metallicity as a function of planetary mass. The bulk metallicity is extracted from our population synthesis shown in Figure \ref{fig:mass_metal}. 
The color denotes the final orbital distance of the planets.
The data points exhibit the normalized bulk metallicity of warm giants derived by \citet{Howard+25} for core+envelope interior structures.
}
\label{fig:mass_bulk}
\end{figure}

\begin{figure*}[t]
{
\centering
\includegraphics[clip,width=\hsize]{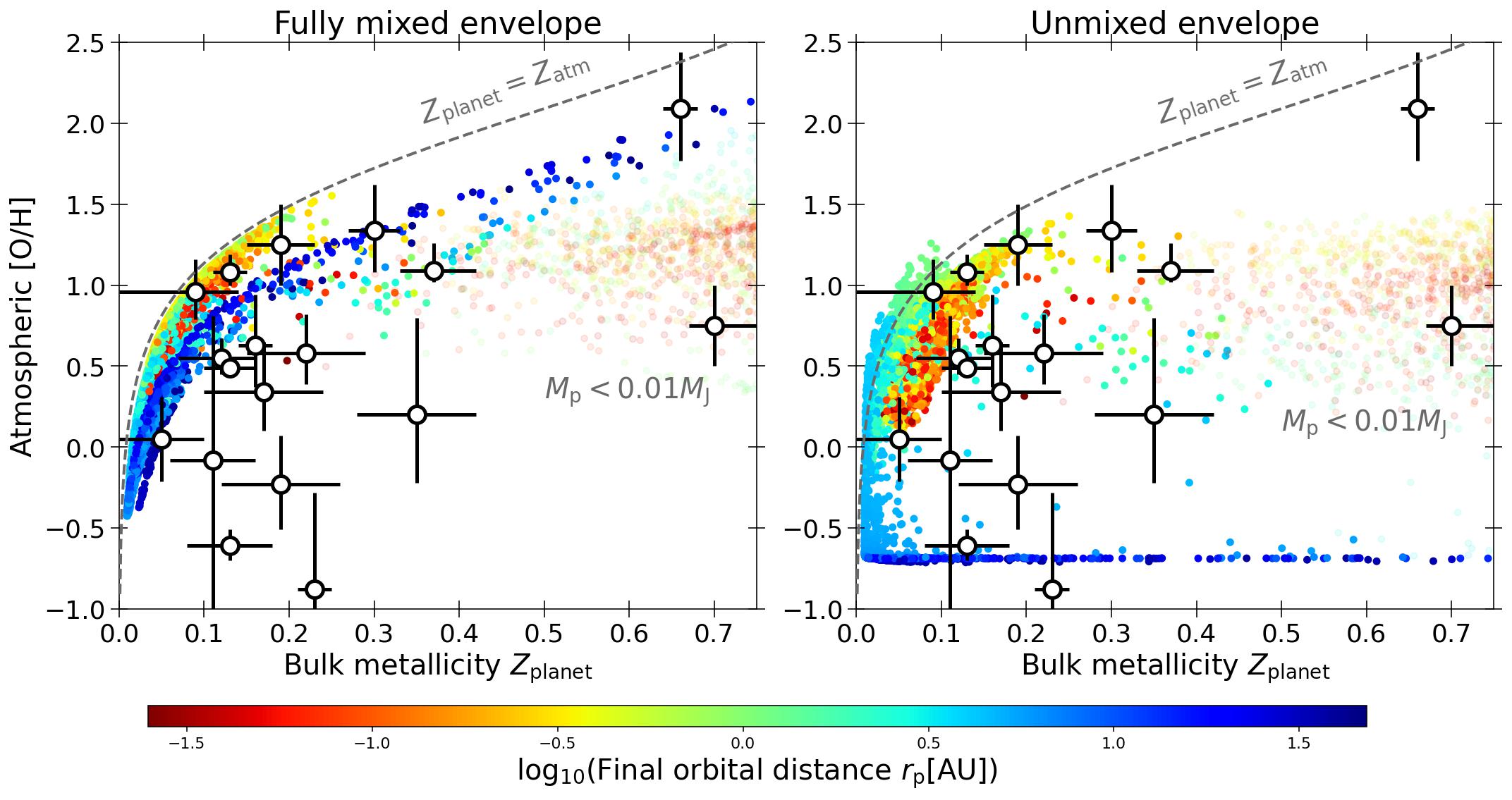}
}
\caption{
Atmospheric metallicity $\log_{\rm 10}[{\rm(O/H)/(O/H)_{star}}]$ as a function of bulk metallicity. 
The colored dots exhibit the synthetic planets for population synthesis shown in Figure \ref{fig:mass_metal}.
The left and right panels exhibit the results for fully mixed and unmixed envelopes, respectively.
The gray dashed line shows the relation of Equation \eqref{eq:Z_convert} for $Z_{\rm planet}=Z_{\rm atm}$ and C/O=0.5. The data points are taken from the atmospheric metallicity reported by JWST along with the bulk metallicity inferred by previous thermal evolution models (for corresponding references, see Table \ref{table:JWST}). The color denotes the final orbital distance of the planets at the timing of disk dissipation.
Planets with masses of $<0.01M_{\rm jup}$ are denoted by transparent dots.
}
\label{fig:bulk_atm}
\end{figure*}


The accretion of vapor-enriched disk gas has been proposed as a possible origin for the high {\it bulk} metallicity of close-in giant planets \citep{Schneider&Bitsch21}. 
Figure \ref{fig:mass_bulk} presents the bulk metallicity of our synthetic giant planets as a function of planetary mass.
In our model, planets in the \rev{I-SL population} exhibit higher bulk metallicity compared to those formed at distant orbits.
This trend indicates that the accretion of vapor-enriched disk gas---the main source of metals of the \rev{I-SL population}---is a more efficient way to acquire heavy elements than forming a massive core, confirming the finding of \citet{Schneider&Bitsch21}.

The bulk metallicity of our synthetic planets decreases with increasing planetary mass, in line with the trends inferred from interior structure modeling \citep{Guillot+06,Miller&Fortney11,Thorngren+16,Muller&Helled23,Howard+25,Chachan+25}.
This mass-bulk metallicity anti-correlation arises from the same physical mechanism that causes the atmospheric metallicity trend: strong disk turbulence reduces the metallicity of disk gas and facilitates the growth of giant planets for the \rev{I-SL population} (see Section \ref{sec:atm_bulk_metal}).

We also compare our simulated bulk metallicity with the bulk metallicity of warm Jupiters estimated by \citet{Howard+25} with a thermal evolution model.
\rev{During the revision of this paper, \citet{Chachan+25} also conducted the comprehensive investigations on the bulk metallicity of warm Jupiters with a greater sample size, but the result is in line with \citet{Howard+25}}.
Overall, our simulated bulk metallicity is broadly consistent with those estimated by \citet{Howard+25}.
Our \rev{I-SL population} reasonably aligns with the systematically high bulk metallicity inferred for warm Jupiters.
Although some planets exhibit bulk metallicity higher than our model prediction, this can be attributed to the fact that we restrict the disk initial mass and radius to $M_{\rm disk}/M_{\rm *}=0.1$ and $r_{\rm c}=148~{\rm au}$) for the synthetic planets shown in Figure \ref{fig:mass_bulk}, and different disk properties could allow a higher bulk metallicity for a given mass.
Although our model can explain the observational mass--bulk metallicity relation, we caution that the estimate of bulk metallicity involves some model-dependent aspects; for example, choice of equation of state and assumptions imposed on a metal distribution can alter the inferred bulk metallicity \citep[e.g.,][]{Thorngren+16,Howard+25}.

\subsection{Bulk-Atmosphere Metallicity Relation}
\citet{Bean+23} recently suggested that atmospheric metallicity correlates more closely with bulk metallicity than with planetary mass.
Figure \ref{fig:bulk_atm} shows atmospheric [O/H] as a function of the bulk metallicity for our synthetic population.
For fully mixed envelopes, the \rev{I-SL population} exhibits a clear correlation between the atmospheric and bulk metallicities.
The correlation lies close to the relation of $Z_{\rm atm}=Z_{\rm planet}$ \footnote{We note that atmospheric metallicity is often reported as the number fraction normalized by solar value, i.e., ${\rm[O/H]=\log_{\rm 10}[(O/H)/(O/H)_{\rm sun}]}$.
This literature metallicity can be translated to the metal mass fraction $Z_{\rm atm}$ as
\begin{equation}
    Z_{\rm atm}\approx\frac{16({\rm O/H})+12({\rm C/H})}{1+4({\rm He/H})+16({\rm O/H})+12({\rm C/H})}
\end{equation}
Alternatively, the literature metallicity can be written as a function of $Z_{\rm atm}$ as
\begin{equation}\label{eq:Z_convert}
    {\rm[O/H]}=\log_{\rm 10}\left[\frac{1+4({\rm He/H})}{(16+12({\rm C/O})){\rm(O/H)_{\rm sun}}}\frac{Z_{\rm atm}}{1-Z_{\rm atm}}\right]
\end{equation}}.
This trend emerges because heavy elements in these planets are primarily delivered through vapor-enriched disk gas that constitute atmospheres, making the atmospheric metallicities nearly equivalent to the bulk metallicity.
Although the bulk-atmosphere metallicity correlation exhibits a greater spread for unmixed envelopes (right panel of Figure \ref{fig:bulk_atm}), the relation still lies around the relation of $Z_{\rm atm}=Z_{\rm planet}$.
It is worth noting that the unmixed envelope allows the presence of a planet whose atmospheric metallicity is higher than bulk metallicity, as the disk gas metallicity at the timing of disk dissipation can be higher than the time averaged gas metallicity.

The bulk--atmosphere metallicity relation of \rev{O-SL population} strongly depends on the inhomogeneity of envelope.
For fully-mixed envelope, the \rev{O-SL population} shows a parallel but offset trend relative to the $Z_{\rm atm}=Z_{\rm planet}$ relation.
This offset arises because distant planets primarily acquire metals during the core formation, and part of metals are sequestered into a solid core part.
Pebble accretion can build a solid core of up to $\sim0.5M_{\rm \oplus}$ \citep{Brouwers+18}, and subsequently accreted pebbles sublimate to yield vapors.
Thus, the total core mass can be split into solid and vapor parts, i.e., $M_{\rm core}=M_{\rm solid}+M_{\rm vapor}$, with only the vapor component contributing to atmospheric metals.
Then, the atmospheric metal mass fraction can be expressed as
\begin{equation}
    Z_{\rm atm}=\frac{M_{\rm core}-M_{\rm solid}}{M_{\rm p}}=Z_{\rm planet}\left(1-\frac{M_{\rm solid}}{M_{\rm core}}\right),
\end{equation}
where we have approximated the bulk metallicity to $Z_{\rm planet}\approx M_{\rm core}/M_{\rm p}$ by ignoring metals delivered by gas accretion. 
Thus, the \rev{O-SL population} shows a bulk--atmosphere metallicity correlation shifted from $Z_{\rm atm}=Z_{\rm planet}$ by a factor of $1-M_{\rm solid}/M_{\rm core}$, as long as the vapors generated by accreted pebbles can uniformly mix in envelopes.
For unmixed envelopes, the \rev{O-SL population} turns out to show atmospheric metallicity lying around [O/H]$\sim-0.7$ because accreting disk gas, which is depleted in metals at distant orbits, dictates the final atmospheric compositions.

For giant planets with high atmospheric metallicity of [O/H]$\gtrsim1.0$, multiple scenarios can reasonably explain the bulk--atmosphere metallicity relation constrained by JWST.
Many such metal-rich giants lie around the simulated bulk--atmosphere metallicity relation of \rev{I-SL population} for both fully-mixed and unmixed envelopes.
The \rev{O-SL population} can also explain these planets if the envelope is fully mixed.
Thus, bulk--atmosphere metallicity relation alone struggles to distinguish these scenarios for giant planets when they have high atmospheric metallicity.

In contrast, our simulations for the unmixed envelopes better explain the bulk--atmosphere metallicity relation for planets with low atmospheric metallicity of [O/H]$\lesssim0.5$. 
The fully mixed scenario fails to explain the combination of low atmospheric metallicity and high bulk metallicity suggested for several giant planets.
This discrepancy occurs because low atmospheric metallicity inevitably accompanies low bulk metallicity for the fully mixed envelope.
By contrast, the unmixed envelope scenario better explains the observed planets with low atmospheric metallicity.
This is because low-metallicity disk gas can accrete on the top of a large core without pollution by a preexisting vapor atmosphere, allowing the formation of a planet with high bulk metallicity but with low atmospheric metallicity.
In other words, different processes independently determine the bulk and atmospheric metallicity for \rev{O-SL population} with unmixed envelopes, allowing flexibility to achieve the various combinations of bulk and atmospheric metallicity.
Our results indicate that the bulk--atmosphere metallicity relation provides a powerful metric to identify giant planets with unmixed envelopes originating from the distant orbits.
\section{Discussion}\label{sec:discussion}
\begin{figure*}[t]

\centering
\includegraphics[clip,width=0.75\hsize]{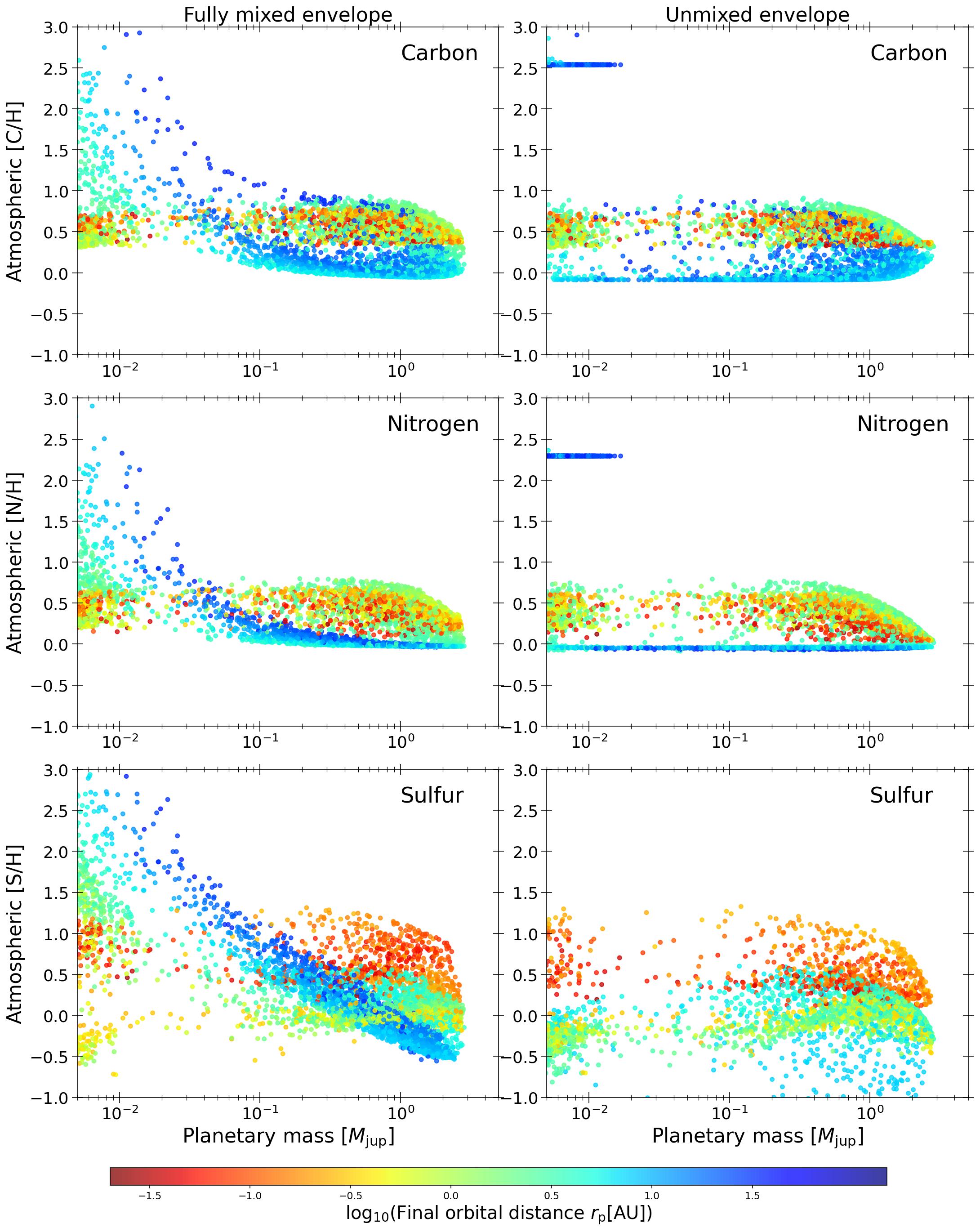}
\caption{
The same as Figure \ref{fig:mass_metal}, but for relations between planetary mass and C/H (upper row), N/H (middle row), and S/H (lower row). 
}
\label{fig:mass-metal_other}
\end{figure*}
\subsection{Mass--metallicity relation for other elements}\label{sec:mass_metal_other}
In this study, we focus on oxygen as a representative heavy element, since its abundance is likely the most readily accessible through atmospheric observations.
Meanwhile, other elements do not necessary follow the trend seen in atmospheric [O/H].
\citet{Welbanks+19} indeed suggested that atmospheric O, Na, and K abundances may follow different anti-correlations with planetary mass from their uniform retrieval analysis.
JWST has begun to detect (possible hints of) molecules containing carbon \citep[e.g.,][]{Bell+23,Madhusudhan+23,Benneke+24,Welbanks+24}, nitrogen \citep[e.g.,][]{Yang+24,Welbanks+24,Mayo+25,Wiser+25}, sulfur \citep[e.g.,][]{Rustamkulov+23,Alderson+23,Dyrek+23,Beatty+24,Fu+24}, and other refractory elements like K and Si \citep{Feinstein+23,Evans+25}.
Motivated by this context, we discuss the possible relations of atmospheric carbon, nitrogen, and sulfur abundances with planetary mass in this section.

\rev{Before introducing the results, we caveat that the present study deliberately ignores the presence of refractory carbons such as CHON organics and amorphous carbon, as the predominant processes destructing refractory carbons remain controversial \citep{Lee+10,Anderson+17,Gail&Trieloff17,Klarmann+18,Binkert&Birnstiel23,Okamoto&Ida24,Vaikundaraman+25}.
Our current model assumes that refractory carbons are destructed in the beginning of the disk evolution, yielding more volatile carbons that tend to stay at distant orbits.
Atmospheres of planets formed at close-in orbits would have higher C/H ratio than what computed here if refractory carbons are mainly decomposed at inner hot orbits---the so-called soot line---and enhance carbon abundance in inner disk gas \citep[][]{Houge+25}.
With keeping some caveats on C reservoirs in mind, our simulations show that} C/H, N/H, and S/H exhibit distinct mass-metallicity relations depending on planetary birthplaces, as similar to O/H, whereas we find that each element shows a different vertical offset.

\subsubsection{Carbon}
Atmospheric C/H shows systematically lower abundances than those of O/H.
The top row of Figure \ref{fig:mass-metal_other} shows the atmospheric C/H as a function of planetary mass, where we analyze the simulations shown in Figure \ref{fig:mass_metal}.
Carbon abundances of \rev{I-SL population} are confined to [C/H]$\sim0$--$1.0$, which is lower than the oxygen abundance found for the \rev{I-SL population} ([O/H]$\sim0.5$--$1.5$, see Figure \ref{fig:mass_metal}).
The systematically low [C/H] stems from the fact that the large fraction of carbon is accommodated into volatile molecules, namely, CO, CH$_4$, and CO$_2$.
Inward drifting pebbles release vapors of these highly volatile molecules much farther away from the central star than H$_2$O snowline, making the carbon enrichment relatively inefficient at inner orbits compared to oxygen.

\rev{The O-SL population} also exhibits relatively low abundances for [C/H].
This is because pebbles tend to lose the highly volatile C-bearing molecules before reaching the core orbit, making the deposition of carbon through the sublimation of accreted pebbles inefficient during the core formation.
Consequently, the \rev{O-SL population} also exhibits relatively low [C/H] compared to [O/H] for low planetary mass, for which the vapor generated by accreted pebbles controls atmospheric elemental abundances.
At high planetary mass, unlike [O/H], [C/H] tends to converge to a near stellar value, since the disk gas at distant orbits can retain relatively abundant CO, CH$_4$, and CO$_2$ vapors compared to H$_2$O.

\begin{figure}[t]
\includegraphics[clip,width=\hsize]{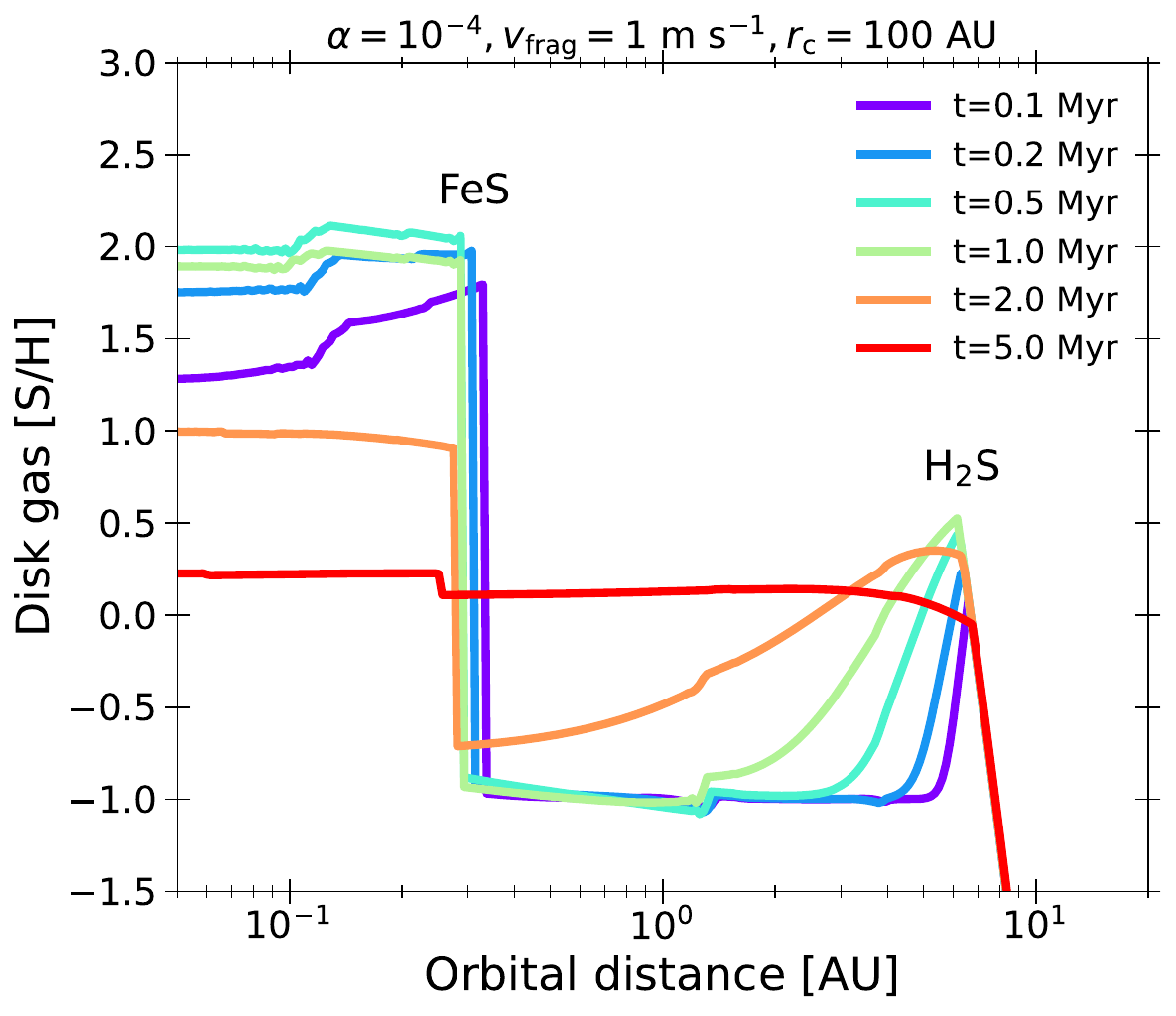}
\caption{
Radial profile of sulfur abundance [S/H] in disk gas. The different colored lines show the profiles at different time. We have set $\alpha={10}^{-4}$, $v_{\rm frag}=1~{\rm m~s^{-1}}$, and $r_{\rm c}=100~{\rm au}$.
}
\label{fig:StoH_example}
\end{figure}

Although we have discussed the potential trend seen in C/H, there are caveats regarding the present results.
This study partitions carbon into CO, CO$_2$, and CH$_4$ (see Appendix \ref{sec:appendix_A}), while observations of comets suggest that approximately half of the carbon is likely present as refractory organics in pristine dust and subsequently converted to gas phase during the disk evolution \citep[e.g.,][]{Bergin+15}.
The process of converting refractory carbons to a gas phase is still under active debate \citep[][]{Lee+10,Anderson+17,Gail&Trieloff17,Klarmann+18,Binkert&Birnstiel23,Okamoto&Ida24,Vaikundaraman+25}. 
Our current setup should be regarded as the limit in which stellar FUV photons quickly decompose refractory carbons to form hydrocarbons such as CH$_4$ even at distant orbits.

\subsubsection{Nitrogen}
The trend of atmospheric N/H resembles that seen in C/H, as shown in the middle row of Figure \ref{fig:mass-metal_other}.
The \rev{I-SL population} exhibits nitrogen abundances confined to [N/H]$\sim0$--$0.7$, and the \rev{O-SL population} exhibits a nearly stellar nitrogen abundance.
The similarities arise because, as similar to carbon, nitrogen is also largely accommodated into a highly volatile molecule, N$_2$.
Thus, inward-drifting pebbles release majority of nitrogen at far away from the central star ($\sim100~{\rm au}$, see Figure \ref{fig:basic}), making nitrogen enrichment inefficient at inner disk gases and thus in the \rev{I-SL population}.
The pebbles at $<100~{\rm au}$ also lack N$_2$ ices, resulting in the inefficient nitrogen deposition through the sublimation of pebbles accreted onto the core for the \rev{O-SL population}.

However, we note caveats for the predictions of nitrogen abundances too.
Recent work by \citet{Nakazawa&Okuzumi25} showed that nitrogen transport by ammonium salts, not included in the present model, can appreciably affect the nitrogen abundance of giant planetary atmospheres.
Furthermore, the interpretation of nitrogen can depend on actual disk temperature structure, as N$_2$ freezes only at $\gtrsim30~{\rm au}$ for passively irradiated disks \citep{Oberg&Wordsworth19,Bosman+19} while the local cold regions due to disk shadow can allow the presence of N$_2$ ice even at $\sim5~{\rm au}$ \citep{Ohno&Ueda21,Notsu+22}.

\subsubsection{Sulfur}
Intriguingly, [S/H] exhibits a relatively well-separated trend as compared to [O/H], [C/H], and [N/H] in our model.
The bottom row of Figure \ref{fig:mass-metal_other} shows atmospheric [S/H] as a function of planetary mass.
The \rev{I-SL population} seen in [O/H], [C/H], and [N/H] can be further split into two sub-groups for [S/H].
Planets with $r_{\rm p}\lesssim0.3~{\rm au}$ exhibit [S/H]$\sim0.5$--$1.5$ that is comparable to [O/H].
The high sulfur abundance is attributed to the sublimation of FeS that greatly enriches the disk gas in sulfur at $\lesssim0.3~{\rm au}$, as illustrated in Figure \ref{fig:StoH_example}.
At farther orbits, atmospheric sulfur abundance drops to stellar to sub-stellar values because of sequestration of sulfur into FeS rock at such orbits. 
At orbits of $\sim10~{\rm au}$, sublimation of accreted pebbles containing FeS deposit abundant sulfur during core formation, leading to cause a clear anti-correlation between planetary mass and [S/H] if the envelope is fully mixed, as similar to the \rev{O-SL population} for [O/H].

\rev{We note that our current model treats the FeS decomposition as a reversible process similar to H$_2$O ices for simplicity (Appendix \ref{sec:appendix_A}).
In reality, FeS decomposition produces volatile molecules such  H$_2$S and S$_2$ \citep{Tachibana&Tsuchiyama98,Steinmeyer+23},
which might diffuse to outer orbits without recondensation to enrich the atmospheres of \rev{O-SL population} in sulfur.
A more detailed approach to deal with the kinetics of FeS-forming reactions \citep[e.g.,][]{Pasek+05,Ciesla+15} would be needed to assess whether such outward sulfur transport occurs.
}

One important finding is that planets at close-in orbits relevant to hot Jupiters ($\lesssim0.3$ au) typically have super-stellar [S/H], though with a large scatter.
We stress that the atmospheric sulfur of thI-SL population comes from the sulfur-rich disk gas enriched by FeS sublimation.
Several recent observations by JWST detected SO$_2$ and H$_2$S in the atmospheres of close-in planets \citep{Rustamkulov+23,Alderson+23,Dyrek+23,Beatty+24,Welbanks+24,Fu+24,Ahrer+25b} that are often interpreted as the sign of planetesimal accretion \citep{Turrini+21,Pacetti+22}.
However, our result suggests that the atmospheric sulfur abundance alone may be insufficient to quantify how much solid materials like planetesimals accreted onto the close-in planets, as their orbits are already inside the FeS sublimation line.
We briefly discuss the possible relation between atmospheric S/O and metallicity (O/H) in Appendix \ref{Appendix_sulfur} but defer further analysis to future studies.
\rev{Note that envelope accretion on low-mass cores of $\lesssim10M_{\rm \oplus}$ is suggested to be inefficient at close-in orbits of $\lesssim0.1~{\rm au}$ due to atmospheric recycling \citep[e.g.,][]{Moldenhauer+22}, which may prevent planets from acquiring sulfur-rich disk gas at close-in orbits.}

\begin{figure*}[t]
\centering
\includegraphics[clip,width=0.9\hsize]{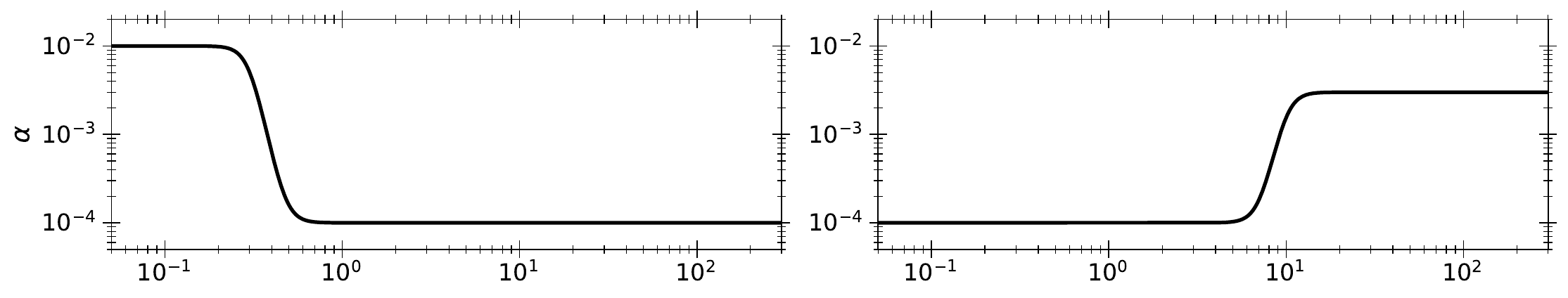}
\includegraphics[clip,width=0.91\hsize]{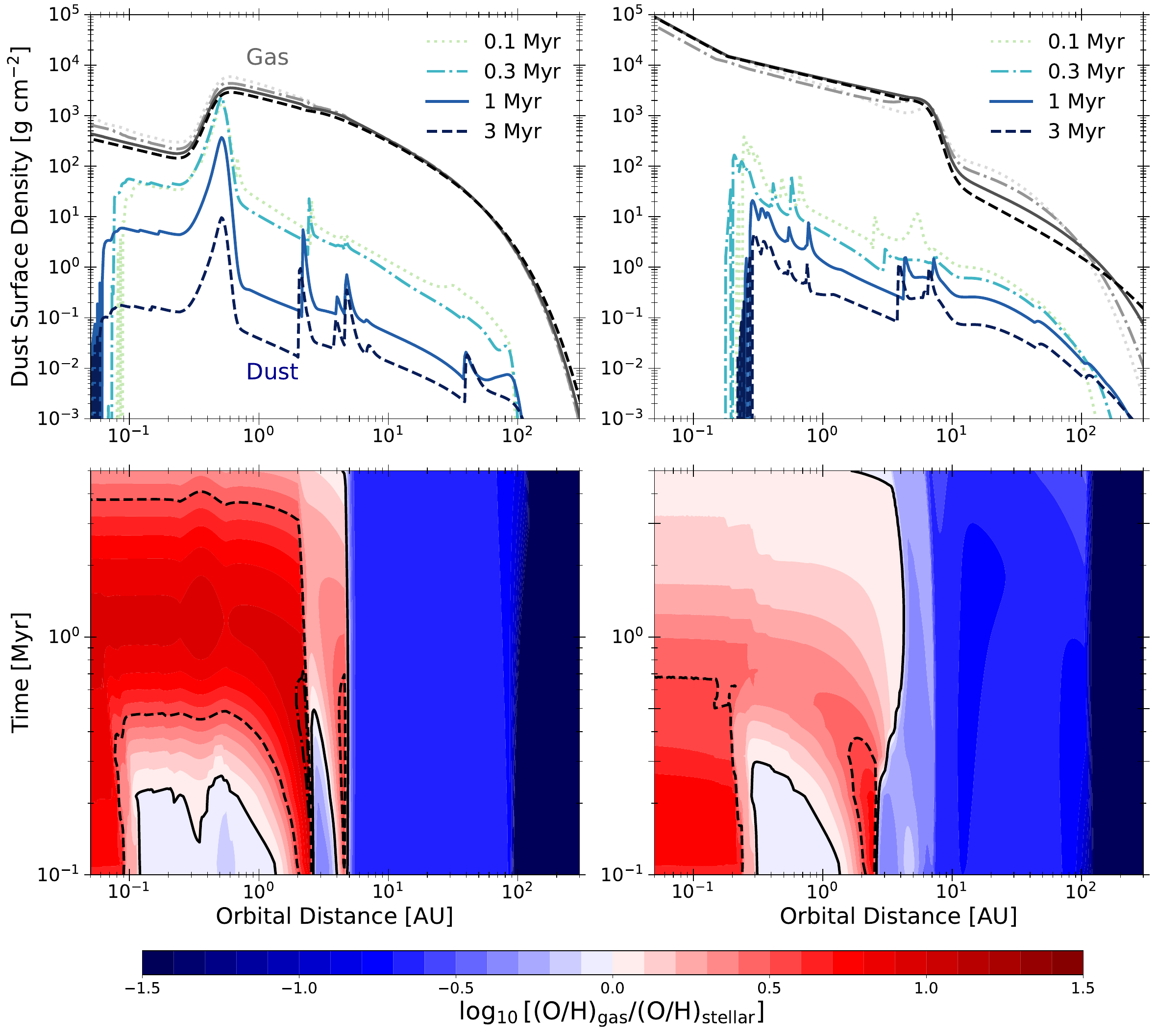}

\caption{
Evolution of surface density (middle) and O/H ratio of disk gas (bottom) under radially-varying turbulence strength $\alpha$ denoted by top panels.
}
\label{fig:alpha_var}
\end{figure*}

\subsection{\rev{Effects of radial variation of turbulence strength}}
\rev{
We have adopted a single value of $\alpha$ to describe turbulence strength across entire disks.
In reality, however, the turbulence strength is expected to vary with orbital distance.
For example, hot disk gasses with $T\gtrsim1000$ K are expected to have a high ionization degree due to thermal ionization of alkali metals, which drives vigorous turbulence at close-in orbits through the magnetorotational instability (MRI) \citep{Gammie96,Desch&Turner15}.
MRI also tends to operate at outer regions of $\gtrsim10~{\rm au}$ because of high ionization degree achieved by cosmic ray attenuation \citep{Delage+22,Delage+23}.
}

\rev{
To examine the impacts of such radially-varying turbulence, we conduct additional simulations of disk evolution that allow enhanced turbulence at either inner or outer orbits.
We introduce a following prescription to mimic the radial variation of turbulence strength.
\begin{equation}
    \alpha(r)=\alpha_{\rm 0}+\frac{1}{2}(\alpha_{\rm MRI}-\alpha_{\rm 0})[1-\tanh{(x\log{(r/r_{\rm MRI})})}],
\end{equation}
where $\alpha_{\rm MRI}$ stands for the enhanced turbulence strength arising from MIR, and $x$ defines the steepness of the transition in $\alpha$ at $r=r_{\rm MRI}$.
We conduct simulations with $(\alpha_{\rm 0},\alpha_{\rm MRI},r_{\rm MRI}~{\rm [au]},x)=(10^{-4},10^{-2},0.3,5)$ to represent the MRI-active regions at close-in orbits and $(\alpha_{\rm 0},\alpha_{\rm MRI},r_{\rm MRI}~{\rm [au]},x)=(10^{-4},3\times10^{-3},10,-5)$ for MRI-active regions at distant orbits, where the parameter choice mimics the previous relevant simulations \citep[e.g.,][]{Ueda+19,Delage+22}.
}

\rev{
We find that inner MRI-active regions barely affects the metallicity evolution of inner disk gas, whereas the outer MRI-active regions act to reduce the metallicity of inner disk gas.
Figure \ref{fig:alpha_var} shows the time evolution of gas and dust surface densities as well as O/H ratio of disk gas.
Inner MRI-active region reduces the gas surface density at close-in orbits and produces a dust pileup around there because of the emergence of local pressure maximum.
Meanwhile, the disk gas at $\lesssim1~{\rm au}$ retains a high O/H ratio which is comparable to that found in the simulation with uniform turbulence of $\alpha={10}^{-4}$ (right column of Figure \ref{fig:basic}).
The comparably high O/H ratio stems from the fact that inner MRI-active region does not affect the pebble-to-gas mass flux ratio from outer regions that dominate the total pebble mass.
Thus, turbulence strength at outer regions plays a role in controlling the metal enrichment of inner disk gas.
This can be seen in the right column of Figure \ref{fig:alpha_var} showing that outer MRI-active regions act to reduce the O/H ratio of inner disk gas, which is the consequence of the enhanced dust fragmentation and reduced pebble flux from outer regions.
}

\subsection{Relation to Previous Studies}
Although our model is similar to several previous studies \citep{Booth+17,Schneider&Bitsch21,Danti+23,Penzlin+24} who also simulated the enrichment of disk gas through pebble drift and atmospheric compositions of giant planets, there are several model differences between previous studies and ours.
\citet{Penzlin+24} adopted a steady-state disk model with constant global accretion rate, dust-to-gas accretion rate ratio, and stokes number, while we simulate the time evolution of gas and dust disks by explicitly calculating pebble growth and drift for given disk parameters (i.e., $\alpha$, $v_{\rm frag}$, etc).
The present model has the advantage of dealing with time-dependent processes such as a transient rise followed by a gradual decay of the disk gas metallicity, which is controlled by pebble drift and exhaustion.
For orbital migration of giant planets, \citet{Booth+17} and \citet{Schneider&Bitsch21} adopted the classical Type II migration \citep[e.g.,][]{Baruteau+14}, whereas this study adopts a recent type II migration model in which the migration speed is scaled by gas surface density at the bottom of gas gap \citet{Kanagawa+18} that is more consistent with recent hydrodynamical simulations \citep[e.g.,][]{Paardekooper+23}.
Consequently, our synthetic gas giants avoid large-scale disk-driven migration and tend to end up in orbits close to where runaway gas accretion sets in, as suggested by \citet{Tanaka+20}.
Despite several differences in modeling detail, our model obtains results broadly consistent with previous studies; for example, atmospheric metallicity of giant planets at ${\lesssim}1~{\rm au}$ reaches [O/H]$\sim0.5$--$1.5$ and reasonably explains the high bulk metallicity of warm Jupiters \citep{Schneider&Bitsch21,Schneider&Bitsch21b}, and our model produces the anti-correlation between atmospheric C/O and O/H ratios in our population synthesis (Appendix \ref{Appendix:CtoO_OtoH}), as similar to \citet{Penzlin+24}.

\rev{Our work is the first study that investigates the atmospheric metallicity trend in the context of pebble accretion paradigm based on a model that simulates dust and gas disk evolution along with giant planet formation in a self-consistent manner.
\citet{Savvidou&Bitsch23} carried out the population synthesis with the time-evolving pebble accretion model of \citet{Schneider&Bitsch21} as similar to our study, while they focused on the conditions of giant planet formation, not the trend of atmospheric compositions.
\citet{Penzlin+24} conducted the atmospheric population synthesis with the model accounting for both planetesimal and pebble accretion, whereas they adopted a steady-state disk model with parameterized gas accretion rate, pebble-to-gas mass flux ratio, and pebble Stokes number.
Although such a parameterized steady model is useful for rapid computation and for controlling key parameters of planet formation, it introduces a risk to cause an inconsistency among model parameters, such as the small Stokes number accompanied with high pebble flux and high pebble flux lasting for very long time.
We also note that our time-evolving disk model does not reach a steady state assumed in \citet{Penzlin+24}, since pebble flux evolves over time in general. 
}
\citet{Danti+23} calculated atmospheric metallicity as a function of planetary masses using an updated model of \citet{Schneider&Bitsch21}. 
Although the presence of atmospheric metallicity trends was ambiguous in \citet{Danti+23} due to the limited number of simulated planets, their mass--metallicity relation for pebble accretion scenario appears to resemble the relation simulated for the \rev{I-SL population} in this study.
On the other hand, their simulation apparently did not exhibit the mass--metallicity anti-correlation for the \rev{O-SL population}.
This difference is presumably attributed to their assumption that only 10\% of the accreted pebbles contribute to primordial vapor atmospheres, which causes the mass--metallicity anti-correlation for distant planets in our model.

\subsection{Relation to present-day planetary orbits}
We note that the final orbital locations of planets in our simulations do not necessarily correspond to their present-day orbits, particularly for those in the \rev{O-SL population}.
While our model accounts for orbital migration driven by planet-disk interactions, it does not include migration processes that occur after disk dispersal.
High-eccentricity migration mechanisms---such as planet-planet scattering or secular interactions---can transport planets formed at distant orbits to close-in regions \citep[for a review, see e.g.,][]{Dawson+18}.
Our results rather suggest that the detailed shape of the mass--metallicity relation, particularly the fraction of planets with sub-stellar atmospheric metallicity, may provide insight into the proportion of close-in planets that completed their formation at distant orbits.

\subsection{Model caveats}\label{sec:model_caveat}


\subsubsection*{Feedback of planet formation on disk evolution}
In this study, we have ignored the feedback of planet formation on disk evolution.
In reality, planet formation would affect disk evolution in multiple ways. 
The envelope accretion of a planet consumes disk gas and reduces the downstream gas surface density \citep[e.g.,][]{Tanigawa&Tanaka16,Rosenthal+20}. 
The gas gap opened by the planet can also trap drifting pebbles, thereby reducing the pebble accretion rate and the disk gas enrichment through pebble sublimation at downstream orbits \citep{Kalyaan+23,Eberlein+24,Easterwood+24}.
Ignoring these effects is justified for single-plane systems, as the planet only affects the downstream gas and pebble flux.
However, the feedback of planet formation can affect the final atmospheric compositions when multiple planets form in the same system.
For example, giant planets formed beyond H$_2$O snowline filter drifting pebbles and may reduce the metallicity of inner disk gas and thus atmospheric metallicity of \rev{I-SL population}, depending on timing of each core formation and orbital configurations.
Further studies will be warranted to understand how the mass--metallicity relation may differ between single- and multiple-systems.

\subsubsection*{\rev{Disk accretion process}}
\rev{
Although we have adopted a conventional viscously accreting disk model, the disk accretion may be primary driven by angular momentum removal by magnetohydrodynamical wind \citep[e.g.,][]{Bai&Stone13,Suzuki+16,Tabone+22}.
Such windy disk introduces qualitative differences in planet formation process, such as disk's thermal structure \citep{Kondo+23,Mori+25}, planet-induced gap profile \citep{Aoyama&Bai23}, and the direction of orbital migration \citep{Kimming+20,W&Chen25}.
Further studies would be warranted to investigate the trends of atmospheric composition of exoplanets formed within windy disks.
}

\subsubsection*{Disk thermal structure}
We calculate the disk temperature structure from stellar radiation and viscous heating as in many previous studies \citep[e.g.,][]{Booth+17,Schneider&Bitsch21,Danti+23}, whereas the actual disk temperature structure may differ from the present model.
Several studies have demonstrated that disk substructures, such as dust pileup, can cast shadows and greatly affect the thermal structures and compositional structures of outer orbits \citep{Ueda+19,Ueda+21,Ohno&Ueda21,Okuzumi+22}.
We anticipate that our predictions for the [O/H] trend would hold for shadowy disks, as the H$_2$O snowline is located in relatively inner orbits where viscous heating plays a role.
In contrast, the disk shadows can affect the interpretation for other elements, especially for nitrogen, as discussed in Section \ref{sec:mass_metal_other}.

\subsubsection*{Disk chemistry}
We have ignored disk chemistry that converts one molecular species to another.
In reality, disk chemistry can alter each molecular abundance; for example, \rev{disk observations have revealed that CO gasses in the disks undergo a factor of $10$--$100$ depletion with a timescale of $\sim1~{\rm Myr}$ \citep[e.g.,][]{Zhang+20_CO}, which is presumably a consequence of the} chemical processing that can convert CO to CO$_2$, CH$_3$OH, and hydrocarbons \citep[e.g.,][]{Furuya&Aikawa14,Bosman+18,Eistrup+18,Notsu+22,Furuya+22}. 
This process alters the CO abundance and would cause a spread in the sub-stellar metallicity of the \rev{O-SL population} with unmixed envelopes, which is concentrated on [O/H]$\sim-0.7$ in the present study (see Figure \ref{fig:mass_metal}).
\rev{Since the CO depletion chemistry affects the carbon reservoirs with a timescale of $\sim1$ Myr, it would also be of great interest for future studies to see whether the CO depletion causes a distinct trend of atmospheric C/O ratios in between early-formed ($\lesssim1$ Myr) and late-formed ($\gtrsim1$ Myr) planets.}
On the other hand, disk chemistry would not greatly affect the [O/H] trend of the \rev{I-SL population}, as chemical processing barely affects H$_2$O abundance---a major contributor to the metallicity of inner disk gas---except for $t\gtrsim5~{\rm Myr}$ \citep{Eistrup+18,Notsu+22}. 

\subsubsection*{Planetesimal accretion}
Although this study has focused on the pure pebble accretion scenario, in reality, pebble accretion would take place in tandem with planetesimal accretion. 
Planetesimal bombardment also deposit heavy elements to the envelope \citep[e.g.,][]{Pollack+86,Fortney+13,Venturini+16}.
Unlike pebbles that cannot accrete once planetary mass exceeds the pebble isolation mass \footnote{Some recent studies argued that pebbles with small Stokes number might accrete onto giant planets together with accreted disk gas even after the planet opens up a gas gap \citep{Morbidelli+23,Bitsch&Mah23}, though we have ignored this possible effect due to uncertainty in how much small pebbles can actually accrete onto the envelope.}, planetesimals can accrete on the envelope even after the onset of runaway gas accretion \citep[e.g.,][]{Shibata&Helled22}.
Several recent works implemented both pebbles and planetesimals in their models \citep{Danti+23,Penzlin+24}, although these studies relied on prescribed radial profiles or the formation rate of planetesimals.
The process of planetesimal formation remains one of the open questions in current planet formation theory, and different formation scenarios lead to distinct radial distributions of planetesimals \citep[for review, see][]{Drakowska+23}.
Further works would be required to understand how planetesimal formation and its interplay with pebbles accretion affect the final compositions of planetary atmospheres.

\subsubsection*{Simplifications for low-mass planets}
Although our model yields high atmospheric metallicity for low-mass ($M_{\rm p}<0.1M_{\rm J}$) exoplanets as suggested by JWST observations, we caveat that the metallicity simulated for such low-mass planets should be regarded as illustrative results, since we have greatly simplified the atmospheric formation for such low-mass planets.
While we switch on the atmospheric accretion only after the core mass exceeds the critical mass, a sub-critical mass core could gradually acquire surrounding disk gas through cooling contraction \citep[e.g.,][]{Ikoma&Hori12,Lee+14,Ginzburg+16,Lee&Chiang16}, \rev{although the rate of gas accretion would be relatively slow due to accretion luminosity from ongoing pebble accretion \citep[e.g.,][]{Ormel+21}.}

Another caveat is that it remains unclear to what extent the sub-Neptunes and super-Earths retain the imprint of planet formation process in their atmospheres. 
Several recent studies argued that the interaction between envelope and magma ocean (e.g., water production through reaction of H$_2$ and FeO in magma, \citealt{Ikoma&Genda06}) may play a major role in dictating atmospheric compositions of super-Earths and sub-Neptunes \citep[e.g.,][]{Kite+20,Seo+24,Ito+25,Heng+25}.
We anticipate that our synthetic low-mass planets with high atmospheric metallicity ([O/H]$\gtrsim1.5$) would be unaffected by magma-envelope interaction, as they typically form at distant orbits and retain icy cores (see Figure \ref{fig:mass_metal}).
On the other hand, our synthetic low-mass planets with low metallicity ([O/H]$\lesssim1.5$) form at close-in orbits and retain rocky cores, which may lead the magma-envelope interaction to alter atmospheric metallicity from those simulated by our model after disk dissipation.


\subsection{\rev{Caveat on the relative occurrence rate}}\label{sec:discussion_tauKH}
\begin{figure}[t]
{
\centering
\includegraphics[clip,width=\hsize]{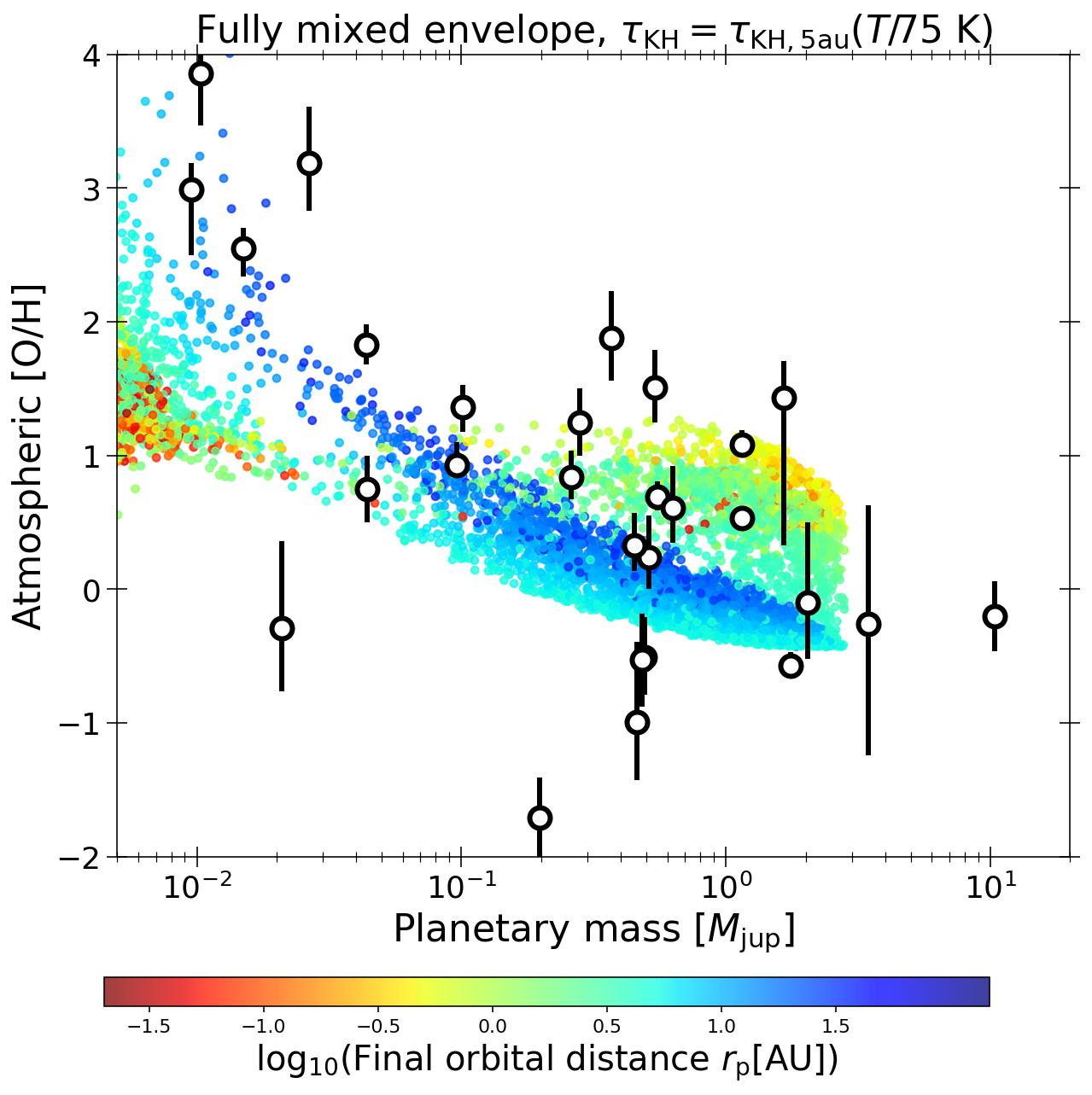}
\includegraphics[clip,width=\hsize]{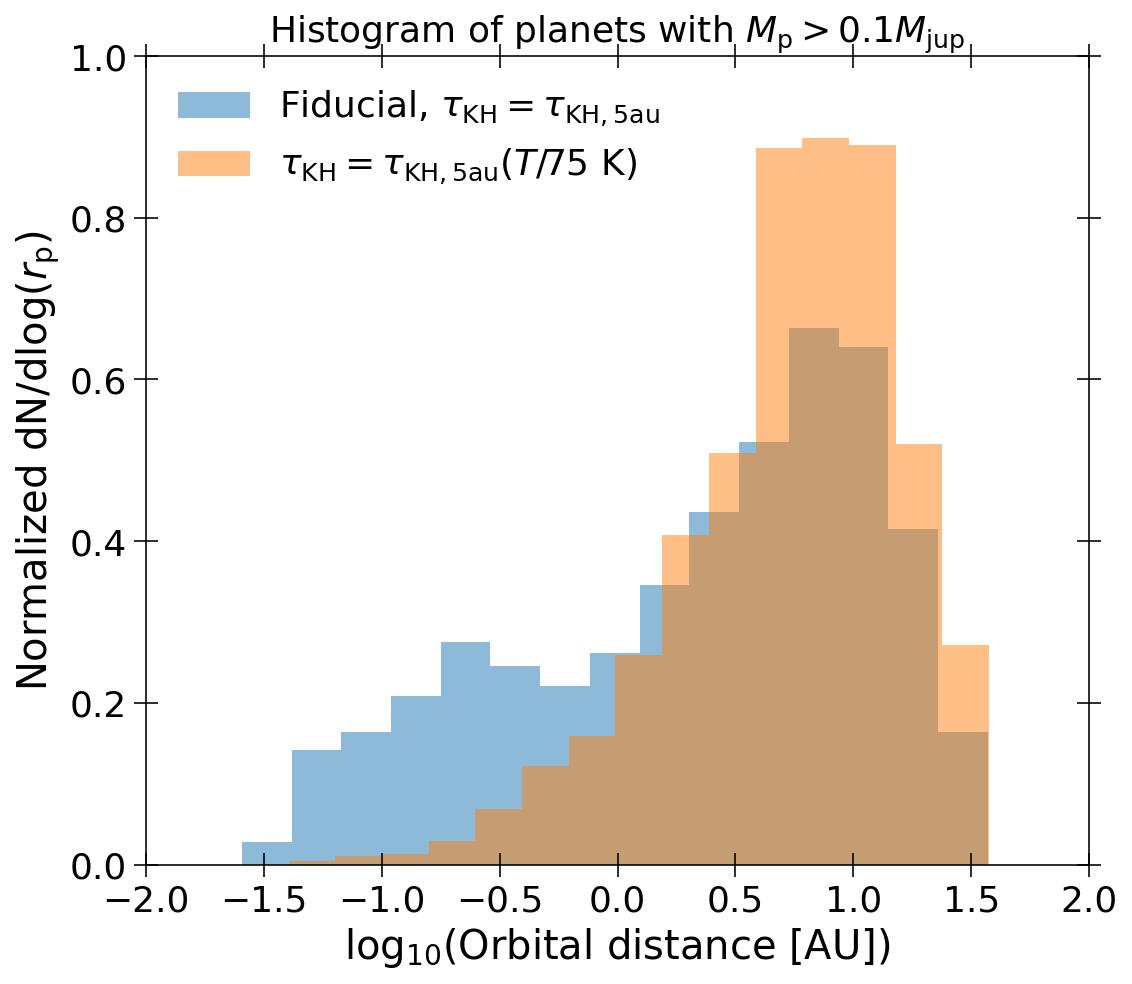}
}
\caption{
(Top) Same as Figure \ref{fig:mass_metal} but for population synthesis where the Kelvin-Helmholtz timescale is proportional to disk temperature (see Section \ref{sec:discussion_tauKH}). (Bottom) Normalized probability density of the final orbital distance of giant planets with $M_{\rm p}>0.1M_{\rm jup}$. The blue and orange histograms show the distributions found in our fiducial population synthesis and population synthesis with $\tau_{\rm KH}=\tau_{\rm KH,5au}(T/75~{\rm K})$, respectively.
}
\label{fig:lowKappa}
\end{figure}

\begin{figure*}[t]
{
\centering
\includegraphics[clip,width=\hsize]{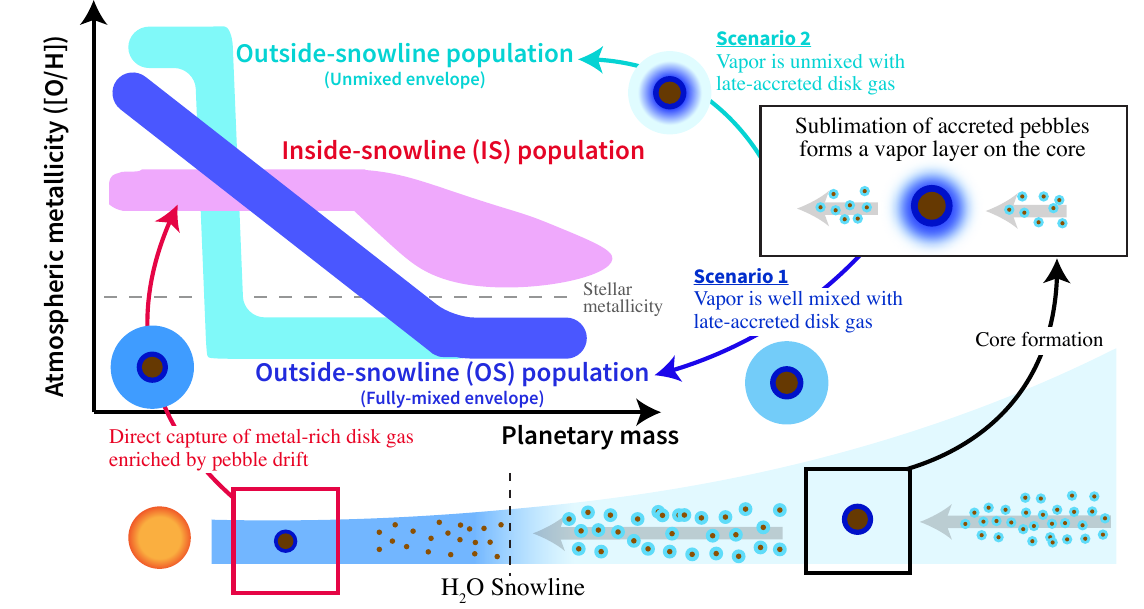}
}
\caption{
Cartoon summarizing the origin of three possible mass--metallicity relations. When planets ended up in close-in orbits inside the H$_2$O snowline (\rev{I-SL population}), atmospheric metals originates from vapor contained in disk gas. Metal enrichment of inner disk gas through pebble drift and sublimation leads to super-stellar metallicity for range of planetary mass. Gas giants in the \rev{I-SL population} exhibits shallow mass--metallicity anti-correlation because of opposite effects of turbulence strength on final planetary mass and disk gas metallicity. When planets ended up in distant orbits beyond H$_2$O snowline (\rev{O-SL population}), atmospheric metals originate from the vapors generated by sublimation of pebbles accreted on the cores during the core formation. If the envelope is fully convective and uniformly mixed (Scenario 1), the late-accreted disk gas gradually dilutes the pre-existing vapor layer on the core, yielding sharp mass--metallicity anti-correlation, as similar to the anti-correlation produced by planetesimals. 
If the envelope is fully unmixed due to inefficient convection (Scenario 2), the pre-existing vapor layer is isolated from upper observable atmosphere. Then, the late-accreted disk gas alone dictates atmospheric metallicity, yielding the swarm of planets with sub-stellar atmospheric metallicity.
Please see Section \ref{sec:result_population} for more in-depth explanations.
}
\label{fig:Cartoon_summary}
\end{figure*}
\rev{
We note a caution that our current model likely overestimates the gas accretion rate onto low-mass cores at close-in orbits, leading to an overestimate of the occurrence of gas giants in the I-SL population.
While our current model adopts the KH contraction timescale from \citet{Ikoma+00}, which is independent of local disk properties, several studies showed that the contraction is slower at inner hotter regions \citep[e.g.,][]{Ikoma&Hori12,Piso&Youdin14,Bitsch+15,Ginzburg+16,Coleman+17}.
The so-called envelope recycling that exchanges the high-entropy disk gases with low-entropy envelope, which is efficient at close-in orbits with short dynamical timescale, is also responsible for suppressing envelope contraction \citep[e.g.,][]{Ormel+15b,Cimerman+17,Lambrechts&Lega17,Lambrechts+19,Ali-Dib+20,Moldenhauer+22}, though how much it actually slows the envelope contraction depends on the penetration depth of the recycling flow and remains controversial \citep{2018MNRAS.479..635K,Savignac&Lee24,Bailey&Zhu24}. 
}

\rev{
To take a glimpse on how the suppressed envelope contraction impacts our results, we run the additional population synthesis by prescribing a disk temperature dependence of the Kelvin-Helmholtz timescale as
\begin{equation}\label{eq:tau_KH2}
    \tau_{\rm KH}=\tau_{\rm KH,5au}\left( \frac{T}{75~{\rm K}}\right)^{\beta},
\end{equation}
where $\tau_{\rm KH,5au}$ is the KH timescale at $5~{\rm au}$ (Equation \ref{eq:Mdot_KH}) obtained by \citet{Ikoma+00}, a normalization constant $75~{\rm K}$ comes from Equation \eqref{eq:Tirr_thick} for $5~{\rm au}$, and here we choose the disk temperature dependence of $\beta=1$.
Previous studies adopted weaker temperature dependence for envelope accretion in quasi-contraction phase \citep[$\beta\sim0.5$--$0.7$,][]{Bitsch+15,Poon+21}, whereas the dependence in runaway accretion phase remains uncertain.
We note that the actual temperature dependence of envelope accretion would be more complicated, especially once the effect of envelope recycling is taken into account.
Thus, Equation \eqref{eq:tau_KH2} should be regarded as a mere crude approximation to qualitatively account for the suppressed KH contraction at hot inner disks.
}

\rev{
We find that the emergence of the distinct mass-metallicity relations for I-SL and O-SL populations is robust to the temperature dependence of KH timescale, whereas the it does affect the relative occurrence of each population.
Figure \ref{fig:lowKappa} shows the simulated mass-metallicity relation for the temperature-dependent KH timescale. 
The two distinct mass-metallicity relations from the in-snowline and O-SL population are persistently present as found in fiducial simulation (Figure \ref{fig:mass_metal}), while the simulations yield less giant planets at close-in orbits.
The lower panel of Figure \ref{fig:lowKappa} shows the normalized orbital distribution of giant planets.
In our model, giant planets preferentially form at distant orbits due to efficient core growth through pebble accretion, whereas the temperature-dependence of KH timescale greatly reduces the relative occurrence of close-in giants compared to distant counterparts, as it prevents small cores migrated to disk inner edge from growing to gas giants.
An advanced envelope accretion model for low-mass cores at close-in orbits will be necessary to quantitatively assess the relative occurrence of the I-SL population, which is left to future studies.
}

\rev{
Although the runaway gas accretion at close-in orbits may be rare occurrence, it is still worth understanding what atmospheric metallicity trend we expect for giant planets formed inside H$_2$O snowline.
In fact, several previous studies argued the possibility of in-situ formation for close-in gas giants \citep{Ikoma+01,Batygin+16,Bailey&Batygin18,Poon+21}.
Conventionally, in-situ formation of giant planets was thought to be difficult because of low isolation mass for core formation \citep[e.g.,][]{Dawson+18}; however, a large core can first form at distant orbits and subsequently migrate to inner regions to serve as a giant planet core, which is also seen in our model (Figure \ref{fig:planet_evolve}).
If the planetesimal accretion is taken into account, it may even be possible to build up a large enough core at close-in orbits, since the pile-up of drifting pebbles potentially form $>10M_{\rm \oplus}$ planetesimals at $\lesssim1~{\rm au}$ in certain cases \citep{Drakowska+16,Ueda+21_DeadZone}.
}

\section{Summary}\label{sec:summary}

In this study, we have investigated the atmospheric metallicity trends of giant exoplanets.
To this end, we have developed a disk model that simulates the pebble radial drift and condensation/sublimation as well as the collision growth and fragmentation to calculate the time evolution of disk compositions.
We have coupled the disk model with a planet formation model that simulates the core formation through pebble accretion, atmospheric accretion onto the core, and orbital migration to calculate atmospheric elemental abundances.
We have conducted atmospheric population-synthesis to investigate possible trends seen in atmospheric metallicity represented by [O/H].
Our main findings are summarized as follows:

\begin{enumerate}
    \item Our model predicts that atmospheric metallicity shows a distinct correlation with planetary masses for the formation locations \rev{inside H$_2$O snowline} ($\lesssim1~{\rm au}$) and \rev{outside the snowline} ($\gtrsim1~{\rm au}$). 
    This is because the source of atmospheric metal is distinct: disk gas enriched by pebble drift controls the atmospheric metallicity for giant planets formed at close-in orbits, while vapors generated by pebbles during core formation turn out to be the main source of atmospheric metals for distant formation orbits (see also Figure \ref{fig:Cartoon_summary} for schematic illustration).
    \revrev{Although our model is incapable of predicting the relative occurrence of each population due to the simplified treatment of envelope accretion at close-in orbits (see Section \ref{sec:discussion_tauKH}), our results suggest that the mass--metallicity relation provides insight into whether the typical birthplace of extrasolar giant planets lies inside or outside the H$_2$O snowline.}

    \item The mass--metallicity correlation arising from distant formation orbits strongly depends on whether the envelope is fully mixed. 
    The fully mixed envelope yields an atmospheric metallicity being inversely proportional to planetary mass, while unmixed envelopes yield sub-stellar metallicity for wide range of planetary mass.     
    In contrast to the distant formation orbits, the mass--metallicity correlation arising from close-in orbits is insensitive to the mixing state of atmospheres.

    \item The offset of the mass--metallicity correlation for close-in orbits depends on the initial disk mass and characteristic radius. A higher initial disk mass leads to higher atmospheric metallicity for a given mass. A smaller disk radius also allows the correlation to extend to a higher mass. 
    
    \item Conversely, fragmentation threshold velocity has minor impacts on the mass--metallicity relation. This is because giant planets preferentially form at disks where pebble drift timescale is shorter than disk lifetime, which weakens the $v_{\rm frag}$ dependence of disk gas metallicity.


    \item The mass--metallicity relation of \rev{the inside-snowline population} can reasonably explain the hot and warm Jupiters with metal-rich atmospheres reported by JWST observations.
    Meanwhile, the presence of giant planets with metal-poor atmospheres possibly indicates that they formed at distant orbits, and convective mixing is not always efficient.
    
    \item Accretion of enriched disk gas alone cannot explain the high atmospheric metallicity of sub-Neptunes. 
    Vapor generation by efficient pebble accretion during core formation can partly explain the high metallicity, although further modeling effort is needed to quantitatively argue the origin of metal-rich atmospheres on sub-Neptunes.

    \item Accretion of enriched disk can explain the high {\it bulk} metallicity of warm Jupiters if they form close-in orbits, confirming the suggestion of \citet{Schneider&Bitsch21}. 
    For a given planetary mass, the bulk metallicity depends on the formation location and tends to decrease with increasing orbital distance.
    
    \item We also investigated the bulk-atmosphere correlation. Our synthetic planets with close-in final orbits exhibit a clear trend that follows the relation of $Z_{\rm atm}{\sim}Z_{\rm bulk}$, as most metals are deposited through the accretion of vapor-enriched disk gasses. Planets with distant final orbits show the correlation being off from $Z_{\rm atm}{\sim}Z_{\rm bulk}$ relation due to the sequestration of some metals into solid cores.

\end{enumerate}

\section*{Acknowledgements}
We thank an anonymous reviewer for many insightful comments that greatly improve this paper.
We also thank Saburo Howard for sharing the bulk metallicity data of \citet{Howard+25} with us.
We also thank Patricia Spalding, Xi Zhang, Akifumi Nakayama, Jonathan Fortney, Kanon Nakazawa, Tamami Okamoto, Yoshitaka Ikeda, and Sharon Xuesong Wang for fruitful discussions.
This work was supported by JSPS Overseas Research Fellowship and the JSPS KAKENHI Grant Number JP23K19072.

\appendix
\section{Recipe of Disk Chemical Composition}\label{sec:appendix_A}

Although this study has mostly focused on oxygen, we have implemented other elements.
Material properties of each chemical species are summarized in Table \ref{table:chem}.
We adopt the solar elemental abundance of \citet{Asplund+21} to calculate each molecular abundance.
In what follows, we introduce the adopted chemical species for each elemental reservoir.

We include refractory rock-forming elements of K, Na, Mg, Si, Fe, Ti, and V in our model.
We consider $\rm NaAlSi_3O_8$ and $\rm KAlSi_3O_8$, $\rm Mg_2SiO_4$, $\rm SiO$, Fe, $\rm TiO$, and $\rm VO$ as the main reservoirs of those rock-forming elements.
These molecules are the first major condensates predicted by the equilibrium condensation model \citep{Woitke+18}.
Although the main reservoir is predicted to be CaTiO$_3$ for Ti \citep{Lodders03,Woitke+18}, we instead assign TiO as a main reservoir of Ti for the sake of simplicity.

Since rock-forming molecules accommodate a fraction of oxygen, we define the oxygen abundance being available for volatile molecules as 
\begin{equation}
    {\rm (O/H)_{volat}} = {\rm (O/H)} -4f_{\rm Mg_2SiO_4}-f_{\rm SiO}-f_{\rm TiO}-f_{\rm VO},
\end{equation}
where we have ignored oxygen included in KAlSi$_3$O$_8$, NaAlSi$_3$O$_8$, and Ca$_5$[PO$_4$]$_3$F, as their abundances are much lower than O.
For reference, the solar elemental abundance of \citet{Asplund+21} yields the volatile oxygen abundance of $\rm(O/H)_{volat}=0.825(O/H)$.
Following \citep{Oberg&Wordsworth19}, we partition a half of the volatile oxygen into H$_2$O and another half into CO and CO$_2$. 
For carbon, refractory organics is suggested to accommodate $60\%$ of the bulk carbon budget \citep{Bergin+15}, which is subsequently converted to volatile carbons.
Since we mainly focus on an oxygen abundance in this study, CH$_4$ accommodates all remaining carbons not partitioned into CO and CO$_2$ for simplicity.
This treatment can be regarded as a limit of efficient decomposition of refractory carbons through photolysis at distant orbits.
For nitrogen, we partition 10\% of the volatile nitrogen into NH$_3$ and the reminder into N$_2$ following \citet{Oberg&Wordsworth19}.

\begin{table}[t]
  \caption{Summary of chemical species implemented in the present model.}\label{table:chem}
  \centering
  \begin{tabular}{l | r r} \hline
     Species & Density [$\rm g~cm^{-3}$] & Abundance $f_{\rm i}$\\ \hline \hline
    H$_2$O & 0.93 & $(1/2)\times{\rm (O/H)_{volat}}$  \\
    CO & 1.14& $(1/4)\times {\rm (O/H)_{volat}}$ \\
    CO$_2$ & 1.98& $(1/8)\times{\rm (O/H)_{volat}}$  \\
    CH$_4$ & 0.66& ${\rm(C/H)}-f_{\rm CO}-f_{\rm CO_2}$ \\
    NH$_3$ & 0.85& $0.1\times{\rm(N/H)_{\rm }}$ \\
    N$_2$ & 0.87& $0.9\times{\rm(N/H)_{\rm }}/2$ \\
    H$_2$S& 0.94& $0.1\times{\rm(S/H)}$ \\
    FeS &  4.83& $0.9\times{\rm(S/H)}$ \\
    KAlSi$_3$O$_8$ & 2.40& ${\rm(K/H)}$  \\
    NaAlSi$_3$O$_8$& 2.40& ${\rm(Na/H)}$ \\
    Fe$_3$P & 6.74 & $0.05\times{\rm(P/H)}$ \\
    Ca$_5$[PO$_4$]$_3$F & 3.19 & $0.95\times{\rm(P/H)/3}$ \\
    Mg$_2$SiO$_4$ & 3.21& ${\rm(Mg/H)}/2$  \\
    SiO &  2.18& ${\rm(Si/H)}-f_{\rm Mg2SiO4}$ \\
    Fe & 7.87& ${\rm(Fe/H)}$ \\
    VO & 5.76& ${\rm(V/H)}$ \\
    TiO & 4.95& ${\rm(Ti/H)}$ 
  \end{tabular}
\end{table}


Most of sulfur ($\gtrsim90$--$99\%$) is likely accommodated into refractory components, such as FeS, as has been suggested by observations of stellar photospheress \citep{Kama+19} and protoplanetary disks by the ALMA MAPS survey \citep{Le-Gal+21}.
The {\it Rosseta} mission to the comet 67P//Churyumov-Gerasimenko found that a small fraction of sulfur is present as a volatile forms, mainly as H$_2$S \citep{Rubin+19}.
Thus, we partition $90\%$ of total sulfur into FeS and remaining 10\% into H$_2$S, which is consistent with \citet{Kama+19} who suggested that $89\pm8\%$ of sulfur is partitioned into the refractory form.

Phosphorus is also likely mostly present as refractory rocks \citep{Krijt+22,Kama+25}. 
\citet{Bergner+22} recently used the ALMA to search for phosphorus molecules in low-mass star forming region and found that PO and PN are present in weakly shocked regions, where SO$_2$ is present, but absent at warmer regions where SiO is present, which indicates that the sublimation temperature of main phosphorus reservoirs is similar to that of the sulfur reservoir.
\citet{Bergner+22} suggested that apatite (Ca$_5$[PO$_4$]$_3$F) may be a main reservoir of phosphorus because of its sublimation temperature close to FeS.
We thus partition 95\% of phosphorus into Ca$_5$[PO$_4$]$_3$F and remaining 5\% into Fe$_3$P.
We have introduced Fe$_3$P because it is predicted by equilibrium condensation model \citep{Lodders03}, and solar abundance of fluorine is not enough to fully convert all phosphorus into Ca$_5$[PO$_4$]$_3$F. 

We note that pure Fe accommodates all Fe element in our model, although FeS and Fe$_3$P apparently accommodate a fraction of Fe.
This is because sublimation of these molecules do not yield Fe vapors; for example, sublimation of FeS proceeds with the following reaction \citep{Tachibana&Tsuchiyama98}
\begin{equation}
    {\rm FeS(s)+H_2(g)}\longrightarrow{\rm H_2S(g)+Fe(s)}.
\end{equation}
Thus, the sublimation of FeS only increases the gas-phase sulfur abundance without changing the gas-phase Fe abundance.
To avoid the overestimation of gas-phase Fe abundance, we accommodate all Fe element in pure Fe.
For calculating atmospheric Fe abundance (Section \ref{sec:method_calc_composition}), we do not count FeS and Fe$_3$P, which can effectively exist as vapors in our model.

Vapor pressure is taken from \citet{LichteneggerKomle91} for H$_2$O, \citet{Fray&Schmitt+09} for CO, NH$_3$, N$_2$, and H$_2$S, \citet{Yu+23} for CO$_2$ and CH$_4$, \citet{Woitke+18} for Fe, VO, TiO, \citet{Gail+13} for SiO, \citet{Nagahara+94} for Mg$_2$SiO$_4$.
When vapor pressure of pure substance is unavailable, we introduce a virtual vapor pressure given by
\begin{equation}
    P_{\rm sat,i}(T)=P_{\rm sat,H_2O}(T)\mathcal{H}(T-T_{\rm sub,i}),
\end{equation}
where $T_{\rm sub}$ is the sublimation temperature, and $\mathcal{H}(x)$ is the step function defined as $\mathcal{H}(x)=0$ for $x<0$ and $\mathcal{H}(x)=1$ for $x\ge0$.
We apply this approximation to FeS, $\rm KAlSi_3O_8$, $\rm NaAlSi_3O_8$, Fe$_3$P, and Ca$_5$[PO$_4$]$_3$F, where the sublimation temperature is taken from \citet{Lodders03}.

\begin{table*}[ht]
\caption{Properties of exoplanets for which JWST has constrained atmospheric metallicity.}\label{table:JWST}
\centering
\begin{threeparttable}
  \centering
  \begin{tabular}{l | l l l l} \hline
    Planet & $M_{\rm p}$ [$M_{\rm J}$] & [O/H]$_{\rm star}$& Atmospheric metallicity [M/H] & Bulk metallicity\\ \hline \hline
    GJ9827d & 0.0095 & $-0.29$ \tnote{a} & $+2.70^{+0.20}_{-0.49}$ \citep{Piaulet+24} & N/A \\
    GJ3090b & 0.010 & $-0.06$ \tnote{a} & $+3.80^{+0.44}_{-0.39}$ \citep{Ahrer+25a} & N/A \\
    TOI-270d & 0.015 & $-0.20$ \tnote{a} & $+2.35^{+0.15}_{-0.21}$ \citep{Benneke+24} & N/A \\
    TOI-421b & 0.021 & $-0.06$ & $-0.35^{+0.65}_{-0.47}$ \citep{Davenport+25} & N/A \\
    GJ1214b & 0.026 & $+0.29$ \tnote{a} & $+3.48^{+0.42}_{-0.36}$ \citep{Ohno+25} & N/A \\
    GJ3470b & 0.043 & $+0.27$ & $+2.1\pm0.15$ \citep{Beatty+24} &  N/A \\
    HIP67522b & 0.044 \tnote{b} & $+0.0$ \tnote{a} & $+0.5$--$1.0$ \citep{Thao+24} & $0.70^{+0.06}_{-0.03}$ \citep{Thao+24} \\
    WASP-107b & 0.096 & +0.16 & $+1.09^{+0.17}_{-0.07}$ \citep{Welbanks+24} & $0.37^{+0.05}_{-0.04}$ \citep{Sing+24}\\
    WASP-166b & 0.101 & $+0.21$ & $+1.57^{+0.17}_{-0.18}$ \citep{Mayo+25} & N/A\\ 
    HAT-P-18b  & 0.197 & $+0.31$ &  $-1.4\pm0.3$\citep{Fournier+24_HAT-P-18b} &$0.10^{+0.05}_{-0.06}$ \citep{Thorngren+16} \\
    WASP-69b  & 0.26 & $+0.12$ &  $+0.96^{+0.20}_{-0.17}$ \citep{Schlawin+24} &$0.09^{+0.05}_{-0.09}$ \citep{Thorngren+16} \\
    WASP-39b  & 0.28 & $+0.0$ & $+1.25^{+0.25}_{-0.25}$ \citep{Feinstein+23} & $0.19\pm0.04$ \citep{Fu+25}\\
    HD149026b & 0.37& $+0.21$ &$+2.09^{+0.35}_{-0.32}$ \citep{Bean+23} & $0.66\pm0.02$ \citep{Bean+23}\\
    WASP-94Ab  & 0.45 & $+0.25$ & $+0.58^{+0.24}_{\rm -0.19}$ \citep{Ahrer+25b} & $0.22\pm0.07$ \citep{ThorngrenFortney19}\\
    WASP-52b  & 0.46 & $+0.11$ & $-0.88^{+0.60}_{\rm -0.44}$ \citep{Fournier+24} & $0.23\pm0.02$ \citep{ThorngrenFortney19} \\
    WASP-96b  & 0.48 & $+0.24$ & $-0.29^{+0.35}_{-0.35}$ \citep{Taylor+23} & N/A\\
    TrES-4b  & 0.494 & +0.28\tnote{a} & $-0.23^{+0.30}_{\rm -0.28}$ \citep{Meech+25} & $0.19\pm0.07$ \citep{ThorngrenFortney19}  \\
    WASP-17b  & 0.512 & $+0.10$ & $+0.34^{+0.31}_{-0.24}$ \citep{Louie+24}  & $0.17\pm0.07$ \citep{ThorngrenFortney19} \\
    WASP-15b  & 0.54 & $-0.17$ & $+1.34^{+0.28}_{-0.26}$ \citep{Kirk+24} & $0.30\pm0.03$ \citep{Kirk+24}\\
    WASP-80b  & 0.55 & $-0.14$\tnote{a} & $+0.55^{+0.12}_{-0.10}$ \citep{Wiser+25} & $0.12^{+0.04}_{-0.05}$ \citep{Thorngren+16}\\
    HD209458b & 0.68 & $+0.02$ & $+0.63^{+0.31}_{-0.26}$\citep{Xue+23} & $0.16\pm0.02$ \citep{Fu+25}\\
    HD189733b & 1.12 & $-0.04$
 &$+0.49^{+0.05}_{-0.05}$ \citep{Fu+24} & $0.13\pm0.03$ \citep{Fu+25}\\
     WASP-121b & 1.15 & $+0.165$\tnote{c}
 &$+1.08^{+0.11}_{-0.08}$ \citep{Evans+25}\tnote{c} & $0.13\pm0.02$ \citep{ThorngrenFortney19}\\
     WASP-178b & 1.66 & $+0.04$ &$+1.47^{+0.28}_{-1.10}$ \citep{Lothringer+25} & N/A\\
    WASP-77Ab & 1.76 & $-0.04$ &$-0.61^{+0.10}_{-0.09}$ \citep{Smith+24} & $0.13\pm0.05$ \citep{ThorngrenFortney19}\\
    WASP-43b & 2.05 & $+0.30$ &$+0.20^{+0.60}_{-0.42}$ \citep{Yang+24} & $0.35\pm0.07$ \citep{ThorngrenFortney19}\\
    HAT-P-14b & 3.44 & $+0.18$ &$-0.08^{+0.89}_{-0.98}$ \citep{Liu+25} & $0.11\pm0.05$ \citep{ThorngrenFortney19}\\
    WASP-18b  & $10.4$& $+0.25$ &$+0.05^{+0.26}_{-0.25}$ \citep{Coulombe+23} & $0.13\pm0.05$ \citep{ThorngrenFortney19} 
  \end{tabular}
  \begin{tablenotes}
      \item[a] Stellar [O/H] is not available and assumed to be [Fe/H].
      \item[b] Unlike other planets, the mass of HIP67522b is obtained from the atmospheric model fit to its transmission spectrum \citep{Thao+24}. 
      \item[c] Stellar [O/H] of WASP-121 is taken from \citet{Evans+25}. Atmospheric metallicity of WASP-121b listed here is calculated relative to stellar O/H in \citet{Evans+25}.
  \end{tablenotes}
\end{threeparttable}
\end{table*}

\section{Summary of JWST Observations}\label{appendix_JWST}

We select planets with published atmospheric metallicity constrained by JWST observations, as summarized in Table \ref{table:JWST}.
We omit several giant planets from our sample: TOI-5205b \citep{Canas+25} and LTT9779b \citep{Radica+24,Coulombe+25} due to orders-of-magnitude uncertainty remaining in atmospheric metallicity, K2-18b \citep{Madhusudhan+23} due to no compelling evidence for the detection of O-bearing species \citep{Schmidt+25}, and GJ436b because only a lower limit could be derived for atmospheric metallicity \citep[$\ge80\times$ solar value,][]{Mukherjee+25}.
For GJ9827d, for which JWST-NIRISS observation suggests a possible steam atmosphere \citep{Piaulet+24}, we calculate atmospheric [O/H] value from solar O/H of \citet{Asplund+21} and atmospheric O/H value by number in \citet{Piaulet+24}.
For atmospheric metallicity, we quote the value derived by physics-based methods, such as the chemically-consistent retrieval, as long as it is available.
If the physics-based approach is not available, we approximately estimate the metallicity from the H$_2$O abundance constrained by free chemistry retrieval, as ${\rm[O/H]}\approx \log_{\rm 10}{({\rm q_{\rm H_2O}})}-\log_{\rm 10}{({\rm q_{\rm H_2O,solar}})}$.
Stellar [O/H] is taken from the HYPATIA catalog linked to the NASA Exoplanet Archive.
For some stars, mainly for M-dwarfs, we could not find stellar [O/H] and hence assume [O/H]$\approx$[Fe/H].

We note a caution that our current dataset was derived by a diverse modeling setup with different model assumptions. 
As argued in \citet{Kawahara+24}, different retrieval frameworks can yield different inferences for atmospheric properties.
The actual uncertainty of atmospheric metallicity may be larger than that reported by literature.
We also note caveats that emission spectroscopy with narrow wavelength coverage can involve a degeneracy between atmospheric metallicity and temperature gradient \citep{Gagnebin+24}, and the application of atmospheric models with uniform terminator can introduce biased inference on atmospheric metallicity if morning and evening terminators have distinct aerosol properties \citep{Line&Parmentier16,Mukherjee+25b}.
In addition, we caveat that the definitions of ``atmospheric metallicity [M/H]'' in literature sometimes have different definition, such as [M/H]=[O/H] or [M/H]=[(C+O)/H].
In this study, we simply assume that the literature metallicity [M/H] is approximately equivalent to [O/H] since the planets currently observed by JWST tend to exhibit low C/O ratio \citep{Kempton&Knutson24,Wiser+25}.
While we anticipate that qualitative comparison is still possible for the current dataset, a uniform atmospheric inference, as done for atmospheric observations with HST \citep[e.g.,][]{Welbanks+19,Changeat+22,Edwards+22}, will be warranted to make a quantitative comparison between atmospheric trends and those predicted by planet formation models.

\begin{figure*}[t]
\includegraphics[clip,width=\hsize]{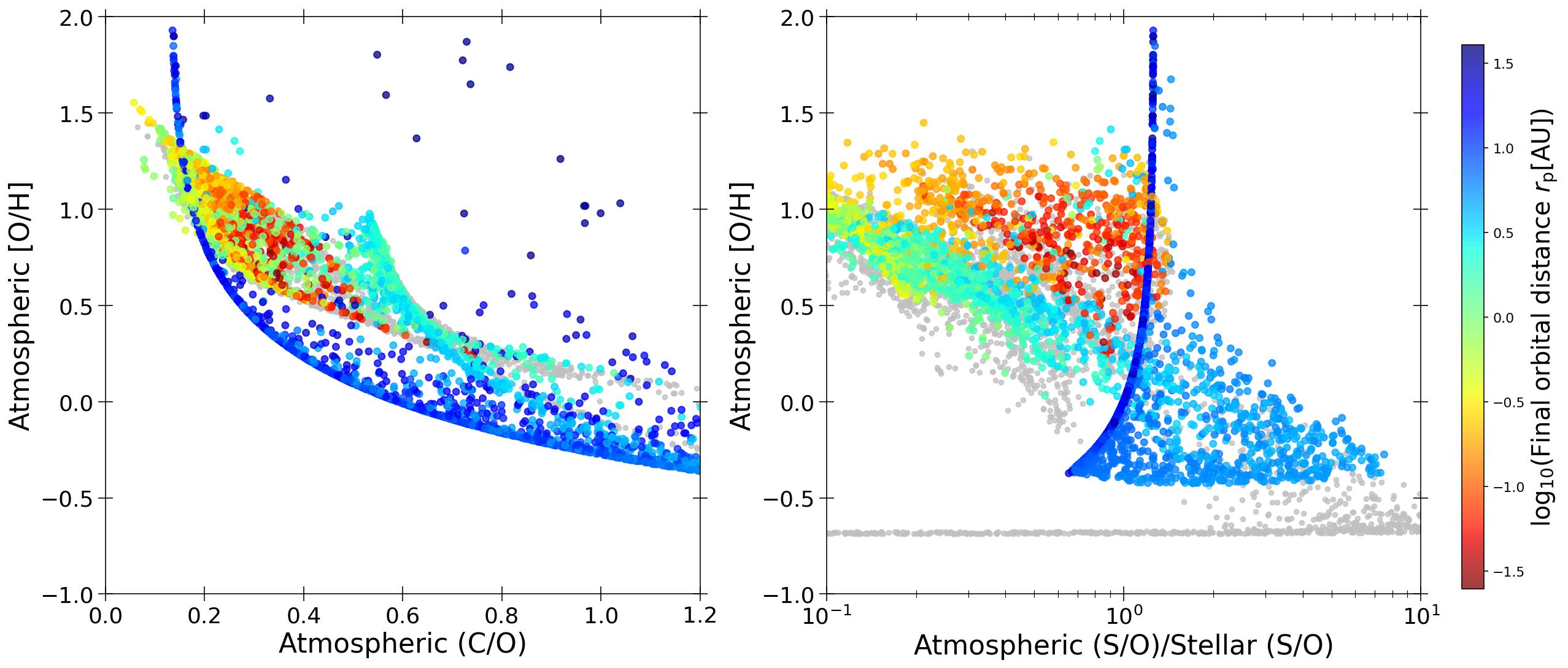}
\caption{
(Left) Radial profile of S/O ratio in disk gas (bluish lines) and dust (black lines), where the ratio is normalized by the stellar value. From thin to thick, each colored line shows the profile for $t=0.1$, $0.3$, $1.0$, $3.0~{\rm Myr}$, respectively. Here we adopt $\alpha={10}^{-3}$, $v_{\rm frag}=1~{\rm m~s^{-1}}$, $r_{\rm c}=100~{\rm au}$. (Right) Relation between the atmospheric [O/H] and S/O ratio in our fiducial population synthesis model ($M_{\rm disk}/M_{\rm *}=0.1$, $r_{\rm c}=148~{\rm au}$, $v_{\rm frag}=1~{\rm m~s^{-1}}$) for fully mixed atmospheres. The color of each dot denotes the final orbital distance of the planets when the gas disk is gone. The gray dots also show the relation for the unmixed atmospheres.
}
\label{fig:StoO}
\end{figure*}

\section{Rrelation between C/O and O/H}\label{Appendix:CtoO_OtoH}
The left panel of Figure \ref{fig:StoO} shows the relation between atmospheric O/H and C/O in our simulation shown in Figure \ref{fig:mass_metal}.
In our model, planets with higher atmospheric metallicity tend to show lower C/O ratio, which is in line with \citet{Penzlin+24}.
The anti-correlation between O/H and C/O appears to be in line with the current JWST observations shown in \citet{Wiser+25}.
However, we defer detailed model-data comparisons to future studies, as the C/O trend can be dependent of several uncertainties such as the abundance of refractory organics and their decomposition processes.

\section{Relation of S/O and O/H}\label{Appendix_sulfur}

The right panel of Figure \ref{fig:StoO} shows the relation between S/O and O/H ratios. 
Planets with different final orbits can be placed in a S/O--O/H phase space.
Planets with close-in final orbits ($\lesssim0.3$ au) tend to have stellar-to-substelar S/O along with super-stellar metallicity of [O/H]$\sim0.5$--$1.5$, as FeS sublimation leads to a nearly stellar S/O ratio in disk gas.
Beyond the FeS sublimation line but inside H$_2$O snowline ($\sim0.3$--$3$ au), planets tend to have highly substellar S/O ratio with keeping high O/H, as most sulfur is locked in FeS rock while disk gas is still enriched in H$_2$O vapor.
For the orbit beyond the H$_2$O snowline ($\sim3$--$10$ au), planets tend to acquire atmospheres with super-stellar S/O ratios with low atmospheric [O/H] due to the depletion of H$_2$O and CO$_2$ vapors. 
For distant orbits where all H$_2$O, CO$_2$, and H$_2$S are in ice phase ($>10$ au), atmospheric O and S are inherited from the sublimated pebbles during the core formation, making near stellar S/O ratio with various atmospheric metallicity.
Atmospheres have slightly super-solar S/O ratio as long as the final orbit is inside the CO snowline, where some fraction of oxygen is still in gas phase.
\bibliographystyle{aa}
\bibliography{references}
\end{document}